%% file: arxiv_correlated_source_journal_for_ERC.tex
\documentclass[12pt,journal,draftclsnofoot,onecolumn,english]{IEEEtran}

\usepackage{comment}

\usepackage{epsfig}
\usepackage{times}
\usepackage{float}
\usepackage{afterpage}
\usepackage{amsmath}
\usepackage{amstext,cite}
\usepackage{amssymb,bm}
\usepackage{latexsym}
\usepackage{color}
\usepackage{graphicx}
\usepackage{amsmath}
\usepackage{amsthm}
\usepackage{graphicx}
\usepackage[center]{caption}
\usepackage{pstricks}
\usepackage{caption}
\usepackage{subcaption}
\usepackage{booktabs}
\usepackage{multicol}
\usepackage{lipsum}% dummy text
\usepackage[T1]{fontenc}
\usepackage{hyperref}
\usepackage{aecompl}
\usepackage{bbm}

\usepackage{empheq}

\allowdisplaybreaks

\input{macros.tex}

\begin{document}

\title{On the Fundamental Limits of Coded Caching with Correlated Files}

\author{
Kai~Wan,~\IEEEmembership{Member,~IEEE,} 
Daniela Tuninetti,~\IEEEmembership{Fellow,~IEEE,}
Mingyue~Ji,~\IEEEmembership{Member,~IEEE,}
and~Giuseppe Caire,~\IEEEmembership{Fellow,~IEEE}
\thanks{
A short version of this paper   was presented at the 2019 IEEE International Symposium on Information Theory. 
}
\thanks{
K.~Wan and G.~Caire are with the Electrical Engineering and Computer
Science Department, Technische Universit\"at Berlin, 10587 Berlin, Germany
(e-mail:  kai.wan@tu-berlin.de; caire@tu-berlin.de). The work of K.~Wan and G.~Caire was partially funded by the European Research Council under the ERC Advanced Grant N. 789190, CARENET.}
\thanks{
D.~Tuninetti is with the Electrical and Computer Engineering Department, University of Illinois at Chicago, Chicago, IL 60607, USA (e-mail: danielat@uic.edu). The work of D.~Tuninetti was supported in part by NSF Award 1527059.}
\thanks{
M.~Ji is with the Electrical and Computer Engineering Department, University of Utah, Salt Lake City, UT 84112, USA (e-mail: mingyue.ji@utah.edu). The work of M.~Ji was supported in part by NSF Awards 1817154 and 1824558.}
}
\maketitle

\begin{abstract}
This paper studies the fundamental limits of the shared-link coded caching problem with correlated files, where a server with a library of $\Nsf$ files communicates with $\Ksf$ users who can locally cache $\Msf$ files. 
Given an integer $\rsf\in[\Nsf]$, correlation is modeled as follows: each $\rsf$-subset of files contains a unique common block. The tradeoff between the cache size and the average transmitted load %over any demand type 
is considered. 
First, a converse bound under the constraint of uncoded cache placement (i.e., each user directly stores a subset of the library bits) is derived.
Then, a caching scheme for the case where every user demands a distinct file (possible for $\Nsf\geq\Ksf$) is shown to be optimal under the constraint of uncoded cache placement. 
This caching scheme is further proved to be decodable and optimal under the constraint of uncoded cache placement when (i) $\Ksf\rsf\Msf\leq 2\Nsf$ or $\Ksf\rsf\Msf\geq (\Ksf-1)\Nsf$ or $\rsf\in \{1,2,\Nsf-1,\Nsf\}$ for every demand type (i.e., when the demanded file are not necessarily distinct),  and (ii) when the number of distinct demanded files is no larger than four. 
Finally, a two-phase delivery scheme based on interference alignment is shown to be optimal to within a factor of $2$ under the constraint of uncoded cache placement for every possible demands. 
As a by-product, the proposed interference alignment scheme is shown to reduce the (worst-case or average) load of state-of-the-art schemes for the coded caching problem where the users can request multiple files.
\end{abstract}

\section{Introduction}
\label{sec:intro}

Cache is a network component that leverages the device memory to transparently store data so that future requests for that data can be served faster. Two phases are included in a caching system: i) cache placement phase: content is pushed into each cache without knowledge of future demands; ii) delivery phase: after each user has made its request and according to the cache contents, the server transmits coded packets in order to satisfy the user demands. The goal is to minimize the number of transmitted bits (or load or rate).

Information theoretic coded caching was originally proposed by Maddah-Ali and Niesen (MAN) in~\cite{dvbt2fundamental} for a shared-link caching systems containing a server with a library of $\Nsf$ equal-length files, which is connected to  $\Ksf$ users through a noiseless shared-link, each of which can store $\Msf$ files in their local cache. Each user demands one file in the delivery phase. The MAN scheme uses a combinatorial design in the placement phase such that during delivery multicast messages simultaneously satisfy the demands of different users. 
Under the constraint of uncoded cache placement (i.e., each user directly caches a subset of the library bits) and for worst-case load, the MAN scheme was proved to be optimal when $\Nsf\geq \Ksf$~\cite{ontheoptimality}.
On the observation that some MAN linear combinations are redundant if there exist files  demanded by several users,  the authors in~\cite{exactrateuncoded} improved the MAN delivery scheme and achieved the optimal worst-case load  under the constraint of uncoded cache placement for any $\Ksf$. The same authors proved in~\cite{yas2} that the multiplicative gap between the optimal caching scheme with uncoded cache placement and any caching scheme with coded cache placement is at most $2$. 

Coded caching strategy was also extended to numerous different models, such as decentralized systems~\cite{decentralizedcoded}, device-to-device (D2D) systems~\cite{d2dcaching}, topological networks~\cite{hierarchicalcaching,cachingJi2015,wan2017combination}, etc. 
The above works aassume that all the files in the library are independent. However, in practice there may be some overlaps among different files (e.g., videos, image streams, etc.). 
In this work, we consider such a coded caching problem with {\it correlated sources}, as originally proposed in~\cite{has2016correctedsource}, where different files have common parts.
In the following, we will review  the literature of coded caching with correlated sources, and introduce our main contributions in this paper.

\subsection{Past Work}

\paragraph*{Coded Caching with Correlated Sources}
In\c~ite{has2016correctedsource}, the authors modeled correlation as each subset of files has an exclusively common part, which is independent to the common parts of other subsets of files. 
By treating the delivery phase as an index coding problem with multiple requests, the authors in~\cite{has2016correctedsource} proposed a delivery scheme based on graph coloring.  In~\cite{ratecorrectedhass2018} a  caching scheme for two-file $\Ksf-$user system and three-file two-user systems based on Gray-Wyner source coding. 

In~\cite{yang2018centralizedcorrected}, the caching problem with correlated files, where the length of the common part among each $\ell\in \{1,\dots,\Nsf\}$ files (referred to as a `$\ell$-block') is the same, was considered;   each file contains $\binom{\Nsf-1}{\ell-1}$ $\ell-$blocks.
By using the MAN cache placement to store each $\ell$-block at the user sides,~\cite{yang2018centralizedcorrected}  proposed a delivery phase which contains $\Nsf$ steps. In step~$\ell$ only $\ell$-blocks are transmitted; thus there are  $\binom{\Nsf-1}{\ell-1}$ rounds for  the transmission of Step~$\ell$. Each round is treated as an MAN caching problem with $\Ksf$ users, each of which should decode exactly one $\ell$-block. The authors then used the caching scheme in~\cite{exactrateuncoded} to transmit packets for each round. 
The caching schemes in~\cite{ratecorrectedhass2018} and~\cite{yang2018centralizedcorrected}, were extended in~\cite{has2018dynamiccorrectedsource} and~\cite{yang2018correctedbroadcast} to  caching problems with 
correlated files where the correlation is dynamic and the channel is a Gaussian broadcast channel, respectively.

\paragraph*{Coded Caching with Multiple Requests}
The caching problem with correlated files in~\cite{yang2018centralizedcorrected} is a special case of coded caching with multiple requests considered in~\cite{ji2015multirequest}, where the library contains $\Nsf$ equal-length and independent files and  each user demands $\Lsf$ files from the library. 
With the MAN placement,  to divide   the delivery phase into $\Lsf$ rounds where in each round the MAN scheme in~\cite{dvbt2fundamental} is used to let each user decode one file, can achieve a generally order optimal worst-case load to within a factor  of $18$~\cite{ji2015multirequest}. By further tightening the converse bound, this order optimality factor was reduced to $11$ in~\cite{Sengupta2017multirequest}.
Instead of using the MAN scheme in each round, the authors in~\cite{multireqWei2017} proposed to use the caching scheme in~\cite{exactrateuncoded} to leverage the multicast opportunities. 

In addition, by considering all the $\Lsf$ rounds, an overall transmission coding matrix can be generated. If the coding matrix is not full-rank, they then take the full-rank sub-matrix. This delivery scheme was proved to be optimal under the constraint of the MAN placement for demands with $\Ksf \leq 4$, $\Msf=\Nsf/\Ksf$, and $\Lsf=2$, with the exception of one demand for $\Ksf=3$ and three demands for $\Ksf=4$. 

Coded caching with multiple requests, where each user demands different number of files, was considered in~\cite{parrinello2018sharedcache,multishared2018Karat,franarxiv,xu2018codedsharedcache}. The caching schemes in~\cite{parrinello2018sharedcache,multishared2018Karat,franarxiv} are based on the round-division strategy as described above while the one in~\cite{xu2018codedsharedcache} considered small memory size regime and used Minimum Distance Separable (MDS) coded cache placement. %Since coded caching with correlated sources is a not special case of coded caching with multiple requests where each user demands different number of files, in the rest of this paper,   when we consider caching problem with multiple requests, the number of demanded files by each user is identical.

Most of the existing works divide the multi-request problem into a sequence of single-request problems (as in~\cite{yang2018centralizedcorrected,ji2015multirequest,Sengupta2017multirequest,multireqWei2017}).
There are three main limitations in dividing the delivery into single-request problems, namley
(1) the same file may exist in different rounds and this round-division method may miss some multicast opportunities, 
(2) even if there does not exist file overlap cross different rounds, this round-division method still cannot fully leverage the multicast opportunities (as illustrated in Example~\ref{ex:scheme 1}), and
(3) %since there are lots of possibilities to divide users' demands into $\Lsf$ round, it is of high 
finding the best division of the users' demands into $\Lsf$ groups is computationally hard. %by greedy search.

\subsection{Contributions}
If one directly considers the most general problem of correlated files, it is very challenging to make general optimality statements. In this paper, we consider a symmetric version of the problem, for which we propose a novel interference alignment based delivery scheme, which jointly serves users' demands instead of dividing the delivery into single-request problems.
The considered model is the following simplification of the model in~\cite{yang2018centralizedcorrected}: we fix $\rsf\in [\Nsf]$ and assume each file only contains $\rsf$-blocks (see Section~\ref{sec:model}). 
Our main contributions are as follows:
\begin{enumerate}

\item We derive a converse bound on the minimal average load among all possible demands under the constraint of uncoded cache placement. We leverage the acyclic index coding converse bound as in~\cite{onthecapacityindex,ourisitinnerbound}.

\item By jointly serving the users' demands, we propose a caching scheme for the case where every user demands a distinct file and whose load matches our proposed converse bound under the constraint of uncoded cache placement.
 
\item By combining the above achievable scheme with an interference alignment idea, we then propose a two-phase delivery scheme for general demands, where the first sub-phase is the same as the one for distinct demand case and the additional second sub-phase is used to align interference at the various users. The two-phase interference alignment delivery scheme is proved to be order optimal within a factor of $2$ for any demand type.

\item By further cancelling interference, we prove that the second sub-phase in the above two-phase delivery is not necessary, thus resulting in exact optimality under the constraint of uncoded cache placement, for
(i) for any demand type if either $\Ksf\rsf\Msf\leq 2\Nsf$ or $\Ksf\rsf\Msf\geq (\Ksf-1)\Nsf$ or $\rsf \in \{1,2,\Nsf-1,\Nsf\}$, and  (ii) when the number of distinct demanded files is no larger than four.  

\item As a by-product,  a modification of  our proposed interference alignment scheme is optimal under the constraint of MAN placement %\dt{?IS MAN placement CRITICAL?} 
   for the four cases left open in~\cite{multireqWei2017} for the  caching problem with multiple requests. % with $\Lsf=2$, $\Ksf \leq 4$, $\Msf=\Nsf/\Ksf$, 
\end{enumerate}

\subsection{Paper Organization}
The rest of the paper is organized as follows.
The system model for the considered  coded caching problem with correlated sources is given in Section~\ref{sec:model}.
In Section~\ref{sec:main}, our main results and some numerical evaluations are presented. 
The proof of the proposed converse bound can be found in Section~\ref{sec:proof of converse bound}, and that of proposed achievable schemes in Section~\ref{sec:achievable scheme}. Section~\ref{sec:conclusion} concludes the paper.  
The proofs of some auxiliary results can be found in Appendix.

\subsection{Notation Convention}
%We use the following notation convention.
Calligraphic symbols denote sets, 
bold symbols denote vectors, 
and sans-serif symbols denote system parameters.
We use $|\cdot|$ to represent the cardinality of a set or the length of a vector;
$[a:b]:=\left\{ a,a+1,\ldots,b\right\}$ and $[n] := [1,2,\ldots,n]$; 
$\oplus$ represents bit-wise XOR.
% $\mathbbm{1}_{\textrm{event}}$ is the indicator function, where   $\mathbbm{1}_{\textrm{event}}=1$ if $\textrm{event}$ is true and  $\mathbbm{1}_{\textrm{event}}=0$ otherwise.

%$i^{\textrm{th}}$ line is $1$ and  $j^{\textrm{th}}$ line is $0$ for all $j\in [n]\setminus\{i\}$. For example, $\ev_{3,1}=[1;0;0]$ , $\ev_{3,2}=[0;1;0]$,  and $\ev_{3,3}=[0;0;1]$.

%We let $\binom{x}{y}=0$ if $x<0$ or $y<0$ or $x<y$.  We use the same convention as that in the literature when it comes to `summing' sets.

\section{System Model}
\label{sec:model}

In a $(\Nsf,\Ksf,\rsf,\Msf)$ shared-link caching problem with correlated files, a server has access to a library of $\Nsf\in\mathbb{N}$ files (each of which contains $\Bsf\in\mathbb{N}$ iid bits) denoted by $\{F_1, \cdots, F_\Nsf\}$. The server is connected to $\Ksf\in\mathbb{N}$ users through a shared error-free link. Each file is composed of $\binom{\Nsf-1}{\rsf-1}$ independent and equal-length blocks, where $\rsf\in[\Nsf]$; we denote 
\begin{align}
F_i=\{W_{\Sc}:\Sc\subseteq [\Nsf], |\Sc|=\rsf, i\in \Sc \}, \  \forall i\in [\Nsf],
\end{align}
where the block $W_{\Sc}$ represents the exclusive common part across the files indexed by $\Sc$. Hence, in the whole library there are $\binom{\Nsf}{\rsf}$ independent blocks, each of which has $\Bsf/\binom{\Nsf-1}{\rsf-1}$ bits. A coded caching scheme has two phases: placement and delivery.

\paragraph*{Placement Phase}
During the cache placement phase, user $k\in[\Ksf]$ stores information about the $\Nsf$ files in its cache of size $\mathsf{MB}$ bits, where $\Msf \in[0,\Nsf/\rsf]$.  This phase is done without knowledge of users' demands. 
We denote the content in the cache of user $k\in[\Ksf]$ by $Z_{k}$ and let $\Zm:=(Z_{1},\ldots,Z_{\Ksf})$. 

\paragraph*{Delivery Phase}
During the delivery phase, user $k\in[\Ksf]$ demands file $d_{k}\in[\Nsf]$.
The demand vector $\dv:=(d_{1},\ldots,d_{\Ksf})$ is revealed to all nodes. 
Given $(\dv,\Zm)$, the server broadcasts a message $X(\dv,\Zm)$
of $\Bsf \Rsf(\dv,\Zm)$ bits to all users.
User $k\in[\Ksf]$ must recover its desired file $F_{d_{k}}$ from $Z_{k}$ and $X(\dv,\Zm)$.

\paragraph*{Load}
For each demand vector $\dv$, we define $\Nc(\Tc):=\{d_k:k\in \Tc\}$ as the set of demanded files by users in $\Tc$, where $\Tc \subseteq [\Ksf]$.
A demand vector $\dv$ is said to be of type $\mathcal{D}_{N_{\textup{e}}(\dv)}$ if it has $N_{\textup{e}}(\dv):=|\Nc([\Ksf])|$ distinct entries.
Based on the uniform demand distribution, the objective is %under the constraint of uncoded cache placement 
to determine the optimal average load among all demands of the same type, that is
%$\mathcal{D}_{s}$ where $s\in[\min \{\Ksf,\Nsf\}]$,
\begin{align}%{rCl}
\Rsf^{\star}(\Msf,s)%_{\mathrm{u}}
:=
\min_{\substack{\Zm}} \ \mathbb{E}_{\dv \in \mathcal{D}_{s}}[ \Rsf(\dv ,\Zm)],
\ \forall s\in[\min \{\Ksf,\Nsf\}],
\label{eq:Ropt(M,s)def}
\end{align}
and the optimal average load among all possible demands is
\begin{align}%{rCl}
\Rsf^{\star}(\Msf)%_\mathrm{u}
:=
\min_{\substack{\Zm}} \ \mathbb{E}_{\dv\in[\Nsf]^{\Ksf}}[\Rsf(\dv ,\Zm)].
\label{eq:Ropt(M)def}
\end{align}
%\dt{
%Notice that the cache placement policies that attain the optimal loads in~\eqref{eq:Ropt(M,s)def} and~\eqref{eq:Ropt(M)def} may be different, and that in general 
%\begin{align}
%\max_{s\in[\min \{\Ksf,\Nsf\}]} \Rsf^{\star}(\Msf,s) %\mathbb{E}_{\dv \in \mathcal{D}_{s}}[ \Rsf(\dv ,\Zm_s^{\star})]
%\leq \Rsf^{\star}(\Msf) \leq \min_{s\in[\min \{\Ksf,\Nsf\}]} \mathbb{E}_{\dv\in[\Nsf]^{\Ksf}}[\Rsf(\dv ,\Zm_s^{\star})],
%\label{eq:Ropt(M) vs Ropt(M,s)}
%\end{align}
%where $\Zm_s^{\star}$ is the policy that attains the optimal loads in~\eqref{eq:Ropt(M,s)def}.
%ARE THE BOUNDS IN \eqref{eq:Ropt(M) vs Ropt(M,s)} ANY GOOD?
%} 
 Note that, with an abuse of notation, $\Rsf^{\star}(\Msf) \not= \mathbb{E}_s[\Rsf^{\star}(\Msf,s)]$ in general, unless the same cache placement policy optimizes the load in~\eqref{eq:Ropt(M,s)def} for all $s\in[\min \{\Ksf,\Nsf\}]$.

\paragraph*{Uncoded Cache Placement}
The cache placement policy is said to be {\em uncoded} if each user directly copies some library bits directly into its cache. Under the constraint of uncoded cache placement, we can partition each block $W_{\Sc}$ is partitioned into sub-blocks as 
\begin{align}
W_{\Sc}=\{W_{\Sc,\Vc}:\Vc \subseteq [\Ksf]\}, \ \forall \Sc\subseteq [\Nsf] : |\Sc|=\rsf,
\end{align}
where $W_{\Sc,\Vc}$ represents the bits of $W_{\Sc}$ which are exclusively cached by users indexed by $\Vc$. 
The optimal loads under the constraint of uncoded cache placement are denoted by $\Rsf^{\star}_{\mathrm{u}}(\Msf,s)$ and $\Rsf^{\star}_{\mathrm{u}}(\Msf)$ and are defined similarly to in~\eqref{eq:Ropt(M,s)def} and~\eqref{eq:Ropt(M)def}, respectively. %and satisfy a relationship as in~\eqref{eq:Ropt(M) vs Ropt(M,s)}.

\paragraph*{Special Cases}
Our model reduces to the MAN coded caching problem with average load when $\rsf=1$, and to the case of a library with a single file when $\rsf=\Nsf$. Both cases are either solved exactly or to within a factor of $2$ in~\cite{yas2}.

\paragraph*{Relation to the More General Coded Caching Problem with Correlated Sources}%General Problem Formulation of in~\cite{yang2018centralizedcorrected}
In this paper, in order to make fundamental progress on the problem of caching correlated content, we simplified the model~\cite{yang2018centralizedcorrected} as follows. In~\cite{yang2018centralizedcorrected} a certain parameters $\ell$ ranges from zero to the number of files in the system (each $\ell_1$ files have a common part, each $\ell_2$ files also have a common part, etc.), %the block size is identical, 
while in our model $\ell$ is fixed to a single value $\rsf$.
Our model is thus a special case of the one in~\cite{yang2018centralizedcorrected}. 
With our models however, we can make conclusive statements (either exact capacity results, or capacity to within a constant multiplicative gap) which eluded in~\cite{yang2018centralizedcorrected}.

\paragraph*{Relation to the Coded Caching Problem with Multiple Requests}
If we identify the ${\Nsf \choose \rsf}$ independent blocks as files of a library, and allow each cache-equipped user to request ${\Nsf - 1 \choose \rsf - 1}$ such blocks/files, the considered caching problem with correlated sources relates to the {\it symmetric} caching problem with multiple requests considered in~\cite{ji2015multirequest}, where `symmetric' means that each user requests the same number of files. There is however a subtle difference between our model and the one in~\cite{ji2015multirequest}: in our model one file corresponds to ${\Nsf-1 \choose \rsf-1}$ {\it distinct} blocks, thus our model corresponds to the one in~\cite{ji2015multirequest} under the constraint that a user has multiple but distinct requests. Moreover, here we consider the average load as our performance metric, while in~\cite{ji2015multirequest} the authors used the worst-case load. Because of these differences, the results in this paper are not special cases of the results in~\cite{ji2015multirequest}.

 The relationship among the two problems can be also explained as follows. 
For the case of multiple requests, assume that the ${\Nsf \choose \rsf}$ independent files are equally popular. On average, each of such independent files will appear on average the same number of times over the ensemble of all possible multiple request configurations. We construct $\Nsf$ such multiple request configurations, each of which is formed by  ${\Nsf - 1 \choose \rsf - 1}$ independent files (in fact, each multiple request configuration corresponds to a ``file'' in the correlated file library of our problem). It follows that each independent file appears on average  $\Nsf {\Nsf - 1 \choose \rsf - 1}  / {\Nsf \choose \rsf} = \rsf$ times in the ensemble of possible multiple requests configurations.  If instead of random multiple requests, we consider the deterministic symmetric case, where the possible multiple requests configurations are all and only those for which each independent files appears exactly $\rsf$ times (and not on average $\rsf$ times), we have the exact equivalence  of our problem with the case of multiple requests of independent files.  
 With this interpretation, the proposed results in this paper also shed light into the very relevant and intricate problem of how to handle optimally the case where each user makes a sequence of requests of independent files (blocks). The fact that there are repeated elements in such sequence of requests is a `fundamental' aspect of caching (also in practice), where one needs to devise schemes that take advantage of previous requests and do not send the same stuff multiple times.

\section{Main Results and Numerical Evaluations}
\label{sec:main}
 
In this section, we state our main results and presents numerical evaluations of the proposed converse and achievable bounds. We shall use the subscripts ``$\text{\rm u,conv}$'' and ``$\text{\rm u,ach}$'' for converse (conv) and achievable (ach) bounds, respectively, under the constraint of uncoded cache placement (u).

\subsection{Converse Bound}
\label{sec:main:converse}

Inspired by~\cite{ourisitinnerbound}, we use the acyclic index coding converse bound from~\cite{onthecapacityindex} to derive the following   converse bound under the constraint of uncoded cache placement for our problem. The proof can be found in Section~\ref{sec:proof of converse bound}.

\begin{thm}[Converse]
\label{thm:converse}
For a $(\Nsf,\Ksf,\rsf,\Msf)$ shared-link caching problem with correlated files, 
$\Rsf^{\star}_\mathrm{u}(\Msf,s), \ s\in[\min \{\Ksf,\Nsf\}],$ is lower bounded by the lower convex envelope
of the following $(\Msf,\Rsf)$ pairs %memory-load 
\begin{align}
\left(
\frac{\Nsf t}{\Ksf \rsf}, c^s_{t} 
\right)_\text{\rm u,conv}, \ \forall  t\in [0:\Ksf], 
\label{eq:converse type s}
\end{align}
where 
\begin{align}
c^s_{t}:= \frac{\sum_{j\in [\min\{s,\Nsf-\rsf+1,\Ksf-t\}]}\binom{\Nsf-j}{\rsf-1}\binom{\Ksf-j}{t}}{\binom{\Nsf-1}{\rsf-1}\binom{\Ksf}{t}}.
\label{eq:ct}
\end{align}
In addition, $\Rsf^{\star}_\mathrm{u}(\Msf)$ is lower bounded by the lower convex envelope
of the following $(\Msf,\Rsf)$ pairs %memory-load 
\begin{align}
\left(
\frac{\Nsf t}{\Ksf \rsf}, \mathbb{E}_{\dv\in[\Nsf]^{\Ksf}}\left[ c^{N_{\textup{e}}(\dv)}_{t} \right]\right)_\text{\rm u,conv}, \ \forall t\in [0:\Ksf].
\label{eq:converse ave}
\end{align}
\end{thm}

Theorem~\ref{thm:converse} for $\rsf=1$ recovers the converse result for the MAN scheme under uncoded placement in~\cite{exactrateuncoded}, in particular the worst-case load is obtained for $s=\min\{\Ksf,\Nsf\}$ in~\eqref{eq:converse type s}, while the average load under uniform demands is given by~\eqref{eq:converse ave}. 
%in this case the converse is tight for any demand type, i..e, any $s\in[\min \{\Ksf,\Nsf\}]$.
%%sanity check
%%\begin{align*}
%%c^s_{t}|_{\rsf=1}
%%  &= \frac{\sum_{j\in [\min\{\Nsf,\Ksf-t,s\}]}\binom{\Ksf-j}{t}}{\binom{\Ksf}{t}}
%%\\&= \frac{\sum_{j\in [\Ksf-t]}\binom{\Ksf-j}{t} - \sum_{j\in [\min\{\Nsf,\Ksf-t,s\}+1:\Ksf-t]}\binom{\Ksf-j}{t}}{\binom{\Ksf}{t}}
%%\\&= \frac{\binom{\Ksf}{t+1} - \binom{\Ksf-\min\{\Nsf,\Ksf-t,s\}}{t+1}}{\binom{\Ksf}{t}}
%%\end{align*}
 Theorem~\ref{thm:converse} for $\rsf=\Nsf$ recovers the converse result for the MAN scheme with a single file, that is, $c^s_{t} = 1-t/\Ksf \Longleftrightarrow \Rsf^{\star}(\Msf)=1-\Msf$ for $\Msf\in[0,1]$.

\subsection{Achievable Scheme} %General 
\label{sec:main:achievable}
 
Let $\Msf=\frac{\Nsf t}{\Ksf \rsf}$ for some integer $t\in[0:\Ksf]$. Recall that
we denoted by $N_{\textup{e}}(\dv)$ the number of distinct files in the demand vector $\dv$, and by $\Lc(\dv)=\{u_1,\ldots,u_{N_{\textup{e}}(\dv)}\}$ the set of chosen leader users.
We propose the following achievable scheme, which is analyzed in Section~\ref{sec:achievable scheme}.
\begin{subequations}
\begin{align}
&\text{\bf Block subdivision:} \ \forall \Sc \subseteq [\Nsf] : |\Sc|=\rsf \ \text{let}\notag 
\\
&W_{\Sc} = \{ W_{\Sc,\Vc} : \forall \Vc \subseteq [\Ksf] : |\Vc|=t \}.
\\
&\text{\bf Placement Phase:} \ \forall k\in[\Ksf] \ \text{let}\notag
\\
&Z_{k} = \{W_{\Sc,\Vc} : \forall \Sc \subseteq [\Nsf] : |\Sc|=\rsf, \ \
                         \forall \Vc \subseteq [\Ksf] : |\Vc|=t, k\in\Vc \}.
\\
&\text{\bf Delivery sub-phase~1:}\notag %for demand $\dv$
  \\& \forall j\in[\min\{N_{\textup{e}}(\dv),\Nsf-\rsf+1,\Ksf-t\}], 
  \\& \quad          \forall \Jc\subseteq [\Ksf]\setminus\{u_1,\ldots,u_{j}\} : |\Jc|=t, 
  \\& \quad\quad     \forall \Bc\subseteq [\Nsf]\setminus\{d_{u_1},\ldots,d_{u_{j}}\} : |\Bc|=\rsf-1, 
  \\& \quad\quad\quad\text{send $C_{\Jc\cup\{u_j\},\Bc}$ as defined in~\eqref{eq:CJi general}}.  \label{eq:masterach:Jc in step 1}
\\
&\text{\bf Delivery sub-phase~2:}\notag
  \\& \forall j\in[\min\{N_{\textup{e}}(\dv),\Nsf-\rsf+1,\Ksf-t\}], 
  \\& \quad               \forall q\in[j+1: \min\{\Nsf-\rsf+2,\Ksf-t+1,N_{\textup{e}}(\dv)\}], 
  \\& \quad\quad          \forall \Jc\subseteq [\Ksf]\setminus\{u_1,\ldots,u_{q}\} : |\Jc|=t-1, \Jc\cap \{u_{q+1},\ldots,u_{N_{\textup{e}}(\dv)}\}\neq \emptyset, %must include some leaders
  \\& \quad\quad\quad     \forall \Bc\subseteq [\Nsf]\setminus\{d_{u_1},\ldots,d_{u_{q}}\} : |\Bc|=\rsf-2, \Bc\cap \Nc([\Ksf])\neq \emptyset, %must include some demanded files
  \\& \quad\quad\quad\quad\text{send $C_{\Jc\cup\{u_j,u_q\},\Bc}$ as defined in~\eqref{eq:CJi general}}.  
\end{align}
\label{eq:masterach}
\end{subequations}
\iffalse
\dt{WHY CANNOT HAVE SUB-PHASE-$q, \ q\in[\min\{t,\rsf-1\}] : |\Jc|=t+1-q, |\Bc|=\rsf-q$ WHERE THE SUB-PHASE FOR $q=2$ NEEDS LESS TRANSMISSIONS THAN WHAT WE HAVE NOW SINCE THERE WILL BE OTHER SUB-PHASES NEXT?}

{\magenta Answer of Kai: this is not equivalent I wrote before. And I confirm that the one you wrote is not decodable.}
\fi

 In the rest of this section we analyze the scheme in~\eqref{eq:masterach} in various settings of increasing order of complexity. Since the scheme is highly combinatorial, we shall start with a case that is the simplest to analyze and that brings to bear some of the key ideas. We shall then show that the a similar analysis applies also to more complex scenarios. In the following, optimality is understood under the constraint of uncoded cache placement. We have: 
\begin{itemize}

\item
In Section~\ref{sub:subphase1 only} we show that the general scheme in~\eqref{eq:masterach} with only the first delivery sub-phase allows each leader user to decode its desired file. We also show the first sub-phase alone is exactly optimal when the users request different files, that is, all users are leaders, which is possible when $N_{\textup{e}}(\dv)=\Ksf \leq \Nsf$. 
%
%\dt{WHY CAN'T WE HAVE THIS AS A CASE~3 IN TH5? MAYBE EVEN MORE GENERAL (AS IT SEEMS TO ME IT HOLDS FOR s=1 AS WELL, FOR EXAMPLE). TH\ref{thm:scheme 1}, OR A CASE~3 IN TH5, SHOUL LIST ALL $s$ FOR WHICH WE HAVE OPTIMALITY FOR $\Rsf^{\star}_\mathrm{u}(\Msf,s)$.
%FOR SURE $s=\Ksf \leq \Nsf$ AS NOW IN TH\ref{thm:scheme 1} (ASLO $s=1$, AND MAYBE $s\leq \min(\Nsf-\rsf+1,\Ksf-t)$ OR $s\geq \min(\Nsf,\Ksf)$).}
%{\magenta Kai's answer: the main reason is the decoding. For the case where users have distinct demands, we only need to use direct decoding. For the cases in Theorem~\ref{thm:exact optimality}, we use interference decoding, which is much more complicated, even if the encoding is the same.   I put these cases in Theorem~\ref{thm:exact optimality} after the general caching scheme, because I want to describe the decoding in the following way: in the general scheme, we transmit some additional packets in sub-phase~2 to align interferences. However, for the cases in Theorem~\ref{thm:exact optimality}, each non-leader can re-construct the transmitted packet in sub-phase~2 by the packets in sub-phase~1, such that we can also ensure the decoding without transmitting packets in sub-phase~2.}

\item
In Section~\ref{sub:subphase1 and2} we show that the scheme in~\eqref{eq:masterach}, with both delivery sub-phases, can satisfy every user regardless of the demand type, where the transmissions in sub-phase~2 are used to cancel the interferences experienced by the non-leader users. We also show its optimality to within a factor of $2$ for any demand type. 

\item
In Section~\ref{sub:subphase1 only again} we show that for some cases (such as, for example, the case of small or large memory size), each non-leader can re-construct the transmitted multicast messages in sub-phase~2 by performing linear combinations of the transmitted multicast messages in sub-phase~1, that is, sub-phase~2 is redundant. For these cases, we show exact optimality.

\item
In Section~\ref{sub:extensions} we show how the scheme in~\eqref{eq:masterach} can be used for other caching problems, by either offering simpler codes for the delivery phase than those known in the literature, or by providing an optimal scheme outperforming known state-of-the-art schemes.

\end{itemize}
In Section~\ref{sub:num} we finally give some numerical evaluations of the proposed bounds.

\subsection{Optimality of~\eqref{eq:masterach} for demand type  $s=\Ksf\leq \Nsf$}
\label{sub:subphase1 only}
Here we consider the case where each user makes a distinct request, which requires $\Ksf \leq \Nsf$ and demand type $s=\Ksf=N_{\textup{e}}(\dv)$.
We propose a caching scheme where we {\it jointly serve the users' demands}. Existing methods approach the problem by serving requests in multiple rounds~\cite{yang2018centralizedcorrected,ji2015multirequest,Sengupta2017multirequest,multireqWei2017}, where each round is a MAN scheme with users having a single request. Our scheme here is as in~\eqref{eq:masterach}, but where only the first sub-phase of the delivery phase takes place. In particular, for $\Msf=\frac{\Nsf t}{\Ksf \rsf}$ and $s=N_{\textup{e}}(\dv)=\Ksf \leq \Nsf$, our proposed delivery phase contains $\min\{\Nsf-\rsf+1,\Ksf-t\}$ steps, where in each step we transmit multicast messages to satisfy one leader user at a time. After all steps are done, the remaining users (who are also leaders, since here we consider a distinct request for each user) can also recover their desired file. The achieved load is presented in the following theorem, whose proof can be found in Section~\ref{sub:scheme 1}.

\begin{thm}[Optimality for Distinct Requests]
\label{thm:scheme 1}
For a $(\Nsf,\Ksf,\rsf,\Msf)$ shared-link caching problem with correlated files,
the loads in Theorem~\ref{thm:converse} are achievable under the constraint of uncoded cache placement when $\Nsf\geq \Ksf=s$ by the scheme in~\eqref{eq:masterach} with only the first delivery sub-phase.
%the lower convex envelope of $\Rsf^{\star}_\mathrm{u}(\Msf,s)$ is equal to the lower convex envelops of $\left(\frac{\Nsf t}{\Ksf \rsf}, c^{s}_{t} \right)$ for $t\in [0:\Ksf]$, where $c^s_{t}$ is defined in~\eqref{eq:ct}.
\end{thm}

\subsection{Performance of~\eqref{eq:masterach} for any demand type}% for General Case}
\label{sub:subphase1 and2}
We analyze here the scheme in~\eqref{eq:masterach} with two sub-phases in the delivery phase, and show that it is able to satisfy general demands. The main ingredients of the schemes are as follows.
We pick a leader user among the users demanding the same file;
in the first delivery sub-phase, we generate multicast messages as in Theorem~\ref{thm:scheme 1} so that each leader user can recover its desired file by the end of this sub-phase;
in the second delivery sub-phase, we transmit some additional multicast messages so that %after the construction of   multicast messages for the two phases,
each non-leader user can cancel all non-intended (aligned interference) sub-blocks from all received multicast messages and thus can eventually recover its desired file.
%Hence, the main strategy can be summarized into the following three  ingredients:
%\begin{enumerate}
%  \item choose a leader for each demanded file; 
%  \item satisfy the demands of leaders subsequently; 
%  \item align the interferences to unsatisfied users. 
%\end{enumerate}
The achieved load is presented in the following theorem, whose proof can be found in Section~\ref{sub:scheme 2}.

\begin{thm}[Interference-Alignment Based Delivery Scheme]%General 
\label{thm:scheme 2}
For a $(\Nsf,\Ksf,\rsf,\Msf)$ shared-link caching problem with correlated files, an achievable memory-load tradeoff for any $s\in[\min \{\Ksf,\Nsf\}]$ is given by the lower convex envelope of the following $(\Msf,\Rsf)$ pairs
\begin{align}
\left(
\frac{\Nsf t}{\Ksf \rsf}, c^s_{t} + e^s_{t}
\right)_\text{\rm u,ach}, \ \forall  t\in [0:\Ksf], 
\label{eq:scheme 2 type s}
\end{align}
where $c^s_{t}$ is defined in~\eqref{eq:ct} and
\begin{align}
e^s_{t}:= \frac{\sum_{j\in [\min\{s,\Nsf-\rsf+1,\Ksf-t\}]} \sum^{\min\{\Nsf-\rsf+2,\Ksf-t+1,s\}}_{q=j+1}  \left(\binom{\Nsf-q}{\rsf-2}-\binom{\Nsf-s}{\rsf-2} \right) \left(\binom{\Ksf-q}{t-1}-\binom{\Ksf-s}{t-1} \right) }{\binom{\Nsf-1}{\rsf-1}\binom{\Ksf}{t}}.  
\label{eq:et}
\end{align}
In addition, $\Rsf^{\star}_\mathrm{u}(\Msf)$ is upper bounded by the lower convex envelope
of the following $(\Msf,\Rsf)$ pairs
\begin{align}
\left(
\frac{\Nsf t}{\Ksf \rsf}, \mathbb{E}_{\dv\in[\Nsf]^{\Ksf}}\left[ c^{N_{\textup{e}}(\dv)}_{t}+e^{N_{\textup{e}}(\dv)}_{t}  \right]\right)_\text{\rm u,ach}, \ \forall t\in [0:\Ksf].
\label{eq:scheme 2 ave}
\end{align}
\end{thm}

\bigskip
By comparing the converse bound in Theorem~\ref{thm:converse} and the achievable bound in Theorem~\ref{thm:scheme 2}, we have the following result, whose proof can be found in Section~\ref{sub:proof thm:order optimality}.

\begin{thm}[Order Optimality for Theorem~\ref{thm:scheme 2}]
\label{thm:order optimality}
For a $(\Nsf,\Ksf,\rsf,\Msf)$ shared-link caching problem with correlated files, the loads in Theorem~\ref{thm:scheme 2} are order optimal to within a factor of $2$ under the constraint of uncoded cache placement.
\end{thm}
%we have
%\begin{enumerate}
%\item \label{item:order 1}
%the average load among all demands in type $s\in [\min\{\Ksf,\Nsf\}]$, achieved by caching scheme in Theorem~\ref{thm:scheme 2}, is order  optimal within a factor of $2$ under the constraint of uncoded cache placement.
%\item \label{item:order 2}
%the average load among all possible demands, achieved by caching scheme in Theorem~\ref{thm:scheme 2}, is order  optimal within a factor of $2$ under the constraint of uncoded cache placement.  
%\end{enumerate}

\subsection{Optimality of~\eqref{eq:masterach} for $\rsf\in \{1,2,\Nsf-1,\Nsf\}$ or  $t\in \{0,1,2,\Ksf-1,\Ksf\}$ or $s\in [\min\{\Nsf,\Ksf,4\}]$} %$\Ksf\rsf\Msf/\Nsf\leq 2$ or $\Ksf\rsf\Msf/\Nsf\geq \Ksf-1$
\label{sub:subphase1 only again}
%In~\eqref{eq:scheme 2 type s} and~\eqref{eq:scheme 2 ave}, $c^s_{t}$ 
In Theorem~\ref{thm:scheme 2}$, c^s_{t}$ in~\eqref{eq:ct} is the load for the first delivery sub-phase while $e^s_{t}$ in~\eqref{eq:et} is the one for the second delivery sub-phase. Hence, compared to the converse bound in Theorem~\ref{thm:converse}, $e^s_{t}$ is the term leading to the sub-optimality. In Theorem~\ref{thm:scheme 1}, where we showed exact optimality for distinct demands, the second sub-phase was not needed. We investigate here other cases where the second sub-phase is not needed. In particular, \emph{we show cases where each non-leader user can re-construct the multicast messages sent in sub-phase~2 by linearly combining multicast messages sent in sub-phase~1}. Since in these cases the second delivery sub-phase is not necessary, we obtain the following exact optimality result proved in Section~\ref{sub:scheme 3 r=2}.

\begin{thm}[Exact Optimality for Some Cases]
\label{thm:exact optimality}
For a $(\Nsf,\Ksf,\rsf,\Msf)$ shared-link caching problem with correlated files, we have that $\Rsf^{\star}_\mathrm{u}(\Msf,s)$ and $\Rsf^{\star}_\mathrm{u}(\Msf)$ are equal to the lower convex envelops of $\left(\frac{\Nsf t}{\Ksf \rsf}, c^s_{t} \right)$ %_\text{\rm u,ach} 
and of $\left(\frac{\Nsf t}{\Ksf \rsf},\mathbb{E}_{\dv\in[\Nsf]^{\Ksf}}\left[ c^{N_{\textup{e}}(\dv)}_{t} \right] \right)$, %_\text{\rm u,ach} %for $t\in [0:\Ksf]$,  
respectively, where $c^s_{t}$ is defined in~\eqref{eq:ct}, in the following cases:
\begin{enumerate}
\item \label{item:optimality 1} %r values, and all (N,K,-,t,s) [$\rsf$ values]
Case~\ref{item:optimality 1} (small or large file correlation): when $\rsf\in \{1,2,\Nsf-1,\Nsf\}$,  where optimality holds for any $s\in [\min\{\Ksf,\Nsf\}]$ and any $t\in [0:\Ksf]$;
\item \label{item:optimality 2} %t values, and all (N,K,r,-,s) [$\tsf$ values]
Case~\ref{item:optimality 2} (small or large cache size): when $t\in \{0,1,2,\Ksf-1,\Ksf\}$,  where optimality holds for any $s\in [\min\{\Ksf,\Nsf\}]$ and any $\rsf\in[\Nsf]$; %when either $\Ksf\rsf\Msf/\Nsf\leq 2$ or $\Ksf\rsf\Msf/\Nsf\geq \Ksf-1$.
\item \label{item:optimality 3}  %s values, [$s$ values] 
%Case~\ref{item:optimality 3} [$s$ values]: when \dt{??s??}.
%\end{itemize}
%In addition, we have that $\Rsf^{\star}_\mathrm{u}(\Msf,s)$, is equal to the lower convex envelope of $\left(\frac{\Nsf t}{\Ksf \rsf}, c^s_{t} \right)$    
%for $t\in [0:\Ksf]$, in the following case:
%\begin{itemize}
%\item \label{item:optimality 3} 
Case~\ref{item:optimality 3} (small number of distinct requests): when $s\in [\min\{\Ksf,\Nsf,4\}]$,  where optimality  holds for any $\rsf\in[\Nsf]$ and any $t\in[0:\Ksf]$.   In this case, no claim can be made on $\Rsf^{\star}_\mathrm{u}(\Msf)$ as only some values of $s$ are exactly characterized. 
\end{enumerate}
\end{thm}

From Theorem~\ref{thm:exact optimality} we immediately have the following corollary, which can be proved straightforwardly by noting that Theorem~\ref{thm:exact optimality}.Case~\ref{item:optimality 3} covers all possible values of $s$ when $\min(\Nsf,\Ksf)\leq 4$.
% and that
%Theorem~\ref{thm:exact optimality}.Case~\ref{item:optimality 1} covers all possible values of $\rsf$ when $3\geq\Nsf-1$, and that
%Theorem~\ref{thm:exact optimality}.Case~\ref{item:optimality 2} covers all possible values of $\Msf$ when $3\geq\Ksf-1$.

\begin{cor}%[Solved Cases Under Uncoded Placement]
\label{cor:solvedfully}
%%%For a $(\Nsf,\Ksf,\rsf,\Msf)$ shared-link caching problem with correlated files with $\min(\Nsf,\Ksf)\leq 4$, we have that $\Rsf^{\star}_\mathrm{u}(\Msf,s)$, for $s\in [\min\{\Ksf,\Nsf\}]$, and $\Rsf^{\star}_\mathrm{u}(\Msf)$ are equal to the lower convex envelops of $\left(\frac{\Nsf t}{\Ksf \rsf}, c^s_{t} \right)$ %_\text{\rm u,ach} 
%%%and of $\left(\frac{\Nsf t}{\Ksf \rsf},\mathbb{E}_{\dv\in[\Nsf]^{\Ksf}}\left[ c^{N_{\textup{e}}(\dv)}_{t} \right] \right)$, %_\text{\rm u,ach}   
%%%for $t\in [0:\Ksf]$, respectively.
%%%%The scheme in Theorem~\ref{thm:scheme 2} is optimal under the constraint of uncoded placement for all $(\Nsf,\Ksf,\rsf,\Msf)$ shared-link caching problems with correlated files with $\min(\Nsf,\Ksf)\leq4$.  
For a $(\Nsf,\Ksf,\rsf,\Msf)$ shared-link caching problem with correlated files, the loads in Theorem~\ref{thm:converse} are achievable under the constraint of uncoded cache placement when $\min(\Nsf,\Ksf)\leq 4$ by the scheme in~\eqref{eq:masterach} with only the first delivery sub-phase.
\end{cor}

\iffalse
\dt{QUESTION: what is the simplest open case? 
\\--From Case1+Case2 I'd say N=K=5, r=t=3 (while all other (r,t) for N=K=5 are covered by our current results). But by Case3 and Thm2, it seems that all cases of s are covered when min(N,K)=5. So is min(N,K)=5 solved too? If so, should be added to the corollary.
\\--If what I said is correct, then N=K=6, either r=3,4 or t=3,4 or s=5 are open. Right?
}

{\magenta Answer of Kai: The simplest case should be $\Nsf=5$, $\Ksf=6$, $t=\rsf=3$.
}
\fi

%{\blue
%For Theorem~\ref{thm:exact optimality} we can also obtain the following corollary, which will be proved in Section~\ref{sec:solved s}. 
%\begin{cor} 
%\label{cor:solved s}
%For a $(\Nsf,\Ksf,\rsf,\Msf)$ shared-link caching problem with correlated files, we have that $\Rsf^{\star}_\mathrm{u}(\Msf,s)$, for $s\in [\min\{\Ksf,\Nsf,4\}]$, is equal to the lower convex envelope of $\left(\frac{\Nsf t}{\Ksf \rsf}, c^s_{t} \right)$    
%for $t\in [0:\Ksf]$.
%\end{cor}
%}
%{\magenta Comments of Kai: I do not want to insert Corollary~\ref{cor:solved s} into Theorem~\ref{thm:exact optimality}, because for the cases in Theorem~\ref{thm:exact optimality} our optimality is for any $s$ and also for the average load over all possible demands. In addition, the results in Corollary~\ref{cor:solved s} are just from the proof of Theorem~\ref{thm:exact optimality}.}

\begin{rem}[Average and Worst-case Loads]
\label{rem:worst-case}
\rm
Our proposed caching scheme and our optimality results directly characterize the optimal worst-case load (and not just the average load), because the worst-case load is the case of demand type $s=\min\{\Nsf,\Ksf\}$.
We note that past works only aimed to design schemes that minimize the worst-case load, % among all possible demands
such as those in~\cite{yang2018centralizedcorrected} (for caching with correlated sources) and in~\cite{ji2015multirequest,Sengupta2017multirequest} (for caching with multiple requests).
Order optimality results (to within factors $11$ and $18$) on the worst-case load were derived in~\cite{ji2015multirequest,Sengupta2017multirequest} for caching with multiple requests;
 to the best of our knowledge, no order optimality results are known specifically for caching with correlated sources. 
Therefore, a major contribution of this paper,  besides sharpening existing results for the worst-case load, it is to have derived (exact or order) optimality results on the average loads for {\it any demand type} and {\it over all possible demands,} under the constraint of uncoded cache placement.
\hfill$\square$
\end{rem}

\subsection{Extensions}
\label{sub:extensions}
Our scheme can be used in models other than the one considered in this paper.
Examples are as follows.

\paragraph*{Extension to the More General Coded Caching Problem with Correlated Sources}
\label{rem:extension various r}
As already mentioned earlier, in this paper
%, in order to make fundamental progress on the problem, 
we simplified the model~\cite{yang2018centralizedcorrected} 
%as follows. In~\cite{yang2018centralizedcorrected} a certain parameters $\ell$ ranges from zero to the number of files in the system (each $\ell_1$ files have a common part, each $\ell_2$ files also have a common part, etc.), %the block size is identical, 
%while in our model $\ell$ is fixed to a single value $\rsf$; with our models, we can make conclusive statements (either exact capacity results, or capacity to within a constant multiplicative gap) which eluded in~\cite{yang2018centralizedcorrected}.
by fixing the parameter $\ell$ in~\cite{yang2018centralizedcorrected} to be equal to $\rsf$ (as opposed to let be to within a range).
We can extend our results to the case where $\ell$ is within a range as follows.
If $\ell$ is in a range as considered in~\cite{yang2018centralizedcorrected}, we can construct a caching scheme by ``memory-sharing'' among the proposed schemes in Theorems~\ref{thm:scheme 1},~\ref{thm:scheme 2}, and~\ref{thm:exact optimality} as follows.
\begin{itemize}

\item
Library.
Assume that the length of each block $W_{\Sc}$, where $\Sc\subseteq [\Nsf]$ and $|\Sc|=\ell$, is $\psf_{\ell} \Bsf/\binom{\Nsf}{\ell}$, where $\psf_{\ell} \in [0,1]$ and $\sum_{\ell \in [\Nsf]}\psf_{\ell}=1$. The values $(\psf_{\ell} : \ell \in [\Nsf])$ are assumed to be fixed system parameters.

\item
Placement.
%The placement is the same as~\cite{yang2018centralizedcorrected}. 
Choose integers  $t_{\ell} \in [\Ksf]$ for $\ell\in [\Nsf]$, %, and denote $\tv:=(t_{\ell} : \ell \in [\Nsf])$. For each $\ell\in [\Nsf]$, 
We partition block $W_{\Sc}$ into $\binom{\Ksf}{t_{|\Sc|}}$ equal-length sub-blocks and denote $W_{\Sc}=\{W_{\Sc,\Vc}:\Vc\subseteq [\Ksf], |\Vc|=t_{|\Sc|} \}$. 
User $k\in [\Ksf]$ caches sub-block $W_{\Sc,\Vc}$ if $k\in \Vc$, which requires a cache of size 
\begin{align}
%{\blue \Msf_{\tv}}
\Msf=\sum_{\ell \in [\Nsf]} \frac{\Nsf t_{\ell} \psf_{\ell}}{\Ksf \ell}.
\label{eq:memory extension}
\end{align} 

\item
Delivery.
For demand vector $\dv$, if $N_{\textup{e}}(\dv)=\Ksf$ (i.e., each user demands a distinct file), we use the  caching scheme in Theorem~\ref{thm:scheme 1} and the achieved load is 
\begin{align}
%{\blue \Rsf_{\tv}(\Msf_{\tv},\Ksf)}=
\Rsf = \sum_{\ell\in [\Nsf]} \psf_{\ell} c^{\Ksf}_{t_{\ell}}.
\label{eq:memory extension load 1}
\end{align}
%In the following, we consider the demand type For each $\ell\in [\Nsf]$, 
If $s=N_{\textup{e}}(\dv)<\Ksf$ we have two cases;
if either $\ell\in\{1,2,\Nsf-1,\Nsf\}$ or $t_{\ell}\in \{0,1,2,\Ksf-1,\Ksf\}$, we use the caching scheme in Theorem~\ref{thm:exact optimality} to encode all blocks; %$W_{\Sc}$ where $|\Sc|=\ell$; 
otherwise, we use the caching scheme in  Theorem~\ref{thm:scheme 2}; %to encode all blocks $W_{\Sc}$ where $|\Sc|=\ell$. 
the achieved load is 
\begin{align}
%{\blue \Rsf_{\tv}(\Msf_{\tv},s)}=
\Rsf = \sum_{\ell_1 \in\{1,2,\Nsf-1,\Nsf\}}  \psf_{\ell_1} c^{s}_{t_{\ell_1}} + \sum_{\ell_2 \in [3:\Nsf-2]} \left( \psf_{\ell_2} c^{s}_{t_{\ell_2}}+ \psf_{\ell_2} e^{s}_{t_{\ell_2}} \mathbbm{1}_{t_{\ell_2}\notin \{0,1,2,\Ksf-1,\Ksf\}}    \right),\label{eq:memory extension load 2}
\end{align}
where $\mathbbm{1}$ is the indicator function, where $\mathbbm{1}_{\textrm{event}}=1$ if $\textrm{event}$ is true and  $\mathbbm{1}_{\textrm{event}}=0$ otherwise. 

\item
The achievable memory-load tradeoff is the lower convex envelope of the above points %$(\Msf_{\tv},\Rsf_{\tv}(\Msf_{\tv},N_{\textup{e}}(\dv)))$ 
for all possible $\tv:=(t_{\ell} : \ell \in [\Nsf])$.

\end{itemize}

\paragraph*{Extension to the Coded Caching Problem with Multiple Requests}
\label{rem:extension multireq}
In this paper, differently from most of the existing work that divides the multi-request problem into a sequence of single-request problems, we can  modify  our proposed interference alignment scheme for the caching problem with correlated sources to  address  the caching problem with multiple requests. 
%We note that in order to  more general multi-request models we could also use the proposed interference alignment strategy. 
By doing so, we can give an optimal scheme for the four cases that were left open in~\cite{multireqWei2017} for the caching problem with multiple requests, where the setting includes up to four users and where each user demands at most two files. 
\iffalse
\dt{COULD WE SOLVE ALSO L=3 OR 4?}. {\magenta Answer of Kai: The problem is that when $\Lsf=3$ or $4$, there are too many possibilities to be checked.}
\fi
 The details  of how to modify our proposed interference alignment scheme (so as to account for the lack of symmetry of the multi-request problem)  are given in Appendix~\ref{sec:codes for extension}. 
%\dt{WHY DO WE NEED A PROOF? DOESN'T IT FOLLOW FROM  Corollary~\ref{cor:solvedfully}/Theorem~\ref{thm:exact optimality}?}
%{\magenta Kai's answer: because these four cases are not totally symmetric, i.e., not equivalent to our considered the caching problem with correlated sources. We need some modification on the delivery scheme.}

\paragraph*{Extension to Distributed Computation}
\label{rem:extension codedcomputing}
When $\Nsf=\Ksf$ and each user demands a distinct file, the $(\Nsf,\Ksf,\rsf,\Msf)$ shared-link caching problem with correlated files is related to the distributed computation problem in~\cite{distributedcomputing}. The only difference  is that in~\cite{distributedcomputing} the link is D2D (i.e., workers/users communication among each other without a central master/server), as opposed to the shared-link case considered here.
In~\cite{distributedcomputing}, the authors proposed an optimal scheme that requires to exchange messages where symbols are from a large finite field size.  In the contrast, for the shared-link caching problem, by using the optimal scheme proposed in this paper, operations are simpler in that they are on the binary field.    
%We can directly extend our proposed shared-link caching scheme  to the D2D case, as done in~\cite{d2dcaching}, so as to achieve the optimal load of~\cite{distributedcomputing} but by only using operations on the binary field. 

\subsection{Numerical Evaluations}
\label{sub:num}
\begin{figure}
    \centering
    \begin{subfigure}[t]{0.5\textwidth}
        \centering
        \includegraphics[scale=0.6]{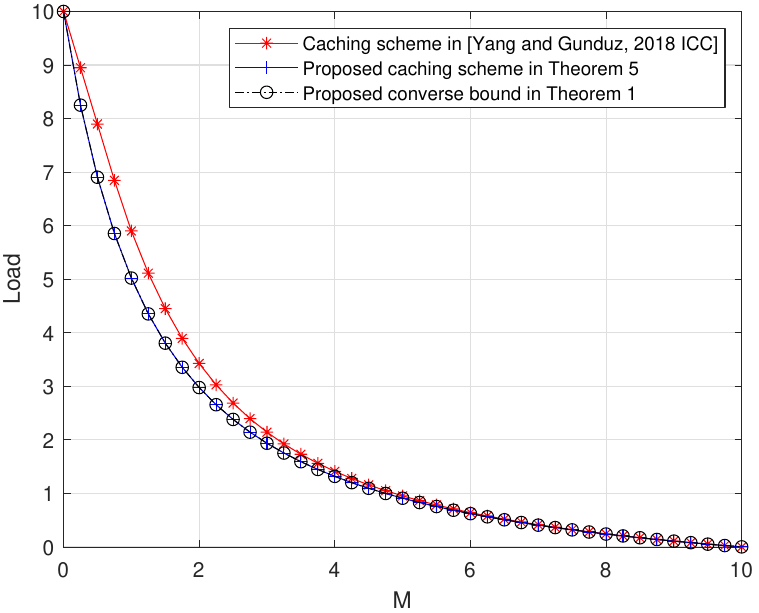}{a}
        \caption{\small $\mathcal{D}_{20}$.}
        \label{fig:numerical 1a}
    \end{subfigure}%
    ~ 
    \begin{subfigure}[t]{0.5\textwidth}
        \centering
        \includegraphics[scale=0.6]{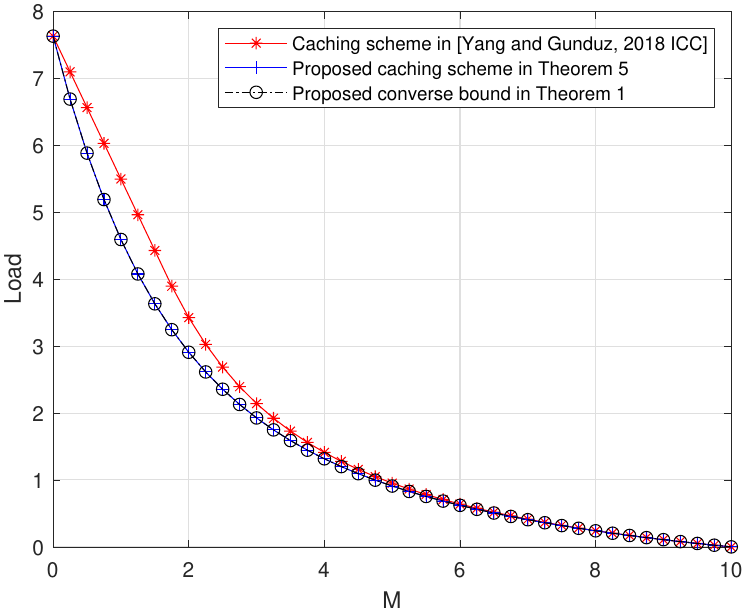}{b}
        \caption{\small $\mathcal{D}_{10}$.}
        \label{fig:numerical 1b}
    \end{subfigure}
    \caption{\small The memory-load trade-off for the  $(\Nsf,\Ksf,\rsf)=(20,40,2)$ shared-link caching problem with correlated files.}
    \label{fig:numerical 1}
\end{figure}

\begin{figure}
    \centering
    \begin{subfigure}[t]{0.5\textwidth}
        \centering
        \includegraphics[scale=0.6]{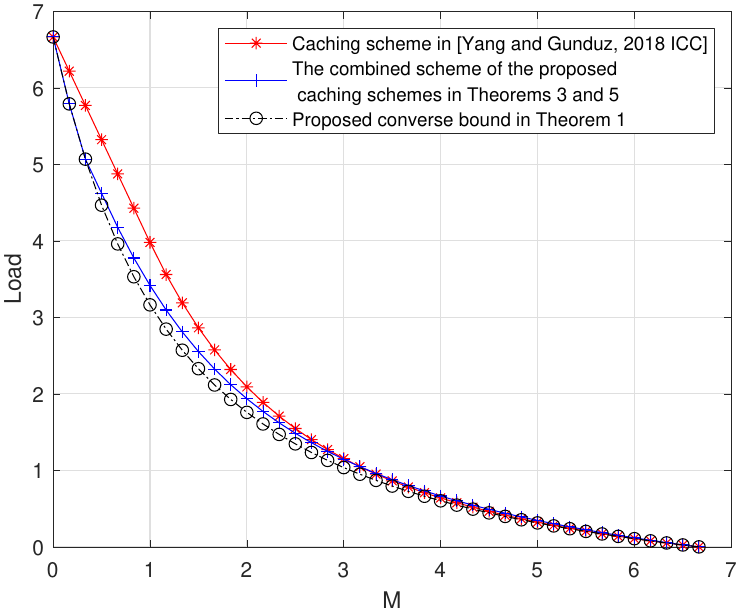}{a}
        \caption{\small $\mathcal{D}_{20}$.}
        \label{fig:numerical 2a}
    \end{subfigure}%
    ~ 
    \begin{subfigure}[t]{0.5\textwidth}
        \centering
        \includegraphics[scale=0.6]{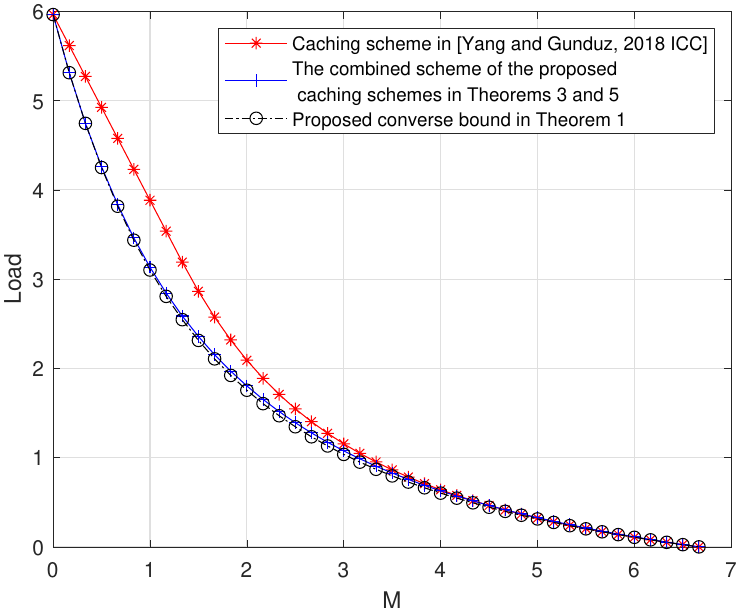}{b}
        \caption{\small $\mathcal{D}_{10}$.}
        \label{fig:numerical 2b}
    \end{subfigure}
    \caption{\small The memory-load trade-off for the  $(\Nsf,\Ksf,\rsf)=(20,40,3)$ shared-link caching problem with correlated files.}
    \label{fig:numerical 2}
\end{figure}

In Figs.~\ref{fig:numerical 1} and~\ref{fig:numerical 2} we  plot the load vs the cache size for demand types $\mathcal{D}_{20}$ (left subfigure) and $\mathcal{D}_{10}$ (right subfigure) for the $(\Nsf,\Ksf)=(20,40)$ shared-link caching problem with correlated files.

In Fig.~\ref{fig:numerical 1} we consider the $\rsf=2$, for which our proposed scheme (with only the first delivery sub-phase) is optimal as stated in Theorem~\ref{thm:exact optimality}.
%we consider the $(\Nsf,\Ksf,\rsf)=(20,40,2)$ shared-link caching problem with correlated files,
%for demand types $\mathcal{D}_{20}$ (in Fig.~\ref{fig:numerical 1a}) and $\mathcal{D}_{10}$ (in Fig.~\ref{fig:numerical 1b}).
%Since $\rsf=2$, we can use our optimal scheme in Theorem~\ref{thm:exact optimality} (only the first delivery sub-phase is necessary).
We also compare %the proposed converse bound in Theorem~\ref{thm:converse} and 
the average loads achieved by our scheme in Theorem~\ref{thm:exact optimality} with that of the suboptimal scheme in~\cite{yang2018centralizedcorrected}. %Fig.~\ref{fig:numerical 1} shows that our proposed converse in Theorem~\ref{thm:converse} and achievable bounds in Theorem~\ref{thm:exact optimality}  coincide. 

In Fig.~\ref{fig:numerical 2} we consider the $\rsf=3$.
%we consider the $(\Nsf,\Ksf,\rsf)=(20,40,3)$ shared-link caching problem with correlated files,
%for demand types $\mathcal{D}_{20}$ (in Fig.~\ref{fig:numerical 2a}) and $\mathcal{D}_{10}$ (in Fig.~\ref{fig:numerical 2b}).
When $t=\Ksf\Msf\rsf/\Nsf \in\{0,1,2,39,40\}$, only the first sub-phase of the delivery scheme is necessary  and the resulting load is optimal as stated in Theorem~\ref{thm:exact optimality}. For other values of the parameter $t$, we use both sub-phases of the delivery scheme as in Theorem~\ref{thm:scheme 2}. Fig.~\ref{fig:numerical 2} shows that our proposed achievable scheme %bound has a lower gap to the proposed converse bound in Theorem~\ref{thm:converse} than the achievable bound 
outperforms the scheme in~\cite{yang2018centralizedcorrected}.

\begin{figure}%[ht]
%\vspace{-2mm}
\centerline{\includegraphics[scale=0.6]{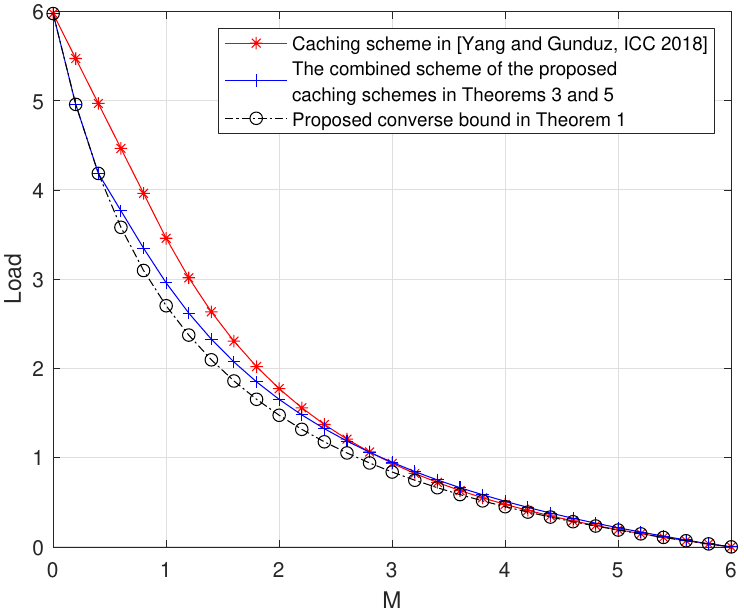}}
\caption{\small The memory-average load over all demands trade-off for the  $(\Nsf,\Ksf,\rsf)=(30,30,5)$ shared-link caching problem with correlated files.}
\label{fig:average all}
\vspace{-5mm}
\end{figure}

%\dt{PLOT A CASE OF LOAD WHEN AVERAGE OVER ALL DEMAND TYPES}
%{\magenta Kai's answer: The difficulty is that to consider the average over all demands, we need to consider $N^K$ demands, which is not computable in computer, unless we use Monte-Carlo simulation.}
In Fig.~\ref{fig:average all}  we plot the average load over all possible demands for the $(\Nsf,\Ksf,\rsf)=(30,30,5)$ shared-link caching problem with correlated files, by   Monte-Carlo simulation with the number of
iteration $10^6$. In each iteration,   the demand of each user is generated  independently at random based on the discrete uniform distribution.
When $t=\Ksf\Msf\rsf/\Nsf \in\{0,1,2,28,29\}$ or $N_{\textup{e}}(\dv)\leq 4$, only the first sub-phase of the delivery scheme is necessary  and the resulting load is optimal as stated in Theorem~\ref{thm:exact optimality}. For other values of the parameters $t$ and $N_{\textup{e}}(\dv)$, we use both sub-phases of the delivery scheme as in Theorem~\ref{thm:scheme 2}. Fig.~\ref{fig:average all} shows that our proposed achievable scheme outperforms the scheme in~\cite{yang2018centralizedcorrected}.

\section{Converse Bound}
\label{sec:proof of converse bound}

\subsection{Proof of Theorem~\ref{thm:converse}}
The delivery phase with uncoded cache placement is equivalent to a multicast  index coding problem~\cite{indexcodingwithsi}. Such a problem can be represented on a directed graph. In this graph, each sub-block demanded  but not cached  by a user is a node; a directed edge exists from node~$a$ to node~$b$ if the user demanding the sub-block represented by node~$b$ has the sub-block represented by node~$a$ in its cache. 
As in~\cite{ourisitinnerbound}, we use the acyclic index coding converse bound from~\cite{onthecapacityindex} to lower bound the number of transmitted bits needed to satisfy all the nodes/users in this index coding problem as follows.

%%%Multi-cast IC where user~$k\in[\Ksf]$ has side information $Z_k$ and demands $\Dc_k(\dv)$
%%%\begin{align}
%%%Z_k &= \{ W_{\Sc,\Vc} : \forall\Sc, k\in\Vc \},  
%%%|Z_k| = \binom{\Nsf}{\rsf} 2^{\Ksf-1}= \left(\binom{\Nsf-1}{\rsf-1}+\binom{\Nsf-1}{\rsf} \right)2^{\Ksf-1}
%%%\\
%%%\Dc_k(\dv) &= \{ W_{\Sc,\Vc} : d_k\in\Sc, k\not\in\Vc \}, |\Dc_k(\dv)| = \binom{\Nsf-1}{\rsf-1} 2^{\Ksf-1}
%%%\\ |\cup_{k\in[\Ksf]} \Dc_k(\dv)| = , %actual message set
%%%\end{align}

For a demand vector $\dv$, with $N_{\textup{e}}(\dv)$ distinct demands, we draw a graph where each sub-block demanded  but not cached by some  of these $N_{\textup{e}}(\dv)$ users is node. % in the sub-graph. 
We then consider a permutation of these $N_{\textup{e}}(\dv)$ users, denoted by $\mathbf{u}=(u_{1},u_{2},...,u_{N_{\textup{e}}(\dv)})$. 
%In Appendix~\ref{sec:proof of acyclicity}, we prove  that 
The set of sub-blocks 
\begin{align} 
 \bigcup_{k\in [\min\{ N_{\textup{e}}(\dv),\Nsf-\rsf+1 \}]}  \
 \bigcup_{\substack{\Sc\subseteq [\Nsf]\setminus \{d_{u_{1}},\ldots,d_{u_{k-1}}\} : \\ |\Sc|=\rsf,d_{u_{k}}\in \Sc}} \
 \bigcup_{\Vc\subseteq [\Ksf]\setminus \{u_{1},\ldots,u_{k}\} }W_{\Sc,\Vc}, 
\label{eq:set of acyclic}
\end{align}
does not contain a directed cycle. 
This can be seen as follows, similarly to~\cite[Lemma 1]{ontheoptimality}. 
We classify the sub-blocks/nodes in the set~\eqref{eq:set of acyclic} into levels. More precisely,
we say that sub-block/node $W_{\Sc,\Vc}$ is in level $i$ if $\Sc\subseteq [\Nsf]\setminus \{d_{u_1},\ldots,d_{u_{i-1}}\}$, $d_{u_i}\in \Sc$ and $\Vc\subseteq [\Ksf] \setminus \{u_{1},\ldots,u_{i}\}$. It is easy to see each node in level $i$ 
 is a sub-block that is demanded but not-cached by user $u_i$ that did not appear in any of the levels with lower index, and corresponds to a user in the index coding problem that has the same side information as user $u_i$ in our caching problem (i.e., each node in level $i$ only knows the nodes $W_{\Sc,\Vc}$ where $u_{i} \in\Vc$). So each node in level $i$ knows neither the nodes in the same level, nor the nodes in the higher levels. As a result, the proposed set in~\eqref{eq:set of acyclic} does not contain a directed cycle.

By the acyclic index coding converse bound, the number of  transmitted bits is not less than total number of bits of the sub-blocks in the set in~\eqref{eq:set of acyclic}, that is, 
\begin{align}
\Rsf^{\star}_\mathrm{u}(\Msf,N_{\textup{e}}(\dv)) \geq  
 \sum_{k\in [\min\{ N_{\textup{e}}(\dv),\Nsf-\rsf+1 \}]}  \
 \sum_{\substack{\Sc\subseteq [\Nsf]\setminus \{d_{u_{1}},\ldots,d_{u_{k-1}}\} : \\ |\Sc|=\rsf,d_{u_{k}}\in \Sc}} \
 \sum_{\Vc\subseteq [\Ksf]\setminus \{u_{1},\ldots,u_{k}\} } \frac{|W_{\Sc,\Vc}|}{\Bsf},
\label{eq:direct from acyclic set}
\end{align}
where $|W_{\Sc,\Vc}|$ represents the length of $W_{\Sc,\Vc}$ in bits.

\begin{subequations}
For a fixed $s\in[\min \{\Ksf,\Nsf\}]$, consider all the demands of type $\Dc_s$, all sets of users with different $s$ distinct demands, and all permutations of those users. By summing together all the resulting inequalities as in~\eqref{eq:direct from acyclic set} we obtain a the following lower bound on $\Rsf^{\star}_\mathrm{u}(\Msf,s)$, 
\begin{align}  
\Rsf^{\star}_\mathrm{u}(\Msf,s)
&\geq 
   \frac{1}{\binom{\Ksf}{s}}\sum_{\Lc \subseteq[\Ksf] : |\Lc| = s} \ % set of s leaders
   \frac{(\Nsf-s)!}{\Nsf! \, s^{\Ksf-s}}\sum_{\dv\in\Dc_s : \Lc \ \text{are leaders}} \
   \frac{1}{s!}\sum_{\usf \in\{ \text{permutations of} \ \Lc\}}  \ %a permutation of the leaders
\\&\sum_{k\in [\min(s, \Nsf-\rsf+1)]}  \
 \sum_{\substack{\Sc\subseteq [\Nsf]\setminus \{d_{u_{1}},\ldots,d_{u_{k-1}}\} : \\ |\Sc|=\rsf,d_{u_{k}}\in \Sc}} \
 \sum _{t=0}^{\Ksf-k} \ \sum_{\Vc\subseteq [\Ksf]\setminus \{u_{1},\ldots,u_{k}\} : |\Vc|=t} \frac{|W_{\Sc,\Vc}|}{\Bsf}.
\end{align}
In such a bound, there are $\sum_{j\in [\min\{s,\Nsf-\rsf+1,\Ksf-t\}]}\binom{\Nsf-j}{\rsf-1}\binom{\Ksf-j}{t}$ sub-blocks known by exactly $t$ users whose coefficient is $1$. We can also note that in general there are $\binom{\Nsf}{\rsf}\binom{\Ksf}{t}$ sub-blocks known by exactly $t$ users. By the symmetry of the problem, in the sum of all these  inequalities from the acyclic index coding converse bound, the coefficient of each sub-blocks stored by exactly $t$ users is identical. 
Hence, we have 
\begin{align}
\Rsf^{\star}_\mathrm{u}(\Msf,s)  &\geq \sum _{t=0}^{\Ksf}   
\frac{\sum_{j\in [\min\{s,\Nsf-\rsf+1,\Ksf-t\}]}\binom{\Nsf-j}{\rsf-1}\binom{\Ksf-j}{t}}{\binom{\Nsf}{\rsf}\binom{\Ksf}{t}} \cdot
\frac{\sum_{\Sc\subseteq [\Nsf]:|\Sc|=\rsf }\sum_{\Vc\subseteq [\Ksf]:|\Vc|=t} |W_{\Sc,\Vc}|}{\Bsf}
\\
&= \sum _{t=0}^{\Ksf}   
\frac{\sum_{j\in [\min\{s,\Nsf-\rsf+1,\Ksf-t\}]}\binom{\Nsf-j}{\rsf-1}\binom{\Ksf-j}{t}}{\binom{\Nsf-1}{\rsf-1}\binom{\Ksf}{t}} \cdot
\frac{\sum_{\Sc\subseteq [\Nsf]:|\Sc|=\rsf }\sum_{\Vc\subseteq [\Ksf]:|\Vc|=t} \rsf |W_{\Sc,\Vc}|}{\Nsf\Bsf} \\
&= \sum_{t=0}^{\Ksf}c^s_{t} \cdot x_{t}, 
\label{eq:n(m)}
\\
&x_{t}:=\sum_{\Sc\subseteq [\Nsf]:|\Sc|=\rsf} \sum_{\Vc\subseteq [\Ksf]:|\Vc|=t} \frac{\rsf|W_{\Sc,\Vc}|}{\Nsf \Bsf}, 
\label{eq:xt}
\\
&c^s_{t}:= \frac{\sum_{j\in [\min\{s,\Nsf-\rsf+1,\Ksf-t\}]}\binom{\Nsf-j}{\rsf-1}\binom{\Ksf-j}{t}}{\binom{\Nsf-1}{\rsf-1}\binom{\Ksf}{t}},  \ \textrm{(as already defined in~\eqref{eq:ct})},
\\
&x_{0}+x_{1}+...+x_{\Ksf}=1, \ \textrm{(file size constraint)},
\label{eq:file size}
\\
&x_{1}+2x_{2}+...+tx_{t}+...+\Ksf x_{\Ksf}\leq \frac{\Ksf\Msf \rsf}{\Nsf},  \ \textrm{(memory size contraint)},
\label{eq:cache size}
\end{align} 
where $x_{t}$ in~\eqref{eq:xt} represent the fraction of all the bits in the library that are cached exactly by $t$ users.
\end{subequations}

As in~\cite{exactrateuncoded}, we can lower bound~\eqref{eq:n(m)} by using Jensen's inequality and the monotonicity of $\textrm{Conv}(c^s_t)$ (i.e., the convex lower envelope of $c^s_t$ in terms of $t$),
\begin{align}
  \Rsf^{\star}_\mathrm{u}(\Msf,s) \geq \textrm{Conv}(c^s_t).
\label{eq:ave on type s}
\end{align}

By considering all the demand types, and from~\eqref{eq:ave on type s}, we also have
\begin{align}
\Rsf^{\star}_\mathrm{u}(\Msf)
&\geq \mathbb{E}_{s\in [\min\{\Nsf,\Ksf\}]}\left[ \Rsf^{\star}_\mathrm{u}(\Msf,s) \right]
 \geq \mathbb{E}_{s\in [\min\{\Nsf,\Ksf\}]}[\textrm{Conv}(c^s_t)].
\label{eq:consider ave}
\end{align}
Since $c^s_t$ is convex in $t$, we can change the order of the expectation and the ``Conv'' in~\eqref{eq:consider ave}. Thus we prove the converse bound in Theorem~\ref{eq:converse ave}.  

Notice that we could also use Fourier-Motzkin elimination to eliminate the parameters $\{x_{t}\}_{t\in[0:\Ksf]}$ in~\eqref{eq:n(m)} and derive the bound in~\eqref{eq:ave on type s}, as done in~\cite{ourisitinnerbound}. %and~\eqref{eq:consider ave}, 

\subsection{Discussion}
\label{subsec:discussionconvese}
 
We conclude this session with some observations on the proposed converse bound, which we shall use as a guideline to design our achievable schemes.
\begin{enumerate}

\item %[Intuition on the placement from the proof of Theorem~\ref{thm:converse}.]
The corner points of our converse bound are of the form $\left(\frac{\Nsf t}{\Ksf \rsf}, c^s_{t} \right)$,  where $c^s_{t}$ is defined in~\eqref{eq:ct}, which may suggest the following placement. 
We partition each block $W_{\Sc}$ into $\binom{\Ksf}{t}$ equal-length sub-blocks of length $\frac{\Bsf}{\binom{\Nsf-1}{\rsf-1} \binom{\Ksf}{t}}$ and indicate $W_{\Sc}=\{W_{\Sc,\Vc}:\Vc\subseteq [\Ksf],|\Vc|=t\}$.
Each user $k\in[\Ksf]$ stores the sub-block $W_{\Sc,\Vc}$ if $k\in \Vc$. 
Hence, user $k\in[\Ksf]$ caches $\frac{\Bsf\binom{\Ksf-1}{t-1}\binom{\Nsf}{\rsf}}{\binom{\Nsf-1}{\rsf-1} \binom{\Ksf}{t}}=\frac{\Bsf \Nsf t}{\Ksf \rsf}$ bits in total.
%this coincides with the way the cache size $\Msf$ is parameterized.

We will use this interpretation to design the file partitioning and the cache placement of our proposed caching schemes, which is the same as in~\cite{yang2018centralizedcorrected}.

\item %[Intuition on the delivery from the proof of Theorem~\ref{thm:converse}.]
If the above placement is used, each sub-block is cached by $t$ users. 
In the proof of Theorem~\ref{thm:converse}, for each %worst-case 
demand $\dv$, we choose a set of leader users (each demanding a differenet file) and consider a permutation $\mathbf{u}=(u_1,\ldots,u_{N_{\textup{e}}(\dv)})$ of these $N_{\textup{e}}(\dv)$ leader users. 
For the permutation $\mathbf{u}$, we find an acyclic set of $\sum_{j\in [\min\{N_{\textup{e}}(\dv),\Nsf-\rsf+1,\Ksf-t\}]}\binom{\Nsf-j}{\rsf-1}\binom{\Ksf-j}{t}$ sub-blocks, and lower bounded the load by the total length of these sub-blocks.
In addition, in this acyclic set of sub-blocks, there are $\binom{\Nsf-j}{\rsf-1}\binom{\Ksf-j}{t}$ sub-blocks desired by user $u_j$ where $j\in[\min\{N_{\textup{e}}(\dv),\Nsf-\rsf+1,\Ksf-t\}]$; these sub-blocks are not cached nor desired by any user $u_{j_1}$ where $j_1< j$. This may suggest a delivery scheme with $\min\{N_{\textup{e}}(\dv),\Nsf-\rsf+1,\Ksf-t\}$ steps, where in Step~$j$ we transmit $\binom{\Nsf-j}{\rsf-1}\binom{\Ksf-j}{t}$ linear combinations such that each linear combination contains one of the $\binom{\Nsf-j}{\rsf-1}\binom{\Ksf-j}{t}$ sub-blocks desired by user $u_j$, and thus at the end of this step user $u_j$ is satisfied. 

We will use this interpretation to design the  first sub-phase of our general delivery scheme  in~\eqref{eq:masterach}, which we shall introduced next in Section~\ref{sec:achievable scheme}. 
\end{enumerate}

\section{Achievable Scheme}
\label{sec:achievable scheme}
In this section, we analyze the achievable scheme in~\eqref{eq:masterach} and prove the statements of  Theorems~\ref{thm:scheme 1},~\ref{thm:scheme 2} and~\ref{thm:exact optimality}. 
Notice that when $\rsf \in \{1,\Nsf\}$, the considered problem is equivalent to the MAN problem (solved under the constraint of uncoded cache placement in~\cite{exactrateuncoded}).  Hence, the novelty of our schemes is for $\rsf \in [2:\Nsf-1]$.
%Hence, in this section, we focus on the case where $\rsf \in [2:\Nsf-1]$. 
The scheme we propose was summarized in~\eqref{eq:masterach}; 
Theorems~\ref{thm:scheme 1} and~\ref{thm:exact optimality} only use the first sub-phase of the delivery, while
Theorem~\ref{thm:scheme 2} uses both sub-phases.

The rest of this section is organized as follows.
In Section~\ref{sub:scheme 1: an example} we give an example of the first sub-phase of the proposed delivery scheme in~\eqref{eq:masterach}; the objective is to highlight how the multicast messages sent in sub-phase~1 enable all leaders to decode their desired file. %when every user requests a distinct file. 
Then in Section~\ref{sub:scheme 1} we show which user can decode which sub-block after receiving the multicast messages in sub-phase~1, regardless of the demand type.
In Section~\ref{sub:scheme 2} we show that every user can decode its desired message by also receiving the multicast messages sent in sub-phase~2. In Section~\ref{sub:scheme 2: an example} we give an example of the second sub-phase of the proposed delivery scheme in~\eqref{eq:masterach}.
In Section~\ref{sub:proof thm:order optimality} we prove the order optimality results in Theorems~\ref{thm:order optimality} for general case.
Finally, in Section~\ref{sub:scheme 3 r=2} we prove the exact optimality results in Theorem~\ref{thm:exact optimality} by observing each non-leader can re-construct the packets of sub-phase~2 by performing linear combinations of the the received packets in sub-phase~1.

\subsection{An example of~\eqref{eq:masterach} with only sub-phase~1 for the delivery scheme}%{Theorem~\ref{thm:scheme 1}: an example}
\label{sub:scheme 1: an example}
%\begin{example}[$\Nsf=\Ksf=4$, $\Msf=1/2$, $\rsf=2$]\rm
\label{ex:scheme 1}

First, we study an example where $\Nsf \geq \Ksf$ and where each user demands a distinct file (i.e., $s=\Ksf$).  
In particular, we consider the $(\Nsf,\Ksf,\rsf,\Msf)=(4,4,2,1/2)$ shared-link caching problem with correlated files.
There are $\binom{\Nsf}{\rsf}= 6$ blocks denoted as $W_{\{1,2\}}$, $W_{\{1,3\}}$,   $W_{\{1,4\}}$, $W_{\{2,3\}}$, $W_{\{2,4\}}$, and $W_{\{3,4\}}$.
%Each file $F_{i}, i\in [\Nsf],$ is composed of $W_{\Sc}$ where $\Sc\subseteq [\Nsf]$, $|\Sc|=\rsf$ and $i\in \Sc$. Hence,
The files are 
\begin{align*}
F_{1}&=\{W_{\{1,2\}},W_{\{1,3\}},W_{\{1,4\}}\},\\
F_{2}&=\{W_{\{1,2\}},W_{\{2,3\}},W_{\{2,4\}}\},\\
F_{3}&=\{W_{\{1,3\}},W_{\{2,3\}},W_{\{3,4\}}\},\\
F_{4}&=\{W_{\{1,4\}},W_{\{2,4\}},W_{\{3,4\}}\}.
\end{align*}

\paragraph*{Block Subdivision} 
Here $t=\frac{\Ksf\Msf \rsf}{\Nsf}=1$. 
We partition each block into $\binom{\Ksf}{t}=4$ equal-length sub-blocks and denote 
$W_{\Sc}
=\{W_{\Sc,\Vc}:\Vc\subseteq [\Ksf], |\Vc|=t=1\}  
=\{W_{\Sc,\{k\}}: k\in [\Ksf]\}$.  
Hence, each sub-block contains $\Bsf/\Big(\binom{\Nsf-1}{\rsf-1} \binom{\Ksf}{t}\Big)=\Bsf/12$ bits.

\paragraph*{Placement Phase} 
The cache placement is inspired by the converse bound (see discussion in Section~\ref{subsec:discussionconvese}). 
%, which is as   the   cache placement in~\cite{yang2018centralizedcorrected}.
User $k\in[\Ksf]$ caches $W_{\Sc,\Vc}$ %for all $\Vc\subseteq [\Ksf]$ of size $|\Vc|=t=1$, 
if $k\in \Vc$, that is, $Z_k = \{W_{\Sc,\{k\}}, \forall \Sc\subseteq [\Nsf] : |\Sc|=\rsf=2\}$.

\paragraph*{Delivery Phase} 
Assume $\dv=(1,2,3,4)$, which has $N_{\textup{e}}(\dv)=4$ distinct demanded files. Pick one user demanding a distinct file, and refer to it as the leader among those users demanding the same file. Since each user has a distinct request in this example, each user is a leader, and the leader set is $[4]$. 
Consider a permutation $\usf$ of the leaders, say $\usf=(1,2,3,4)$. %, where $u_1=1$, $u_2=2$, $u_3=3$, and $u_4=4$.

Our proposed first sub-phase of the general delivery scheme contains $\min\{N_{\textup{e}}(\dv),\Nsf-\rsf+1,\Ksf-t\}=3$ steps;
%(corresponding to $c^{N_{\textup{e}}}_t$ in~\eqref{eq:converse type s} for the converse bound) and 
after the $j^{\textrm{th}}$ step, the $j^{\textrm{th}}$ element/leader in the permutation can decode its desired file;  
after finishing all steps, the remaining leaders can also decode their desired file.
We next describe, one by one, the three steps in the delivery phase for this example, 
where each step we send multicast messages of the type
\begin{align}
C_{\Jc,\Bc}:=
\underset{k\in \Jc}{\oplus } 
\underset{\substack{\Sc\subseteq  \Nc(\Jc)\cup \Bc :\\ |\Sc|=\rsf,\Bc \subseteq \Sc, d_{k}\in \Sc }}{\oplus } W_{\Sc,\Jc\setminus\{k\}},
\ \text{for $\Jc \subseteq [\Ksf] : |\Jc|=t+1$ and $\Bc \subseteq  [\Nsf]$,}
\label{eq:CJi general}
\end{align}
where $\Nc(\Jc)$ is the set of demanded files by the users in $\Jc$.
%For each set of users $\Tc \subseteq [\Ksf]$ where $|\Tc|=t+1=2$, and each set of files $\Hc \subseteq  [\Nsf]$, let
In plain words, the multicast message $C_{\Tc,\Hc}$ in~\eqref{eq:CJi general} is the binary sum of each sub-block desired by one user in $\Jc$  and known by all the other users in $\Jc$. 
Note that, when $\rsf=1$ (in which case our model reduces to the MAN system in~\cite{dvbt2fundamental}), 
$C_{\Tc,\emptyset}$ in~\eqref{eq:CJi general} is equivalent to the MAN multicast message
 %$M_{\Jc}$ proposed  in~\cite{dvbt2fundamental} given by
\begin{align}
%M_{\Tc}=
C_{\Jc,\emptyset} = \underset{k\in \Jc}{\oplus } W_{\{d_k\},\Jc \setminus \{k\}}.
\label{eq:MAN multicast message}
\end{align}

{\it Delivery Sub-Phase~1.Step~$1$.} 
In this step we aim to satisfy leader user~$u_1=1$, who misses three sub-blocks of the three blocks the made up the first file, that is, user~$1$ must recover nine sub-blocks. 
Each time we consider one set of users $\Jc\subseteq [\Ksf]$ where $|\Jc|=t+1=2$ and $u_1\in \Jc$  (recall that $u_1=1$), and one set of files $\Bc\subseteq [\Nsf]\setminus \{d_{u_1}\}$ (recall that $d_{u_1}=1$) where $|\Bc|=\rsf-1=1$. 
\begin{subequations}
For example, for $\Jc=\{1,2\}$ and $\Bc=\{2\}$, we transmit 
\begin{align}
%C_{\Jc,\Bc}=
C_{\{1,2\},\{2\}}=W_{\{1,2\},\{2\}} \oplus W_{\{1,2\},\{1\}}.\label{eq:example 1 C_{12,2}}
\end{align}
In $C_{\{1,2\},\{2\}}$, user~$1$ knows $W_{\{1,2\},\{1\}}$ and can thus decode $W_{\{1,2\},\{2\}}$. 
Similarly, user~$2$ knows $W_{\{1,2\},\{2\}}$ and can thus decode $W_{\{1,2\},\{1\}}$.
Similarly, we transmit
\begin{align}
&C_{\{1,2\},\{3\}}=W_{\{1,3\},\{2\}} \oplus W_{\{2,3\},\{1\}};\label{eq:example 1 C_{12,3}}\\
&C_{\{1,2\},\{4\}}=W_{\{1,4\},\{2\}} \oplus W_{\{2,4\},\{1\}};\label{eq:example 1 C_{12,4}}\\
&C_{\{1,3\},\{2\}}=W_{\{1,2\},\{3\}} \oplus W_{\{2,3\},\{1\}};\label{eq:example 1 C_{13,2}}\\
&C_{\{1,3\},\{3\}}=W_{\{1,3\},\{3\}} \oplus W_{\{1,3\},\{1\}};\label{eq:example 1 C_{13,3}}\\
&C_{\{1,3\},\{4\}}=W_{\{1,4\},\{3\}} \oplus W_{\{3,4\},\{1\}};\label{eq:example 1 C_{13,4}}\\
&C_{\{1,4\},\{2\}}=W_{\{1,2\},\{4\}} \oplus W_{\{2,4\},\{1\}};\label{eq:example 1 C_{14,2}}\\
&C_{\{1,4\},\{3\}}=W_{\{1,3\},\{4\}} \oplus W_{\{3,4\},\{1\}};\label{eq:example 1 C_{14,3}}\\
&C_{\{1,4\},\{4\}}=W_{\{1,4\},\{4\}} \oplus W_{\{1,4\},\{1\}}.\label{eq:example 1 C_{14,4}}
\end{align}
\label{eq:example 1 allC step1}
\end{subequations}

From~\eqref{eq:example 1 allC step1} %{eq:example 1 C_{12,2}} to~\eqref{eq:example 1 C_{14,4}} 
and its cached content, user~$u_1=1$ can recover $W_{\{1,2\}}$, $W_{\{1,3\}}$, and $W_{\{1,4\}}$.
 User~$u_1=1$ is satisfied after this first step (i.e., it has recovered the missing nine sub-blocks from the nine received multicast messages in the first step). 

Let us then focus on user~$u_2=2$. User~$2$ can directly recover $W_{\{1,2\},\{1\}}$ from~\eqref{eq:example 1 C_{12,2}},   $W_{\{2,3\},\{1\}}$ from~\eqref{eq:example 1 C_{12,3}},   $W_{\{2,4\},\{1\}}$ from~\eqref{eq:example 1 C_{12,4}}. Since user~$2$ has recovered  $W_{\{2,3\},\{1\}}$, it  then can recover $W_{\{1,2\},\{3\}}$ from~\eqref{eq:example 1 C_{13,2}}. Since user~$2$ has recovered   $W_{\{2,4\},\{1\}}$, it  then can recover $W_{\{1,2\},\{4\}}$ from~\eqref{eq:example 1 C_{14,2}}. In conclusion, after Step~$1$, user~$2$ can recover $W_{\{1,2\}}$ and also recover $W_{\{2,3\},\{1\}}$ and $W_{\{2,4\},\{1\}}$. 
User~$u_2=2$ after this first step still misses four sub-blocks, namely $\{W_{\{2,3\},\{k\}}, W_{\{2,4\},\{k\}} : k\in[3,4]\}$. 

Similarly to user~$u_2=2$, each user $k\in \{2,3,4\}$ can recover $W_{\Sc}$ where $\{d_{u_1},d_k\}\subseteq \Sc$, and can also recover $W_{\Sc_1,\Vc_1}$ where $d_k \in \Sc_1$ and $u_1 \in \Vc_1$, after Step~$1$. 
Each of these users still misses four sub-blocks.

{\it Delivery Sub-Phase~1.Step~$2$.} 
In this step we aim to satisfy leader user~$u_2=2$. 
Each time we consider one set of users $\Jc\subseteq ([\Ksf]\setminus\{u_1\})$ where $|\Jc|=t+1$ and $u_2\in \Jc$, and one set of files $\Bc\subseteq ([\Nsf]\setminus \{d_{u_1},d_{u_2}\})$ where $|\Bc|=\rsf-1=1$ (recall that $u_1=d_{u_1}=1, u_2=d_{u_2}=2$). 
\begin{subequations}
For example, we for $\Jc=\{2,3\}$ and $\Bc=\{3\}$, we transmit 
\begin{align}
%C_{\Jc,\Bc}=
C_{\{2,3\},\{3\}}=W_{\{2,3\},\{3\}} \oplus W_{\{2,3\},\{2\}}.
\label{eq:example 1 C_{23,3}}
\end{align}
From~\eqref{eq:example 1 C_{23,3}} user~$2$ can recover $W_{\{2,3\},\{3\}}$ and user~$3$ can recover $W_{\{2,3\},\{2\}}$. 
Similarly, we transmit
\begin{align}
&C_{\{2,3\},\{4\}}=W_{\{2,4\},\{3\}} \oplus W_{\{3,4\},\{2\}};\label{eq:example 1 C_{23,4}}\\
&C_{\{2,4\},\{3\}}=W_{\{2,3\},\{4\}} \oplus W_{\{3,4\},\{2\}};\label{eq:example 1 C_{24,3}}\\
&C_{\{2,4\},\{4\}}=W_{\{2,4\},\{4\}} \oplus W_{\{2,4\},\{2\}}.\label{eq:example 1 C_{24,4}}
\end{align}
\label{eq:example 1 allC step2}
\end{subequations}

From~\eqref{eq:example 1 allC step2} %{eq:example 1 C_{23,3}} to~\eqref{eq:example 1 C_{24,4}},
user~$u_2=2$ can recover the desired sub-blocks that were not recovered from Step~$1$. 
User~$u_2=2$ is satisfied after this second step (i.e., it has recovered the missing four sub-blocks from the four received multicast messages in the second step). 

Let us then focus on user~$u_3=3$. User~$3$ can directly recover $W_{\{2,3\},\{2\}}$ from~\eqref{eq:example 1 C_{23,3}} and $W_{\{3,4\},\{2\}}$ from~\eqref{eq:example 1 C_{23,4}}. Since user~$3$ has recovered $W_{\{3,4\},\{2\}}$, it then can recover $W_{\{2,3\},\{4\}}$ from~\eqref{eq:example 1 C_{24,3}}. 
 User~$u_3=3$ after this second step still misses $W_{\{3,4\},\{4\}}$. 

Similarly to user~$u_3=3$, at the end of Step~$2$, each user $k\in \{3,4\}$ can recover $W_{\Sc}$ where $d_k \in \Sc$, $\{d_{u_1},d_{u_2}\}\cap \Sc\neq \emptyset$, and also recover $W_{\Sc_1,\Vc_1}$ where $d_k \in \Sc_1$ and $\{u_1,u_2\}\cap  \Vc_1 \neq \emptyset$. 
Each of these users still misses one sub-block.

{\it Delivery Sub-Phase~1.Step~$3$.} 
In this step we aim to satisfy leader user~$u_3=3$.
Each time we consider one set of users $\Jc\subseteq ([\Ksf]\setminus\{u_1,u_2\})$ where $|\Jc|=t+1$ and $u_3\in \Jc$, and one set of files $\Bc\subseteq ([\Nsf]\setminus \{d_{u_1},d_{u_2},d_{u_3}\})$ where $|\Bc|=\rsf-1=1$ (recall that $u_1=d_{u_1}=1, u_2=d_{u_2}=2, u_3=d_{u_3}=3$). 
Hence, at this point there is one possibility, $\Jc=\{3,4\}$ and $\Bc=\{4\}$, for which we transmit
\begin{align}
%C_{\Jc,\Bc}=
C_{\{3,4\},\{4\}}=W_{\{3,4\},\{4\}} \oplus W_{\{3,4\},\{3\}}.\label{eq:example 1 C_{34,4}}
\end{align}
From~\eqref{eq:example 1 C_{34,4}}, user~$3$ can recover $W_{\{3,4\},\{4\}}$, and user~$4$ can recover $W_{\{3,4\},\{3\}}$. 
 Hence, at the end this third step, users $3$ and $4$ are satisfied (i.e., they recovered the missing sub-block from the received multicast message in the third step).

\paragraph*{Performance} 
Based on the above placement and delivery scheme, all users are able to decode their desired blocks.
We sent %$\binom{\Nsf-1}{\rsf-1}\binom{\Ksf-1}{t}+\binom{\Nsf-2}{\rsf-1}\binom{\Ksf-2}{t}+\binom{\Nsf-3}{\rsf-1}\binom{\Ksf-3}{t}$
$\sum_{j} \binom{\Nsf-j}{\rsf-1}\binom{\Ksf-j}{t}
=14$ linear combinations, each of length $\Bsf/12$ bits. So the  load is $7/6$, which coincides with the converse bound in Theorem~\ref{thm:converse} for $s=4$.

\paragraph*{Comparison with state-of-the-art `round-division' schemes}
Let us then consider the round-division methods  in~\cite{yang2018centralizedcorrected,ji2015multirequest,Sengupta2017multirequest,multireqWei2017}. It is obvious that if there exists some sub-block appearing in different rounds, a round-division strategy that treats each round as an independent MAN caching problem with a single request may miss some multicast opportunities.   Here we show that a round-division strategy is sub-optimal even if we can divide users' demands into multiple rounds such that there does not exist any sub-block appearing in different rounds.
More precisely, since each user demands $3$ blocks, we can divide the delivery into the following three rounds:
\begin{itemize}

\item 
Round~1:
In the first round, users~$1$ and~$2$ demand $W_{\{1,2\}}$, and users~$3$ and~$4$ demand $W_{\{3,4\}}$. This is equivalent to the MAN caching problem with $4$ users and $2$ files. By using the optimal caching scheme under the constraint of uncoded cache placement in~\cite{exactrateuncoded}, we need to transmit $\binom{\Ksf}{t}-\binom{\Ksf-\Nsf}{t}=\binom{4}{2}-\binom{2}{2}=5$ linear combinations, each of which contains $\Bsf/12$ bits, in order to satisfy these requests.

\item 
Round~2:
In the second round, users~$1$ and~$3$ demand $W_{\{1,3\}}$, and users~$2$ and~$4$ demand $W_{\{2,4\}}$. By using the caching scheme  in~\cite{exactrateuncoded}, we need to transmit $5$ linear combinations to satisfy these requests.

\item
Round~3:
In the third round, users~$1$ and~$4$ demand $W_{\{1,4\}}$, and users~$2$ and~$3$ demand $W_{\{2,3\}}$. By using the caching scheme  in~\cite{exactrateuncoded}, we need to transmit $5$ linear combinations to satisfy these requests.

\end{itemize} 
Hence, by this round-division strategy, the load is $15/12 > 7/6$, which is strictly sub-optimal.
In conclusion, in order to achieve optimality in this example, we need to jointly serve users' demands (as proposed in this paper) in order to fully leverage all multicast opportunities.
%\hfill$\square$
%\end{example}

\subsection{Proof of Theorem~\ref{thm:scheme 1}} 
\label{sub:scheme 1}
 
Here we shall prove that after the first sub-phase of the delivery scheme in~\eqref{eq:masterach} every leader user is able to decode its desired file (as in the example in Section~\ref{sub:scheme 1: an example}), and that the load of the first sub-phase matches the load of the converse bound in~\eqref{eq:ct}. Thus, for the case where every user is a leader (i.e., every user demands a distinct file, as in the example in Section~\ref{sub:scheme 1: an example}), we have proved the exact optimality under the constraint of uncoded cache placement of the proposed achievable scheme as claimed in Theorem~\ref{thm:scheme 1}.

\paragraph*{Decodability after delivery sub-phase~1}
We need to establish which user can decode which sub-block at each step of delivery sub-phase~1.
The following Lemma~\ref{lem:Jc}, which is proved by induction in Appendix~\ref{sec:proof of direct lemma}, describes the decoding procedure for delivery sub-phase~1 for general demands. Lemma~\ref{lem:Jc} is the most technical (i.e., highly combinatorial) contribution in this paper. 
%Lemma~\ref{lem:Jc} describes which desired sub-blocks can be decoded by each user after the transmission of the multicast messages in sub-phase~1 of the general delivery scheme.

\begin{lem}[Decoding after sub-phase~1] %of the general delivery scheme]
\label{lem:Jc}
In the first sub-phase of the proposed delivery scheme in~\eqref{eq:masterach} with leader set $\{u_1,\ldots,u_{N_{\textup{e}}(\dv)}\}$, 
in Step~$j\in [\min\{N_{\textup{e}}(\dv),\Nsf-\rsf+1,\Ksf-t\}]$, 
for each set of users $\Jc$\footnote{Please note that here we write the set in the first subscript of $C_{\Jc,\Bc}$ is an equivalent but slightly different form compared to~\eqref{eq:masterach:Jc in step 1}.}
where $\Jc\subseteq [\Ksf]\setminus\{u_1,\ldots,u_{j-1}\}$ such that $|\Jc|=t+1$ and $u_j\in \Jc$, and 
for each set of files $\Bc\subseteq [\Nsf]\setminus \{d_{u_1},\ldots,d_{u_{j}}\}$ where $|\Bc|=\rsf-1$,  we transmit $C_{\Jc,\Bc}$ as defined in~\eqref{eq:CJi general}.
Note that by construction (i.e., $|\Bc|=\rsf-1$ and $d_{u_j}\notin \Bc$), $C_{\Jc,\Bc}$ contains only one sub-block desired by user $u_j$ (which is $W_{\{d_{u_j}\}\cup \Bc, \Jc\setminus \{u_j\} }$), while all other sub-blocks are cached by user $u_j$.

Let $u_{g(i)}$ represent the leader user who demands file $i\in\Lc(\dv)$.
At the end of the first delivery sub-phase, we have:
\begin{enumerate}

\item \label{item:lem JC1}
For a $(\Jc,\Bc)$, each user in $\Jc$ can recover all the sub-blocks in $C_{\Jc,\Bc}$. %(by combining its cached content with the received multicast messages until Step~$j$).
%\dt{HOW ABOUT THE TERM FOR USER $k\in\Jc : d_{k}\in\Bc$, i.e., $\underset{ i \in \Nc(\Jc)\setminus \Bc }{\oplus } W_{\Bc\cup\{i\},\Jc\setminus\{k\}}$?}

\item \label{item:lem JC2}
%For each user $k\in [\Ksf]$,
At the end of Step~$j \in [\min\{g(d_{k}), \Nsf-\rsf+1,\Ksf-t\}]$, user $k\in [\Ksf]$ can recover $W_{\Sc_1,\Vc_1}$
 if  $d_{k}\in \Sc_1$ and $\{u_{1},\ldots,u_j\} \cap \Vc_1 \neq \emptyset$.

\item \label{item:lem JC3}
%For each user $q\in [\Ksf]$,
At the end of Step~$j \in [\min\{g(d_{q})-1, \Nsf-\rsf+1,\Ksf-t\}]$, user $q\in [\Ksf]$ can recover  $W_{\Sc}$
 if  $d_{q}\in \Sc$ and $\{d_{u_1},\ldots,d_{u_j}\} \cap \Sc \neq \emptyset$.

\end{enumerate}
\end{lem}

\paragraph*{Decodability for leader users after sub-phase~1}
%In order to prove Theorem~\ref{thm:scheme 1}, 
We use Lemma~\ref{lem:Jc} to show that every leader user is able to recovered its demanded file after delivery sub-phase~1. Indeed, for any system parameters, for leader user $u_p$, where $p\in [N_{\textup{e}}(\dv)]$, we have:
%\dt{WHY ARE WE ASSUMING $\min\{\Nsf-\rsf+1,\Ksf-t\} \leq N_{\textup{e}}(\dv)$? THIS SEEMS WRONG, SO I CHANGED THE FOLLOWING.}
\begin{itemize}

\item
Case $p\leq \min\{ N_{\textup{e}}(\dv),\Nsf-\rsf+1,\Ksf-t\}$.

From Lemma~\ref{lem:Jc}.Item~\ref{item:lem JC3} , %it can be seen that 
user $u_p$ can recover $W_{\Sc}$, where  $d_{u_p}\in \Sc$ and $\{d_{u_1},\dots,d_{u_{p-1}}\} \cap \Sc \neq \emptyset$, at the end of Step~$p-1$. 

In addition, from Lemma~\ref{lem:Jc}.Item~\ref{item:lem JC2} , user $u_p$ can also recover $W_{\Sc_1,\Vc_1}$, where  $d_{u_p}\in \Sc_1$ and $\{u_{1},\ldots,u_{p-1}\} \cap \Vc_1 \neq \emptyset$, at the end of Step~$p-1$. 

Hence, user $u_p$ still needs to recover $W_{\Sc_2,\Vc_2}$, where $d_{u_p}\in \Sc_2$, $\{d_{u_1},\ldots,d_{u_{p-1}}\}\cap \Sc_2=\emptyset$ and $\{u_1,\ldots,u_{p}\} \cap  \Vc_2=\emptyset$, at the end of Step~$p-1$ (recall that $W_{\Sc_2,\Vc_2}$ where $u_p \in \Vc_2$  is cached by user $u_p$). Such a $W_{\Sc_2,\Vc_2}$ appears in $C_{\Vc_2\cup\{u_p\}, \Sc_2\setminus \{d_{u_p}\}},$ which is sent in Step~$p$. 
Hence, from Lemma~\ref{lem:Jc}.Item~\ref{item:lem JC1} , %we prove that 
user $u_p$ can recover $W_{\Sc_2,\Vc_2}$  at the end of Step~$p$. 

\item 
 Case $N_{\textup{e}}(\dv) > \min\{\Nsf-\rsf+1,\Ksf-t\}$ and $\min\{\Nsf-\rsf+1,\Ksf-t\} < p \leq N_{\textup{e}}(\dv)$. 

We distinguish two cases:
\begin{itemize}
\item
If $\Nsf-\rsf+1\leq \Ksf-t$, it can be seen that for each desired block of user $u_p$ (assumed to be $W_{\Sc}$), we have $\Sc\cap \{d_{u_1},\ldots,d_{u_{\Nsf-\rsf+1}}\}\neq \emptyset$, and thus from Lemma~\ref{lem:Jc}.Item~\ref{item:lem JC3} , user $u_p$ can recover $W_{\Sc}$.
\item
If $\Nsf-\rsf+1 > \Ksf-t$, it can be seen that for each desired sub-block of user $u_p$ (assumed to be $W_{\Sc_1,\Vc_1}$), we have $\Vc_1\cap \{u_1,\ldots,u_{\Nsf-\rsf+1}\}\neq \emptyset$, and thus from Lemma~\ref{lem:Jc}.Item~\ref{item:lem JC2} , user $u_p$ can recover $W_{\Sc_1,\Vc_1}$.
\end{itemize}

\end{itemize}
This proves that each leader can recover its demanded file after sub-phase~1.

\iffalse
==================
\dt{WHY THE CASE OF LARGE s (LIKE $s\geq \min\{\Nsf-\rsf+1,\Ksf-t\}$) CANNOT BE SOLVED, BUT ONLY $s=K\geq N$ IN THM1? HAVE TO THINK ABOUT WHAT KAI SAYS NEXT:}
{\magenta Kai's answer: we cannot say we solve the case  $N_{\textup{e}}(\dv) \geq \min\{\Nsf-\rsf+1,\Ksf-t\}$  for $\rsf \geq 3$. If $\rsf=1$, we can say it because if $N_{\textup{e}}(\dv) \geq  \Ksf-t $, each $\Jc$ where $|\Jc|=t+1$ is transmitted such that each user can recover all its demanded subfiles. However, when $\rsf \geq 3$, even if all $\Jc$'s are transmitted, it is still hard to let each non-leader successfully decode. It is because, the number of transmitted packets for each $\Jc$ is different. For example, N=5, K=6, r=3, t=3, and d=(1,2,3,4,5,1).  If the leader permutation is (1,2,3,4,5). 
For $\Jc=\{3,4,5,6\}$, We only transmit one packet
$C_{3456,45}=W_{345,456}+W_{345,356}+W_{345,346}+W_{145,345}$.
 Hence, from this packet, user 6 can only recover directly one sub-block, which is $W_{145,345}$. However, there is no $\Jc$ including user 6 and also including $W_{123,345}$, nor $W_{124,345}$, nor $W_{125,345}$.}
==================
\fi

\paragraph*{Load of sub-phase~1}
%The file subdivision, the placement phase and the delivery phase are as in~\eqref{eq:masterach}. 
This proposed sub-phase~1 of the delivery scheme contains $\binom{\Nsf-j}{\rsf-1}\binom{\Ksf-j}{t}$ multicast messages in Step~$j\in [\min\{N_{\textup{e}}(\dv),\Nsf-\rsf+1,\Ksf-t\}]$, which follows the intuition from the proof of our converse bound (see discussion in Section~\ref{subsec:discussionconvese}). Thus, by summing over all steps in sub-phase~1, we get that the load of this delivery sub-phase matches the load of the converse bound in~\eqref{eq:ct}. 
%\footnote{
%Note that, if $\rsf=1$, we have $|\Bc|=\rsf-1=0$ and thus $C_{\Jc,\emptyset}$ is equivalent to the MAN multicast message defined in~\eqref{eq:MAN multicast message}.  Hence, the proposed scheme is equivalent to the optimality caching scheme under the constraint of uncoded cache placement in~\cite{exactrateuncoded} for the MAN caching problem with any demand, which transmits % $M_{\Jc}$ 
%$C_{\Jc,\emptyset}$ for each $\Jc\subseteq [\Ksf]$ where $|\Jc|=t+1$ and \dt{??WHY THIS TOO $\Jc\cap \Lc\neq \emptyset$??}.
%{\magenta Kai's answer: it is because in each Step~$j$, $\Jc$ contains the leader $u_j$.}
%}  

\paragraph*{Optimality for the case of distinct demands} 
From the above reasoning, when all users are leaders, that is for the case $\Nsf \geq \Ksf$ and demand type $s=\Ksf$, the claim of Theorem~\ref{thm:scheme 1} is proved, i.e., every user is satisfied at the end of sub-phase~1, whose load matches the converse bound.

\subsection{Proof of Theorem~\ref{thm:scheme 2}} %: General Two-phase-delivery Interference Alignment Scheme}
\label{sub:scheme 2}
Here we shall prove that after the two sub-phases of the delivery scheme in~\eqref{eq:masterach} %from Section~\ref{ex:scheme 2} 
every user is able to decode its desired file. This requires showing that after the second sub-phase the demands of all non-leader users are satisfied. Sub-phase~2 of the delivery scheme in~\eqref{eq:masterach} is a form of interference alignment.

The block split and the cache placement phase are as described in~\eqref{eq:masterach}. 
The delivery phase contains two sub-phases, where the first sub-phase is the same as in Section~\ref{sub:scheme 1}, and the second sub-phase is such that non-leader can align or cancel the non-demanded sub-blocks and eventually decode their demanded file. We specify next what each user can decode at the end of each step.
%In particular:

\paragraph*{First Delivery Sub-Phase} 
In Step~$j\in [\min\{N_{\textup{e}}(\dv),\Nsf-\rsf+1,\Ksf-t\}]$ of the first sub-phase, for each  set of users $\Jc\subseteq [\Ksf]\setminus\{u_1,\ldots,u_{j-1}\}$ where $|\Jc|=t+1$ and $u_j\in \Jc$, and each set of files $\Bc\subseteq [\Nsf]\setminus \{d_{u_1},\ldots,d_{u_{j}}\}$ where $|\Bc|=\rsf-1$, we transmit $C_{\Jc,\Bc}$ as defined in~\eqref{eq:CJi general}.
As shown in Section~\ref{sub:scheme 1}, at the end of this sub-phase, each leader can recover its desired file.  

In addition, from Lemma~\ref{lem:Jc},  recalling that $u_{g(i)}$ represent the leader user who demands file $i$,
each non-leader user $k\in[\Ksf]\setminus\Lc(\dv)$ can decode   $W_{\Sc}$, where $d_k\in \Sc$ and $\{d_{u_1},\ldots,d_{u_{g(d_k)-1}} \}\cap \Sc\neq \emptyset$, and can decode $W_{\Sc_1,\Vc_1}$ where $d_k\in \Sc_1$ and $\{u_1,\ldots,u_{g(d_k)}\}\cap \Vc_1 \neq \emptyset$.

The non-leader users are thus not yet satisfy, and thus we proceed to send further multicast messages in sub-phase~2.

%%%can decode $W_{\Sc}$, where $d_k\in \Sc$ and $\{d_{u_1},\ldots,d_{u_{g(d_k)-1}} \}\cap \Sc\neq \emptyset$, and can decode $W_{\Sc_1,\Vc_1}$ where $d_k\in \Sc_1$ and $\{u_1,\ldots,u_{g(d_k)}\}\cap \Vc_1 \neq \emptyset$ by direct decoding.
%%%%
%%%It will be shown later, user $k$ will decode the remaining sub-blocks by {\bf interference alignment decoding} from the transmitted messages in both two sub-phases. For a non-leader $k$, we now summarize which transmitted messages in the first sub-phase are useful to user $k$, and which ones contain interferences to it.

%\begin{itemize}

%\item $C_{\Jc,\Bc}$ transmitted in  Step~$j<g(d_{k})$ \dt{NO $\min\{g(d_{k}), \Nsf-\rsf+1,\Ksf-t\}$ AS IN LEMMA\ref{lem:Jc}?}. If $k\in\Jc$ or $d_k \in \Bc$,  $C_{\Jc,\Bc}$ is useful to user $k$ and in  $C_{\Jc,\Bc}$, each sub-block is either cached or desired by it; otherwise, user $k$ discards $C_{\Jc,\Bc}$ / treat it as noise.

%\item $C_{\Jc,\Bc}$ transmitted in  Step~$g(d_{k})$.  $C_{\Jc,\Bc}$ is useful to user $k$. If $k\in\Jc$, the sub-blocks in $C_{\Jc,\Bc}$ are either cached or desired by it; otherwise, in $C_{\Jc,\Bc}$ only the sub-blocks of $W_{\Bc\cup \{d_k\}}$ are desired by user $k$ and other sub-blocks  are interferences to it.

%\item $C_{\Jc,\Bc}$ transmitted in  Step~$j> g(d_{k})$. If $k\in\Jc$,  $C_{\Jc,\Bc}$ is useful to user $k$ and in  $C_{\Jc,\Bc}$, each sub-block is either cached or desired by it; otherwise, user $k$ discards $C_{\Jc,\Bc}$ / treat it as noise.

%\end{itemize}

\paragraph*{Second Delivery Sub-Phase}  
The second sub-phase also contains $\min\{N_{\textup{e}}(\dv),\Nsf-\rsf+1,\Ksf-t\}$ steps. In Step~$j$, each time we focus on one integer $q\in [j+1: \min\{\Nsf-\rsf+2,\Ksf-t+1,N_{\textup{e}}(\dv)\}]$. 
For each $\Jc^{\prime}\subseteq [\Ksf]\setminus\{u_1,\ldots,u_{q}\}$ where $|\Jc^{\prime}|=t-1$ and $\Jc^{\prime}\cap \{u_{q+1},\ldots,u_{N_{\textup{e}}(\dv)}\}\neq \emptyset$, and each $\Bc^{\prime}\subseteq [\Nsf]\setminus \{d_{u_1},\ldots,d_{u_{q}}\}$ where $|\Bc^{\prime}|=\rsf-2$ and $\Bc^{\prime}\cap \Nc([\Ksf])\neq \emptyset$, we transmit $C_{\Jc^{\prime}\cup\{u_j,u_q\},\Bc^{\prime}}$ as defined in~\eqref{eq:CJi general}.
 We describe next how each non-leader users can recover the demanded file by combining the multicast messages from both sub-phases. The decoding is rather involved, thus we break down the key steps into lemmas that are proved in Appendix.  
 
%Each transmitted message in Step~$j$ of the second sub-phase (assumed to be $C_{\Jc^{\prime}\cup\{u_j,u_q\},\Bc^{\prime}}$), is useful to all non-leaders whose demanded file is in $\Bc^{\prime} \cup  \Nc(\Jc^{\prime}\cup\{u_j,u_q\})$, while all other users discard it / treat it as noise. For each non-leader user $k$, with $k\in\Jc^{\prime}$, all sub-blocks in $C_{\Jc^{\prime}\cup\{u_j,u_q\},\Bc^{\prime}}$ are either cached or desired; otherwise, in $C_{\Jc^{\prime}\cup\{u_j,u_q\},\Bc^{\prime}}$ the sub-blocks  which are not contained in $F_{d_k}$, are interference to user $k$.

In Step~$j$ of the second sub-phase, the transmitted multicast message $C_{\Jc^{\prime} \cup\{u_j,u_q\},\Bc^{\prime} }$  by construction satisfies  $\Jc^{\prime}\cap \{u_{q+1},\ldots,u_{N_{\textup{e}}(\dv)}\}\neq \emptyset$; 
 however, non-leader user $k$ also needs multicast message such that $\Jc^{\prime}\cap \{u_{q+1},\ldots,u_{N_{\textup{e}}(\dv)}\} = \emptyset$. 
It is proved in Appendix~\ref{sec:proof of general scheme lemma}
that each user $k$  who demands $F_{d_{u_j}}$ can reconstruct $C_{\Jc^{\prime} \cup\{u_j,u_q\},\Bc^{\prime} }$, where $\Jc^{\prime}\cap \{u_{q+1},\ldots,u_{N_{\textup{e}}(\dv)}\}= \emptyset$ by using previously received multicast messages, as formalized in the next lemma.
%%%{\blue  e.g., please see that user $1$ can reconstruct $C_{\{1,2,9,10\},\{3\}}$ in the example in Section~\ref{sub:scheme 2: an example}.}  

\begin{lem}
\label{lem:additional decoding}
In Step~$j\in[\min\{N_{\textup{e}}(\dv),\Nsf-\rsf+1,\Ksf-t\}]$ of sub-phase~2, for any integer $q\in [j+1: \min\{\Nsf-\rsf+2,\Ksf-t+1,N_{\textup{e}}(\dv)\}]$, each $\Jc^{\prime}\subseteq [\Ksf]\setminus\{u_1,\ldots,u_{N_{\textup{e}}(\dv)}\}$ where $|\Jc^{\prime}|=t-1$, and each $\Bc^{\prime}\subseteq [\Nsf]\setminus \{d_{u_1},\ldots,d_{u_{q}}\}$ where $|\Bc^{\prime}|=\rsf-2$ and $\Bc^{\prime}\cap \Nc([\Ksf])\neq \emptyset$, user $k$ who demands $d_{u_j}$ can  obtain $C_{\Jc^{\prime}\cup\{u_j,u_q\},\Bc^{\prime}}$ by making linear combinations of already received multicast messages. 
\end{lem}

\iffalse
{\magenta Comments of Kai: the proof of Lemma~\ref{lem:additional decoding} is the most combinatorial proof in our paper. So I suggest to continue your reading and focus on it in the end.}
 \fi

The following Lemma~\ref{lem:lemma for interference alignment}, whose proof is in Appendix~\ref{sec:proof of transition lemma},   specifies some properties of the linear combinations $C_{\Tc,\Hc}$ defined in~\eqref{eq:CJi general}.  
%we introduce some properties of $C_{\Tc,\Hc}$ defined in~\eqref{eq:CJi general}.

\begin{lem}[Properties of function $C_{\Tc,\Hc}$ defined in~\eqref{eq:CJi general}]
\label{lem:lemma for interference alignment}
For each $\Jc\subseteq [\Ksf]$ where $|\Jc|=t+1$, and each $\Bc\subseteq [\Nsf]$ where $|\Bc|=\rsf-1$, we have
\begin{align}
C_{\Jc,\Bc}=\underset{k\in\Jc}{\oplus } C_{(\Jc\setminus \{k\})\cup  \{u_{g(i)}\},(\Bc\setminus\{i\})\cup \{d_k\}},\label{eq:interference transition} 
\end{align}
for any $i\in \Bc$ where $u_{g(i)}\notin \Jc$. In addition, for each   $\Jc_1\subseteq [\Ksf]$ where $|\Jc_1|=t+1$, and each $\Bc_1\subseteq [\Nsf]$ where $|\Bc_1|=\rsf-1$ and $\Nc(\Jc_1)\cap \Bc_1\neq \emptyset$, we have
\begin{align}
C_{\Jc_1,\Bc_1}=\underset{i\in \Nc(\Jc_1)\setminus \Bc_1}{\oplus } C_{\Jc_1,(\Bc_1\setminus\{i_1\})\cup \{i\}},\label{eq:interference transition2} 
\end{align}
for any $i_1 \in \Nc(\Jc_1)\cap \Bc_1$.
\end{lem}
%%%{\blue
%%%In the example  in Section~\ref{sub:scheme 2: an example}, we prove  (with $(d_1,\ldots,d_5)=(1,\ldots,5)$ and $i=u_{g_i}=1$)
%%%$$
%%%C_{\{2,3,4,5\},\{1, 3\}}=C_{\{1,3,4,5\},\{2,3\}}\oplus C_{\{1,2,3,5\},\{3,4\}} \oplus C_{\{1,2,3,4\},\{3,5\}} \oplus  C_{\{1,2,4,5\},\{ 3\}}
%%%$$
%%%which shows the property in~\eqref{eq:interference transition}, and prove 
%%%$$
%%%C_{\{1,2,3,4\},\{1,3\}}=C_{\{1,2,3,4\},\{2,3\}}  \oplus C_{\{1,2,3,4\},\{3,4\}},
%%%$$
%%%which shows the property in~\eqref{eq:interference transition2}.
%%%}

From Lemma~\ref{lem:lemma for interference alignment}, we prove the following Lemma~\ref{lem:lemma for Bc} (whose proof is in Appendix~\ref{sec:proof of lemma Bc}), which is the key result for our interference alignment based delivery scheme. Recall that $\Lc(\dv)$ denotes the set of leader users.

\begin{lem}[Interference alignment lemma]
\label{lem:lemma for Bc}
For each $j\in [\min\{N_{\textup{e}}(\dv),\Nsf-\rsf+1,\Ksf-t\}]$ and each $i\in\{d_{u_1},\ldots,d_{u_{j}}\}$,
any non-leader $k\in[\Ksf]\setminus \Lc(\dv)$ can  reconstruct  $C_{\Jc_2\cup\{u_j\},\Bc_2\cup\{i\}}$ where $\Jc_2\subseteq [\Ksf]\setminus \{u_{1},\ldots,u_{j}\}$, $|\Jc_2|=t$, $\Bc_2\subseteq [\Nsf]\setminus \{d_{u_1},\ldots,d_{u_{j}}\}$, $|\Bc_2|=\rsf-2$, and $\Nc(\Jc_2\cap \Lc)\setminus \Bc_2 \neq \emptyset$.
\end{lem}
%%%{\blue In the example  in Section~\ref{sub:scheme 2: an example} with $j=1$, $u_1=d_{u_{1}}=1$, and $\Lc=[5]$,    we prove that  each user can reconstruct  $C_{\{1,2,3,4\},\{1,3\}}=C_{\{1,2,3,4\},\{2,3\}}  \oplus C_{\{1,2,3,4\},\{3,4\}}$, where $C_{\{1,2,3,4\},\{2,3\}}$ and $ C_{\{1,2,3,4\},\{3,4\}}$ are transmitted in Step~$1$ of sub-phase~1.}

%\paragraph*{Outline of interference alignment decoding}
Lemma~\ref{lem:lemma for Bc} can be understood as follows.
After the first sub-phase, the remaining sub-blocks to be decoded for each non-leader $k \in [\Ksf]\setminus \Lc(\dv)$ are  $W_{\Sc_2,\Vc_2}$ where $d_k \in \Sc_2$, $\{d_{u_1},\ldots,d_{u_{g(d_k)-1}}\} \cap \Sc_2=\emptyset$ and $\{k,u_1,\ldots,u_{g(d_k)}\}\cap \Vc_2 = \emptyset$.
In Step~$g(d_k)$ of the first sub-phase, the transmitted message $C_{\Jc  ,\Bc }$ should satisfy $d_{g(d_k)}\notin \Bc$. From Lemma~\ref{lem:lemma for Bc}, we show user $k$ can also reconstruct  $C_{\Jc^{\prime}  ,\Bc^{\prime}}$ where $d_{g(d_k)}\in  \Bc^{\prime}$. Since $d_{u_{g(d_k)}}\in \Bc^{\prime}$, each sub-block in $C_{\Jc^{\prime},\Bc^{\prime}}$ is desired or cached by user $k$ who demands $F_{k}$. In other words, in order to reconstruct  $C_{\Jc^{\prime}  ,\Bc^{\prime}}$, we align/cancel the interferences to user $k$. By induction, all sub-blocks except one in $C_{\Jc^{\prime}  ,\Bc^{\prime}}$ have been already recovered or cached by user $k$ such that it can recover that sub-block. The details of the decodability proof is presented in Appendix~\ref{sec:decodability general scheme}. 
An example of how the interference alignment scheme works is given in Section~\ref{ex:scheme 2}.

%{\magenta Comments of Kai: the proof in Appendix~\ref{sec:decodability general scheme} is the most important in our paper. So I suggest to pay more attention on it among all appendices.}

\paragraph*{Performance}
As we showed in Section~\ref{sub:scheme 1}, in the first sub-phase  we transmit $c^{s}_t$ bits, with $s=N_{\textup{e}}(\dv)$.  
In Step~$j\in [\min\{s,\Nsf-\rsf+1,\Ksf-t\}], \ s=N_{\textup{e}}(\dv),$ of the second sub-phase, the number of transmitted bits is 
\begin{align}
\frac{\sum^{\min\{\Nsf-\rsf+2,\Ksf-t+1,s\}}_{q=j+1}  \left(\binom{\Nsf-q}{\rsf-2}-\binom{\Nsf-s}{\rsf-2} \right) \left(\binom{\Ksf-q}{t-1}-\binom{\Ksf-s}{t-1} \right) }{\binom{\Nsf-1}{\rsf-1}\binom{\Ksf}{t}}\Bsf.
\end{align} 
Hence, by summing the number of transmitted bits in each step of sub-phase~2 and the number of transmitted bits in sub-phase~1, the load equals $e^{s}_t+c^{s}_t$ as defined in~\eqref{eq:ct} and~\eqref{eq:et},  with $s=N_{\textup{e}}(\dv)$.  

This concludes the proof of Theorem~\ref{thm:scheme 2}.

\subsection{An example of sub-phase~2 in~\eqref{eq:masterach}}%{Theorem~\ref{thm:scheme 2}: an example}
\label{sub:scheme 2: an example}
We will use the following example to illustrate our interference alignment scheme.
%\begin{example}[$\Nsf=5$, $\Ksf=15$, $\rsf=t=3$]
\label{ex:scheme 2}
%\rm

Consider an $(\Nsf,\Ksf,\Msf,\rsf)=(5,10,1/2,3)$ shared-link caching problem with correlated files.
There are $\binom{\Nsf}{\rsf}=10$  blocks, $W_{\Sc}$ where $\Sc\subseteq [5]$ and $|\Sc|=\rsf=3$.
The files are 
\begin{align*}
F_{1}&=\{W_{\{1,2,3\}},W_{\{1,2,4\}},W_{\{1,2,5\}},W_{\{1,3,4\}},W_{\{1,3,5\}},W_{\{1,4,5\}}\},\\
F_{2}&=\{W_{\{1,2,3\}},W_{\{1,2,4\}},W_{\{1,2,5\}}, W_{\{2,3,4\}}, W_{\{2,3,5\}}, W_{\{2,4,5\}}\}, \\
F_{3}&=\{W_{\{1,2,3\}},W_{\{1,3,4\}},W_{\{1,3,5\}}, W_{\{2,3,4\}}, W_{\{2,3,5\}}, W_{\{3,4,5\}}\},\\
F_{4}&=\{W_{\{1,2,4\}},W_{\{1,3,4\}},W_{\{1,4,5\}}, W_{\{2,3,4\}}, W_{\{2,4,5\}}, W_{\{3,4,5\}}\},\\
F_{5}&=\{W_{\{1,2,5\}},W_{\{1,3,5\}},W_{\{1,4,5\}}, W_{\{2,3,5\}}, W_{\{2,4,5\}}, W_{\{3,4,5\}}\}.
\end{align*}

\paragraph*{Placement Phase} 
%We use the   cache placement in~\cite{yang2018centralizedcorrected}. 
Here $t=\frac{\Ksf\Msf \rsf}{\Nsf}=3$. 
We partition each block into $\binom{\Ksf}{t}=120$ equal-length sub-blocks and denote $W_{\Sc}=\{W_{\Sc,\Vc}:\Vc\subseteq [\Ksf], |\Vc|=t=3\}$. 
Each user $k\in[\Ksf]$ caches $W_{\Sc,\Vc}$ %for all $\Vc\subseteq [\Ksf]$ of size $|\Vc|=t=3$
if $k\in \Vc$.

\paragraph*{Delivery Phase} 
Assume $\dv=(1,2,3,4,5,1,2,3,4,5)$, which has $N_{\textup{e}}(\dv)=5$ distinct demanded files. 
We choose as leaders the users in $\uv = (1,2,3,4,5)$. %permutation as

\paragraph*{First delivery sub-phase} 
In Step    $j\in [\min\{N_{\textup{e}}(\dv),\Nsf-\rsf+1,\Ksf-t\}]=[3]$ of the first sub-phase, for each   set of users $\Jc\subseteq [\Ksf]\setminus [j-1]$ where $|\Jc|=t+1=4$ and $j\in \Jc$, and each set of files $\Bc\subseteq [\Nsf]\setminus [j]$ where $|\Bc|=\rsf-1=2$, we transmit $C_{\Jc,\Bc}$.
 
At the end of the first sub-phase, as shown in Section~\ref{sub:scheme 1}, each leader user can recover its desired file.  

For the non-leaders, we focus on user~$6$. From Lemma~\ref{lem:Jc}, %by direct decoding 
 user~$6$ can decode $W_{\Sc_1,\Vc_1}$ where $1\in \Sc_1$ and $1\in \Vc_1$. Hence, after the first sub-phase, user~$6$ still needs to recover  $W_{\Sc_2,\Vc_2}$ where $1\in \Sc_2$ and $\{1,6\}\cap \Vc_2=\emptyset$.   Similarly, each non-leader user $k\in[6:10]$ still needs to recover  $W_{\Sc_2,\Vc_2}$ where $d_k \in \Sc_2$, $\{1,\ldots, d_k-1\}\cap \Sc_2 =\emptyset$, and $\{k, 1,\ldots, d_k\}\cap \Vc_2=\emptyset$ (recall that $g(d_k)=d_k$  in this example).

\paragraph*{Second delivery sub-phase} 
In Step~$j\in [3]$ of the second sub-phase, for each $q\in [j+1: 4]$,  each $\Jc^{\prime}\subseteq [\Ksf]\setminus [q]$ where $|\Jc^{\prime}|=t-1=2$ and $\Jc^{\prime}\cap [q+1:5]\neq \emptyset$, and each $\Bc^{\prime}\subseteq [\Nsf]\setminus [q]$ where $|\Bc^{\prime}|=\rsf-2=1$, we transmit
$C_{\Jc^{\prime}\cup\{j,q\},\Bc^{\prime}}$.

We now prove the decodability of user $6$. Observe that leader $g(d_6)=1$ also demands $F_1$, we show the decodability of user $6$ by induction. For each $j\in [g(d_6)+1:N_{\textup{e}}(\dv)]=[2:5]$,  we prove user $6$ can recover its  desired  sub-block $W_{\Sc_2,\Vc_2}$ where $d_{u_j} \in \Sc_2$ or $u_{j}\in \Vc_2$. 

We start from $j=2$.
In the following, we show user~$6$ can recover $W_{\{1,2,3\},\Vc_2}$  where  $\{1,6\}\cap \Vc_2=\emptyset$ by interference alignment decoding (i.e., $W_{\{1,2,3\},\{2,3,4\}}$,  $W_{\{1,2,3\},\{2,3,5\}}$,  $W_{\{1,2,3\},\{2,4,5\}}$, and  $W_{\{1,2,3\},\{3,4,5\}}$).  A similar argument applies to every  non-leader user $k\in[6:10]$. 

 We first focus on  $W_{\{1,2,3\},\Vc_2}$ where  $\{g(d_6),6\}=\{1,6\} \notin \Vc_2$ and $u_j=2\in \Vc_2$, e.g.,  $W_{\{1,2,3\},\{2,3,4\}}$.   In Step~$1$ of the first sub-phase, user~$6$ receives
\begin{align}
C_{\{1,2,3,4\},\{2,3\}}&=W_{\{1,2,3\},\{2,3,4\}} \oplus W_{\{1,2,3\},\{1,3,4\}} \oplus W_{\{1,2,3\},\{1,2,4\}}\nonumber\\ & \oplus W_{\{2,3,4\},\{1,3,4\}} \oplus W_{\{2,3,4\},\{1,2,4\}} \oplus W_{\{2,3,4\},\{1,2,3\}};\label{eq:C 1234 23}\\
 C_{\{1,2,3,4\},\{3,4\}}&=W_{\{1,3,4\},\{2,3,4\}} \oplus W_{\{1,3,4\},\{1,2,4\}} \oplus W_{\{1,3,4\},\{1,2,3\}}\nonumber\\ & \oplus W_{\{2,3,4\},\{1,3,4\}} \oplus W_{\{2,3,4\},\{1,2,4\}} \oplus W_{\{2,3,4\},\{1,2,3\}}.\label{eq:C 1234 34} 
\end{align}
By summing~\eqref{eq:C 1234 23} and~\eqref{eq:C 1234 34}, we can obtain 
\begin{align}
C_{\{1,2,3,4\},\{1,3\}}&=W_{\{1,2,3\},\{2,3,4\}} \oplus W_{\{1,2,3\},\{1,3,4\}} \oplus W_{\{1,2,3\},\{1,2,4\}}\nonumber\\  &\oplus W_{\{1,3,4\},\{2,3,4\}} \oplus W_{\{1,3,4\},\{1,2,4\}} \oplus W_{\{1,3,4\},\{1,2,3\}}
\label{eq:C 1234 13} \\
&= C_{\{1,2,3,4\},\{2,3\}}  \oplus C_{\{1,2,3,4\},\{3,4\}},\label{eq:C 1234 13, property} 
\end{align}
which shows the property in~\eqref{eq:interference transition2} in Lemma~\ref{lem:lemma for Bc}.
It can be seen by summing~\eqref{eq:C 1234 23} and~\eqref{eq:C 1234 34}, we cancel the interferences from the sub-blocks of $W_{\{2,3,4\}}$ to user~$6$.  From   Lemma~\ref{lem:Jc}, user~$6$ can     decode $W_{\Sc_1,\Vc_1}$ where $1\in \Sc_1$ and $1\in \Vc_1$. In addition, in 
\begin{align}
C_{\{2,3,4,6\},\{3,4\}}&=W_{\{1,3,4\},\{2,4,6\}} \oplus W_{\{1,3,4\},\{2,3,6\}} \oplus W_{\{1,3,4\},\{2,3,4\}}\nonumber\\ & \oplus W_{\{2,3,4\},\{ 3,4,6\}} \oplus W_{\{2,3,4\},\{ 2,4,6\}} \oplus W_{\{2,3,4\},\{ 2,3,6\}},\label{eq:C 2346 34} 
\end{align}
which is transmitted in Step~$2$ of the first sub-phase, user~$6$ caches all except $W_{\{1,3,4\},\{2,3,4\}}$ such that it can recover $W_{\{1,3,4\},\{2,3,4\}}$   by directly reading off. 
Hence, user~$6$ has decoded all except $W_{\{1,2,3\},\{2,3,4\}}$ in~\eqref{eq:C 1234 13} such that it can recover  $W_{\{1,2,3\},\{2,3,4\}}$. 

{\bf By similar steps, for each desired sub-block  $W_{\{1,2,3\},\Vc_2}$ where  $\{1,6\}\cap \Vc_2=\emptyset$ and $u_j= 2\in \Vc_2$, user~$6$ first  reconstructs  $C_{\Vc_2\cup \{1\},\{1,2,3\}\setminus \{2\}}$ and then recovers $W_{\{1,2,3\},\Vc_2}$ from  $C_{\Vc_2\cup \{1\},\{1,2,3\}\setminus \{2\}}$}.

 We then focus on  $W_{\{1,2,3\},\Vc_2}$ where     $\{g(d_6),6\}\cup\{u_j\}=\{1,2,6\} \cap \Vc_2=\emptyset$, e.g.,
$W_{\{1,2,3\},\{3,4,5\}}$.  In Step~$1$ of the first sub-phase, user~$6$ receives 
\begin{align}
C_{\{1,3,4,5\},\{2,3\}}&=W_{\{1,2,3\},\{3,4,5\}} \oplus W_{\{1,2,3\},\{1,4,5\}} \oplus W_{\{2,3,4\},\{1,4,5\}}\nonumber\\ & \oplus W_{\{2,3,4\},\{1,3,5\}} \oplus W_{\{2,3,5\},\{1,4,5\}} \oplus W_{\{2,3,5\},\{1,3,4\}};\label{eq:C 1345 23}\\
C_{\{1,2,3,5\},\{3,4\}}&=W_{\{1,3,4\},\{2,3,5\}} \oplus W_{\{1,3,4\},\{1,2,5\}} \oplus W_{\{2,3,4\},\{1,3,5\}}\nonumber\\ & \oplus W_{\{2,3,4\},\{1,2,5\}} \oplus W_{\{3,4,5\},\{1,2,5\}} \oplus W_{\{3,4,5\},\{1,2,3\}};\label{eq:C 1235 34}\\
C_{\{1,2,3,4\},\{3,5\}}&=W_{\{1,3,5\},\{2,3,4\}} \oplus W_{\{1,3,5\},\{1,2,4\}} \oplus W_{\{2,3,5\},\{1,3,4\}}\nonumber\\ & \oplus W_{\{2,3,5\},\{1,2,4\}} \oplus W_{\{3,4,5\},\{1,2,4\}} \oplus W_{\{3,4,5\},\{1,2,3\}}. \label{eq:C 1234 35}
\end{align}
In Step~$1$ of the second sub-phase (with $j=1$, $q=2$, $\Jc^{\prime}=\{4,5\}$, $\Bc^{\prime}=\{3\}$), user~$6$ receives 
\begin{align}
C_{\{1,2,4,5\},\{ 3\}}&=W_{\{1,2,3\},\{2,4,5\}} \oplus W_{\{1,2,3\},\{1,4,5\}} \oplus W_{\{1,3,4\},\{2,4,5\}}  \oplus W_{\{1,3,4\},\{1,2,5\}} \nonumber\\ & \oplus W_{\{1,3,5\},\{2,4,5\}} \oplus W_{\{1,3,5\},\{1,2,4\}} \oplus W_{\{2,3,4\},\{1,4,5\}}  \oplus W_{\{2,3,4\},\{1,2,5\}}  \nonumber\\ & 
 \oplus W_{\{2,3,5\},\{1,4,5\}} \oplus W_{\{2,3,5\},\{1,2,4\}} \oplus W_{\{3,4,5\},\{1,2,5\}}  \oplus W_{\{3,4,5\},\{1,2,4\}}. \label{eq:C 1245 3} 
\end{align}
By  summing~\eqref{eq:C 2346 34}-\eqref{eq:C 1245 3}, we have 
\begin{align}
C_{\{2,3,4,5\},\{1, 3\}}&= W_{\{1,2,3\},\{3,4,5\}} \oplus W_{\{1,2,3\},\{2,4,5\}} \oplus W_{\{1,3,4\},\{2,4,5\}}\nonumber\\ & \oplus W_{\{1,3,4\},\{2,3,5\}} \oplus W_{\{1,3,5\},\{2,4,5\}} \oplus W_{\{1,3,5\},\{2,3,4\}}\label{eq:C 2345 13}\\
&=C_{\{1,3,4,5\},\{2,3\}}\oplus C_{\{1,2,3,5\},\{3,4\}} \oplus C_{\{1,2,3,4\},\{3,5\}} \oplus  C_{\{1,2,4,5\},\{ 3\}},\label{eq:sum 4}
\end{align}
which shows the property in~\eqref{eq:interference transition} in Lemma~\ref{lem:lemma for Bc}.
Hence, by~\eqref{eq:sum 4}, user~$6$ can reconstruct $C_{\{2,3,4,5\},\{1, 3\}}$  while cancelling the interferences in~\eqref{eq:C 2346 34}-\eqref{eq:C 1245 3}, coinciding with Lemma~\ref{lem:lemma for Bc}. We then focus on each sub-block in $C_{\{2,3,4,5\},\{1, 3\}}$. $W_{\{1,2,3\},\{2,4,5\}}$ can be recovered by user~$6$ as we showed previously for $W_{\{1,2,3\},\{2,3,4\}}$. For $W_{\{1,3,4\},\{2,4,5\}}$,  in 
\begin{align}
C_{\{2,4,5,6\},\{3,4\}}&=W_{\{1,3,4\},\{2,5,6\}} \oplus W_{\{1,3,4\},\{2,4,5\}} \oplus W_{\{2,3,4\},\{4,5,6\}}\nonumber\\ & \oplus W_{\{2,3,4\},\{ 2,5,6\}} \oplus W_{\{3,4,5\},\{ 2,5,6\}} \oplus W_{\{3,4,5\},\{ 2,4,6\}},\label{eq:C 2456 34} 
\end{align}
which is transmitted in Step~$2$ of the first sub-phase, user~$6$ caches all except $W_{\{1,3,4\},\{2,4,5\}}$ such that it can recover $W_{\{1,3,4\},\{2,4,5\}}$   by directly reading off. Similarly, user~$6$ can recover  $ W_{\{1,3,4\},\{2,3,5\}}$, $W_{\{1,3,5\},\{2,4,5\}}$, and $ W_{\{1,3,5\},\{2,3,4\}}$ from Step~$2$ of the first sub-phase  by directly reading off. Hence, in $C_{\{2,3,4,5\},\{1, 3\}}$, user~$6$ has  recovered all except  $W_{\{1,2,3\},\{3,4,5\}}$ such that user~$6$ can recover $W_{\{1,2,3\},\{3,4,5\}}$.

Finally, we consider $W_{\{1,2,3\},\{3,9,10\}}$, where $d_9=4$ and $d_{10}=5$.
Notice that, $C_{\{1,2,9,10\},\{ 3\}}$ 
is not transmitted in the second sub-phase, because both of users $9,10$  are not leaders,
which contradicts the constraint on the transmission of the second sub-phase ($\Jc^{\prime}\cap [q+1:5]\neq \emptyset$ with $q=2$ and $\Jc^{\prime}=\{9,10\}$).
However, it can be seen that if user~$6$ can reconstruct  $C_{\{1,2,9,10\},\{ 3\}}$, by the same decoding procedure as $W_{\{1,2,3\},\{ 3,4,5\}}$, user~$6$ can recover  $W_{\{1,2,3\},\{3,9,10\}}$. So in the following, we prove user~$6$ can reconstruct $C_{\{1,2,9,10\},\{ 3\}}$, as described in  Lemma~\ref{lem:additional decoding}. 

Notice that $C_{\{1,2,4,10\},\{2, 3\}}$ and $C_{\{1,2,4,10\},\{ 3\}}$ are transmitted in Step~$1$ of the first and second sub-phases, respectively. Hence, user~$6$ can obtain 
\begin{align}
C_{\{1,2,4,10\},\{2, 3\}}\oplus C_{\{1,2,4,10\},\{ 3\}}&= 
         W_{\{1,3,4\},\{2,4,10\}} \oplus  W_{\{1,3,4\},\{1,2,10\}} \oplus  W_{\{1,3,5\},\{2,4,10\}}\nonumber\\
&\oplus  W_{\{1,3,5\},\{1,2,4\}}  \oplus  W_{\{3,4,5\},\{1,2,10\}} \oplus  W_{\{3,4,5\},\{1,2,4\}}.
\label{eq:sum 12410 23 3}
\end{align}
On the RHS of~\eqref{eq:sum 12410 23 3}, $ W_{\{1,3,4\},\{2,4,10\}}$ and $W_{\{1,3,5\},\{2,4,10\}}$ can be   recovered by user~$6$ from $C_{\{2,4,6,10\},\{3,4\}}$ and $C_{\{2,4,6,10\},\{3,5\}}$ transmitted in Step~$2$ of the first sub-phase, respectively (by directly reading off). $ W_{\{1,3,4\},\{1,2,10\}}$ and $ W_{\{1,3,5\},\{1,2,4\}}$ can be recovered by user~$6$ because they are cached by user $1$ and thus we can use Lemma~\ref{lem:Jc}.Item~\ref{item:lem JC2}. Hence, from~\eqref{eq:sum 12410 23 3}, user~$6$ can recover 
\begin{align}
 W_{\{3,4,5\},\{1,2,10\}} \oplus  W_{\{3,4,5\},\{1,2,4\}}.\label{eq:inter1}
\end{align}
Similarly, user~$6$ can recover 
\begin{align}
 &W_{\{3,4,5\},\{1,2,5\}} \oplus  W_{\{3,4,5\},\{1,2,4\}},\label{eq:inter2} \textrm{and } \\
 &W_{\{3,4,5\},\{1,2,5\}} \oplus  W_{\{3,4,5\},\{1,2,9\}},\label{eq:inter3}
\end{align}
from $C_{\{1,2,4,5\},\{2, 3\}}\oplus C_{\{1,2,4,5\},\{ 3\}}$ and $C_{\{1,2,8,5\},\{2, 3\}}\oplus C_{\{1,2,8,5\},\{ 3\}}$, respectively. By summing~\eqref{eq:inter1}-~\eqref{eq:inter3}, user~$6$ can obtain
\begin{align}
 W_{\{3,4,5\},\{1,2,10\}} \oplus W_{\{3,4,5\},\{1,2,9\}}.\label{eq:345 120}
\end{align}
 
Similar to~\eqref{eq:sum 12410 23 3}, we have 
\begin{align}
C_{\{1,2,9,10\},\{2, 3\}}&=C_{\{1,2,9,10\},\{ 3\}}\oplus  W_{\{1,3,4\},\{2,9,10\}} \oplus  W_{\{1,3,4\},\{1,2,10\}}\oplus  W_{\{1,3,5\},\{2,9,10\}}\nonumber\\
&\oplus  W_{\{1,3,5\},\{1,2,9\}} \oplus  W_{\{3,4,5\},\{1,2,10\}} \oplus  W_{\{3,4,5\},\{1,2,9\}}.\label{eq:sum 12910 23 3}
\end{align} 
On the RHS of~\eqref{eq:sum 12910 23 3},   $C_{\{1,2,9,10\},\{2, 3\}}$ is transmitted in Step~$1$ of the first sub-phase.
In addition, $ W_{\{1,3,4\},\{2,9,10\}}$ and $W_{\{1,3,5\},\{2,9,10\}}$ can be   recovered by user~$6$ from $C_{\{2,6,9,10\},\{3,4\}}$ and $C_{\{2,6,9,10\},\{3,5\}}$ transmitted in Step~$2$ of the first sub-phase, respectively (by directly reading off). $ W_{\{1,3,4\},\{1,2,10\}}$ and $ W_{\{1,3,5\},\{1,2,9\}}$ can be recovered by user~$6$ because they are cached by user $1$ and thus we can use Lemma~\ref{lem:Jc}.Item~\ref{item:lem JC2} . We also proved in~\eqref{eq:345 120} that $ W_{\{3,4,5\},\{1,2,10\}} \oplus  W_{\{3,4,5\},\{1,2,9\}}$ can be recovered by user~$6$.
Hence, user~$6$ can reconstruct $C_{\{1,2,9,10\},\{ 3\}}$ and thus it can recover $W_{\{1,2,3\},\{3,9,10\}}$.
%Similarly, we prove that user~$6$  can decode $W_{\{1,2,3\},\Vc_2}$ where  $\{1,2,6\}\cap \Vc_2=\emptyset$, and thus user~$6$ can recover $W_{\{1,2,3\}}$.

{\bf By similar steps, for each desired sub-block  $W_{\{1,2,3\},\Vc_2}$ where $\{1,2,6\}\cap \Vc_2=\emptyset$, user~$6$ first reconstructs $C_{\Vc_2\cup \{2\},\{1,2,3\}\setminus \{2\}}$ and then recovers $W_{\{1,2,3\},\Vc_2}$ from  $C_{\Vc_2\cup \{2\},\{1,2,3\}\setminus \{2\}}$}.

Hence, we prove that   user~$6$ can recover $W_{\{1,2,3\}}$.
Similarly, we can prove user $6$ can recover $W_{\Sc_2}$ where     $\{d_k, d_{u_j}\}=\{1,2 \} \subseteq \Sc_2$.

 For each sub-block $W_{\Sc_2,\Vc_2}$ where $d_6=1 \in \Sc_2$, $d_{u_j}=2 \notin \Sc_2$, $6\notin \Vc_2$, and  $u_{j}=2\in \Vc_2$, user $6$ can recover $W_{\Sc_2,\Vc_2}$
from $C_{\Vc_2\cup \{6\}, \Sc_2\setminus \{1\}}$ by directly reading off. Hence, we finish the proof of the decodability of user $6$ for $j=2$.  

By the induction method,
other desired blocks can also be recovered by user~$6$ with the above decoding procedures.

\paragraph*{Performance}
The achieved load is $31/30\approx 1.033$ while the converse bound in Theorem~\ref{thm:converse} is $707/720\approx 0.982$ and the achieved load in~\cite{yang2018centralizedcorrected} is $7/6\approx 1.167$.

\subsection{Proof of Theorem~\ref{thm:order optimality}}
\label{sub:proof thm:order optimality}
%\begin{IEEEproof}
For type $s\in [\min\{\Ksf,\Nsf\}]$ and each corner point $\Msf=\frac{\Nsf t}{\Ksf \rsf}$ where $t\in [0:\Ksf]$, from Theorem~\ref{thm:scheme 2}, we can achieve the load %$\Rsf^s_{\eqref{eq:scheme 2 type s}}$ defined as
\begin{subequations}
\begin{align}
%\Rsf^s_{\eqref{eq:scheme 2 type s}}&:= 
&c^s_{t} + e^s_{t} \notag\\
&= \sum_{j\in [\min\{s,\Nsf-\rsf+1,\Ksf-t\}]}\frac{\binom{\Nsf-j}{\rsf-1}\binom{\Ksf-j}{t}+\sum^{\min\{\Nsf-\rsf+2,\Ksf-t+1,s\}}_{q=j+1}  \left(\binom{\Nsf-q}{\rsf-2}-\binom{\Nsf-s}{\rsf-2} \right) \left(\binom{\Ksf-q}{t-1}-\binom{\Ksf-s}{t-1} \right) }{\binom{\Nsf-1}{\rsf-1}\binom{\Ksf}{t}}\\
&\leq \sum_{j\in [\min\{s,\Nsf-\rsf+1,\Ksf-t\}]}\frac{ \binom{\Nsf-j}{\rsf-1}\binom{\Ksf-j}{t}+\sum^{\min\{\Nsf-\rsf+2,\Ksf-t+1,s\}}_{q=j+1} \binom{\Nsf-q}{\rsf-2}\binom{\Ksf-q}{t-1} }{\binom{\Nsf-1}{\rsf-1}\binom{\Ksf}{t}}\\
&\leq \sum_{j\in [\min\{s,\Nsf-\rsf+1,\Ksf-t\}]}\frac{ \binom{\Nsf-j}{\rsf-1}\binom{\Ksf-j}{t}+\sum^{\min\{\Nsf-\rsf+2,\Ksf-t+1,s\}}_{q=j+1} \binom{\Nsf-q}{\rsf-2}\binom{\Ksf-j-1}{t-1} }{\binom{\Nsf-1}{\rsf-1}\binom{\Ksf}{t}}\\
&\leq \sum_{j\in [\min\{s,\Nsf-\rsf+1,\Ksf-t\}]}\frac{ \binom{\Nsf-j}{\rsf-1}\binom{\Ksf-j}{t}+  \binom{\Nsf-j}{\rsf-1}\binom{\Ksf-j-1}{t-1} }{\binom{\Nsf-1}{\rsf-1}\binom{\Ksf}{t}}\label{eq:from pascal}\\
&\leq 2\sum_{j\in [\min\{s,\Nsf-\rsf+1,\Ksf-t\}]}\frac{\binom{\Nsf-j}{\rsf-1}\binom{\Ksf-j}{t} }{\binom{\Nsf-1}{\rsf-1}\binom{\Ksf}{t}}\\
&=2 c^s_{t}\label{eq:order optimal derivation}, 
\end{align}
\end{subequations}
where~\eqref{eq:from pascal} comes from the Pascal's triangle. Hence, from~\eqref{eq:order optimal derivation} and the converse bound in Theorem~\ref{thm:converse}, we proved the proposed caching scheme in Theorem~\ref{thm:scheme 2} is order optimal to within a factor of $2$ under the constraint of uncoded cache placement for demand type $s$.

Similarly, we can prove that the average load among all possible demands in Theorem~\ref{thm:scheme 2} is order optimal to within a factor of $2$ under the constraint of uncoded cache placement. 
%\end{IEEEproof}

\subsection{Proof of Theorem~\ref{thm:exact optimality}}%Optimal Scheme in
\label{sub:scheme 3 r=2}
 
%\begin{rem}
%\label{rem:r=2 deco}
%\rm
From the proof of the decodability in Appendix~\ref{sec:decodability general scheme}, we have the following observations (Observations~2 and~3 are proved in Appendix~\ref{sec:decodability general scheme}), which will help us to further reduce the load for some special cases:
\begin{enumerate}

\item \label{item:obs1}
Observation~\ref{item:obs1}:  when $\rsf=2$,  the transmission of the second sub-phase does not exist because $|\Bc^{\prime}|=\rsf-2=0$ and $\Bc^{\prime}\cap \Nc([\Ksf])\neq \emptyset$ can not hold simultaneously. When $t=1$,
 the  transmission of the second sub-phase does not exist because $|\Jc^{\prime}|=t-1=0$ and $\Jc^{\prime}\cap \{u_{q+1},\ldots,u_{N_{\textup{e}}(\dv)}\}\neq \emptyset$, can not hold simultaneously.
In other words, each non-leader can recover all its desired files from the first sub-phase if $\rsf=2$ or $t=1$.

\item \label{item:obs2}
Observation~\ref{item:obs2}: for a non-leader $k$, to decode $W_{\Sc_2,\Vc_2}$ where $d_k \in \Sc_2$, $\{d_{u_1},\ldots,d_{u_{g(d_k)-1}}\} \cap \Sc_2=\emptyset$ and 
 $\{k,u_1,\ldots,u_{g(d_k)}\}\cap \Vc_2 = \emptyset$, if there is no user in $\Vc_2$ whose demanded file is in $\{d_{u_1},\ldots,d_{u_{g(d_k)-1}}\}$,   user $k$ only needs to use the transmission of the first sub-phase, Step~$g(d_k)$ of the second sub-phase and  Step~$g(d_k)$ in  Lemma~\ref{lem:additional decoding}. 

\item \label{item:obs3}
Observation~\ref{item:obs3}: for a non-leader $k$, to decode $W_{\Sc_2,\Vc_2}$ where $d_k \in \Sc_2$, $\{d_{u_1},\ldots,d_{u_{g(d_k)-1}}\} \cap \Sc_2=\emptyset$,
 $\{k,u_1,\ldots,u_{g(d_k)}\}\cap \Vc_2 = \emptyset$, and  $(\cup_{k_1\in \Vc_2} \{d_{k_1}\}) \cap (\Sc_2\setminus \{d_k\})=\emptyset$, user $k$ only needs the transmission of the first sub-phase.

\end{enumerate}
%\hfill$\square$
%\end{rem}

In the following, we will show if $\rsf \in \{1, 2,\Nsf -1 ,\Nsf \}$  or $t \in \{0,1,2,\Ksf-1,\Ksf\}$ or   $s \in [\min\{\Ksf,\Nsf,4\}]$, the transmission of the second sub-phase is not needed. Notice that the transmitted load of the first sub-phase coincides with the proposed converse bound in Theorem~\ref{thm:converse}. Hence, for the above cases, the transmission of the first sub-phase is optimal under the constraint of uncoded cache placement.

When $\rsf \in \{1,\Nsf\}$, the considered problem is equivalent to the MAN caching problem, the first sub-phase is equivalent to the caching scheme in~\cite{exactrateuncoded}, which is optimal under the constraint of uncoded cache placement.

When $t\in \{0,\Ksf\}$, it is trivial to achieve the optimality by transmitting all demanded files  or nothing. 

When $\rsf=2$ or $t=1$, as shown in Observation~\ref{item:obs1}, each non-leader can   recover its desired files from the transmission of the first sub-phase.

When $t=\Ksf-1$, there is only one step in the first sub-phase. From Lemma~\ref{lem:Jc}.Item~\ref{item:lem JC2} , it can be seen that any non-leader can recover its desired blocks from Step~$1$ of the first sub-phase. Hence, the second sub-phase is not necessary.

We now consider  $\rsf=\Nsf-1$ or $t=2$ and let each non-user $k$ recover $W_{\Sc_2,\Vc_2}$ where $d_k \in \Sc_2$, $\{d_{u_1},\ldots,d_{u_{g(d_k)-1}}\} \cap \Sc_2=\emptyset$ and 
 $\{k,u_1,\ldots,u_{g(d_k)}\}\cap \Vc_2 = \emptyset$, by the transmission of the first sub-phase.
The main reason that the first sub-phase is enough for these two cases, is that Step~$g(d_k)$ of the second sub-phase could be reconstructed by user $k$ from the first sub-phase.  Consider one message
$C_{\Jc^{\prime}\cup\{u_{g(d_k)},u_q\},\Bc^{\prime}}$ which is transmitted in the second sub-phase. Notice that $\Bc^{\prime} \subseteq   [\Nsf]\setminus \{d_{u_1},\ldots,d_{u_{q}}\}$. If $t=2$, we have $|\Jc^{\prime}|=1$. If $\rsf=\Nsf-1$, we have $|\Bc^{\prime}|=\rsf-2=\Nsf-3$. Hence, for the case  $\rsf=\Nsf-1$ or $t=2$, all  interferences  in  $C_{\Jc^{\prime}\cup\{u_{g(d_k)},u_q\},\Bc^{\prime}}$ to user $k$ whose demands $F_{d_{u_{g(d_k)}}}$,  are from one block (assuming this block is $W_{\Bc^{\prime}\cup \{i\}}$). Hence, the binary sum of these interferences is equal to the sum of the interferences in $C_{\Jc^{\prime}\cup\{u_{g(d_k)},u_q\},\Bc^{\prime}\cup\{i\}}$ or $C_{\Jc^{\prime}\cup\{u_{g(d_k)},u_q\},\Bc^{\prime}\cup \{d_{u_q}\} }$. It will be proved in Appendix~\ref{sec:r=N-1 t=2} that user $k$ can recover this sum of interferences from the first sub-phase and then it can reconstruct $C_{\Jc^{\prime}\cup\{u_{g(d_k)},u_q\},\Bc^{\prime}}$. 

Hence, from Observation~\ref{item:obs2}, user $k$ can recover $W_{\Sc_2,\Vc_2}$ if there is no user in $\Vc_2$ whose demanded file is in $\{d_{u_1},\ldots,d_{u_{g(d_k)-1}}\}$.
It will be proved in Appendix~\ref{sec:r=N-1 t=2}, if  there is some user in $\Vc_2$ whose demanded file is in $\{d_{u_1},\ldots,d_{u_{g(d_k)-1}}\}$, for the case  $\rsf=\Nsf-1$ or $t=2$, user $k$ can also recover $W_{\Sc_2,\Vc_2}$ from the reconstruction of Step~$g(d_k)$ of the second sub-phase.
 
In conclusion, for the cases where $\rsf \in \{1,2,\Nsf-1,\Nsf\}$ or $t\in \{1,2,\Ksf-1,\Ksf\}$, we prove that from the first delivery sub-phase, each user can recover its desired file.  
Comparing  the converse bound in Theorem~\ref{thm:converse} and the achieved load  (given in  Section~\ref{sub:scheme 3 r=2}), we   have the optimality for Case 1 where  $\rsf\in \{1,2,\Nsf-1,\Nsf\}$.  The optimality for Case 2 where either $\Ksf\rsf\Msf/\Nsf\leq 2$ or $\Ksf\rsf\Msf/\Nsf\geq \Ksf-1$,  is  due to the fact that in the converse bound~\eqref{eq:converse ave}, $c^{N_{\textup{e}}(\dv)}_{t}$ is  convex   in terms of $t$ and when $t\in \{0,1,2,\Ksf-1,\Ksf\}$, our proposed scheme is optimal.

Finally, we prove the optimality of  $\Rsf^{\star}_\mathrm{u}(\Msf,s)$ for Case 3 where $s\in [\min\{\Ksf,\Nsf,4\}]$. We consider the following two cases.
\begin{enumerate}
\item $\min\{\Ksf,\Nsf\} \leq 4$.  
Theorem~\ref{thm:exact optimality}.Case~\ref{item:optimality 1} covers all possible values of $\rsf$ when $3\geq\Nsf-1$, and
Theorem~\ref{thm:exact optimality}.Case~\ref{item:optimality 2} covers all possible values of $\Msf$ when $3\geq\Ksf-1$. Hence, when $\min\{\Ksf,\Nsf\} \leq 4$, we can prove the optimality.   
\item  $\min\{\Ksf,\Nsf\} > 4$.  In this case, $s=|\Nc([\Ksf])|\leq 4$.   
For each subset of files $\Tc\subseteq [\Nsf]\setminus \Nc([\Ksf])$ where $\rsf- 4 \leq |\Tc|<\rsf$, 
we can gather all blocks $W_{\Sc}$ where $\Sc\subseteq [\Nsf]$, $|\Sc|=\rsf$, $\Sc \setminus \Nc([\Ksf])=\Tc$. The proposed first delivery sub-phase on these blocks is equivalent to the first delivery sub-phase 
for $\Nc_{\text{eq}}([\Ksf])=\Nsf_{\text{eq}}=s$, $\Ksf_{\text{eq}}=\Ksf$, $\rsf_{\text{eq}}=\rsf- |\Tc|$, and $t_{\text{eq}}=t$. Since we proved the decodability of the proposed first delivery sub-phase  for the system including up to $4$ files, we can prove the    blocks in this group can be recovered by the demanding users. Hence, we prove that each user can recover its desired file from the  first delivery sub-phase.
\end{enumerate}
As a result, we prove when $s\in [\min\{\Ksf,\Nsf,4\}]$, each user can recover  its desired file from the  first delivery sub-phase, and thus we prove the optimality for Theorem~\ref{thm:exact optimality}.Case~\ref{item:optimality 3}.

\section{Conclusions}
\label{sec:conclusion}
In this paper, we studied the coded caching problem with correlated sources.
We proposed a converse bound under the constraint of uncoded cache placement and two-phase delivery scheme.
For any demand type, under the constraint of uncoded cache placement, our caching scheme is optimal to within a factor of $2$. For the case where each user has a distinct request,  or the case with any demand type with either $\rsf \in \{1,2,\Nsf-1,\Nsf\}$  or $\Ksf\rsf\Msf\leq 2\Nsf$  or $\Ksf\rsf\Msf\geq (\Ksf-1)\Nsf$ or   $s \in [\min\{\Ksf,\Nsf,4\}]$, the second sub-phase is not necessary and thus the proposed scheme is optimal under the constraint of uncoded cache placement.
As a by-product, we also showed that the proposed scheme reduces the load of existing schemes for the caching problem with multiple requests.

\appendices

%\section{Proof of~\eqref{eq:set of acyclic}}%on the Acyclicity
%\label{sec:proof of acyclicity}
%The proof is based on the proof of~\cite[Lemma 1]{ontheoptimality}.
%For a  $\mathbf{u}=(u_{1},u_{2},...,u_{N_{\textup{e}}(\dv)})$, we classify the sub-blocks/nodes in the set~\eqref{eq:set of acyclic} into $N_{\textup{e}}(\dv)$ levels. More precisely,
%we say that sub-blocks/nodes $W_{\Sc_i,\Vc_{i}}$, for all $\Sc_i\subseteq ([\Nsf]\setminus \{d_{u_1},\ldots,d_{u_{i-1}}\})$, $d_{u_i}\in \Sc_i$ and $\Wc_{i}\subseteq ([\Ksf] \setminus \{u_{1},\ldots,u_{i}\})$, are in level $i$. 
%It is easy to see each node in level $i$ has the same side information as user $u_i$ in the caching problem (i.e., each node in level $i$
%only knows the sub-blocks $W_{\Sc,\Vc}$ where $u_{i} \in\Vc$). So each node in level $i$ knows neither the sub-blocks in the same level, nor the sub-blocks in the higher levels. As a result, in the proposed set~\eqref{eq:set of acyclic}, there is no sub-set containing a directed cycle.

\section{Proof of Lemma~\ref{lem:Jc}}
\label{sec:proof of direct lemma}

For a given demand vector $\dv$, let $s := N_{\textup{e}}(\dv)$, $j_{\max} := \min\{s,\Nsf-\rsf+1,\Ksf-t\}$, and order the leader users as $(u_1,\ldots,u_{s})$. 
Recall that in step~$j\in [j_{\max}]$ of delivery sub-phase~1 of the scheme in~\eqref{eq:masterach}  we satisfy the demand of leader user $u_j$ as follows: for each set of users $\Jc\subseteq [\Ksf]\setminus\{u_1,\ldots,u_{j-1}\}$ such that $|\Jc|=t+1$ and $u_j\in \Jc$, and for each set of files $\Bc\subseteq [\Nsf]\setminus \{d_{u_1},\ldots,d_{u_{j}}\}$ such that $|\Bc|=\rsf-1$, we transmit the multicast message in~\eqref{eq:CJi general}, which we re-write as
\begin{subequations}
\begin{align}
C_{\Jc,\Bc}^{j}
  &=\underset{k\in \Jc : d_k\not\in \Bc}{\oplus } W_{\Bc\cup\{d_k\},\Jc\setminus\{k\}}
\label{eq:eq:CJi general expanded1}
\\&
   +\underset{k\in \Jc : d_k\in \Bc}{\oplus } \left(W_{\Bc\cup\{d_{u_j}\},\Jc\setminus\{k\}} +
\underset{i\in \Nc(\Jc)\backslash (\Bc\cup\{d_{u_j}\})}{\oplus } W_{\Bc\cup\{i\},\Jc\setminus\{k\}}
\right),
\label{eq:eq:CJi general expanded2}
\end{align}
\label{eq:eq:CJi general expanded}
\end{subequations}
where we introduced the superscript $j$ to indicate the leader user for whom the multicast message $C_{\Jc,\Bc}^{j}$ has been ``designed,'' by which we mean that by construction (i.e., $d_{u_j}\notin \Bc$), $C_{\Jc,\Bc}^{j}$ in~\eqref{eq:eq:CJi general expanded} contains only one sub-block desired by user $u_j$ (which is $W_{\Bc\cup\{d_{u_j}\}, \Jc\setminus \{u_j\} }$), while all other sub-blocks in $C_{\Jc,\Bc}^{j}$ are in its cache. Based on this observation, we introduce the following terminology:

\emph{Directly read off.}
The observation made for leader user $u_j$ actually holds for every user $k\in \Jc$ such that $d_k\not\in \Bc$ (i.e., term in~\eqref{eq:eq:CJi general expanded1}). Thus, we say that user  $k$ `directly reads off' its desired sub-block $W_{\Bc\cup\{d_k\},\Jc\setminus\{k\}}$ from the multicast message $C_{\Jc,\Bc}^{j}$. Here we use ``directly'' to mean that it is enough to remove the contribution of cached sub-blocks in order to recover a desired sub-block.

\emph{Indirectly read off.}
For user $k\in \Jc$ such that $d_k\in \Bc$, its desired sub-blocks appear in $C_{\Jc,\Bc}^{j}$ as the linear combination $\underset{i\in \Nc(\Jc)\backslash \Bc}{\oplus } W_{\Bc\cup\{i\},\Jc\setminus\{k\}}$ (i.e., term in~\eqref{eq:eq:CJi general expanded2}). Evidently, in~\eqref{eq:eq:CJi general expanded2}, the user who desires file  $i\not\in\Bc\cup\{d_{u_j}\}$ is in $\Jc$ and is not $u_j$, thus $W_{\Bc\cup\{i\},\Jc\setminus\{k\}}$ can be `directly read off' from $C_{\Jc,\Bc\backslash\{d_k\} \cup\{i\}}^{j}$. Thus, we say that user  $k$ `indirectly reads off' its desired sub-block $W_{\Bc\cup\{d_{u_j}\},\Jc\setminus\{k\}}$ from the multicast message $C_{\Jc,\Bc}^{j}$. Here we use ``indirectly'' to mean that it is not enough to remove the contribution of cached sub-blocks in order to recover a desired sub-block, but in addition one has to remove the contribution of sub-blocks that have been `directly read off' from some other multicast messages.

%Also, by the end of step~$j$, the collection of multicast messages  $\{ C_{\Jc,\Bc}^{i} : i\in[j]\}$ has `spanned' all sets $\Jc\subseteq [\Ksf]$ such that $|\Jc|=t+1$ and $\{u_1,\ldots,u_{j}\}\cap\Jc\not=\emptyset$;  therefore, future steps strictly beyond step~$j$ `consider' the sets $\Jc\subseteq [\Ksf]\setminus\{u_1,\ldots,u_{j}\}$ such that $|\Jc|=t+1$. %and $\{u_1,\ldots,u_{j}\}\cap\Jc=\emptyset$. At the end of step $j_{\max}$, there are still sets $\Jc\subseteq [\Ksf]\setminus\{u_1,\ldots,u_{j_{\max}}\}$ such that $|\Jc|=t+1$, only if $j_{\max} = \min\{s,\Nsf-\rsf+1\} < \Ksf-t.$

Lemma~\ref{lem:Jc} is proved by induction.

\subsection{Step~$1$}
\label{sub:step 1 of proof lem jc}
\paragraph*{Lemma~\ref{lem:Jc}.Item~\ref{item:lem JC1} }
We focus on one set of users $\Jc\subseteq [\Ksf]$ where $|\Jc|=t+1$ and $u_1\in \Jc$, and one set of files $\Bc\subseteq [\Nsf]\setminus \{d_{u_1}\}$ where $|\Bc|=\rsf-1$.
We will prove that from  Step~$1$, each user in $k_1\in\Jc$ can recover all sub-blocks in $C_{\Jc,\Bc}$. We consider two cases:
\begin{itemize}
\item $d_{k_1} \notin \Bc$: in $C_{\Jc,\Bc}$ user $k_1$ caches all sub-blocks except $W_{\Bc\cup\{d_{k_1}\},\Jc\setminus \{k_1\}}$. Hence, user $k_1$ can recover $W_{\Bc\cup\{d_{k_1}\},\Jc\setminus \{k_1\}}$ by directly reading off.
 \item $d_{k_1} \in \Bc$: in $C_{\Jc,\Bc}$ user $k_1$ caches all sub-blocks except  $W_{\Bc\cup\{i\},\Jc\setminus \{k_1\}}$, where $i \in \Nc(\Jc)\setminus \Bc$.  
   \begin{itemize}
   \item If $i \neq d_{u_1}$, user $k_1$ can recover  $W_{\Bc\cup\{i\},\Jc\setminus \{k_1\}}$ from $C_{\Jc,(\Bc\cup \{i\})\setminus \{d_{k_1}\}}$ by directly reading off  as the similar reason described in the above case. %Hence,   in $C_{\Jc,\Bc}$ user $k_1$ has cached or recovered all sub-blocks except $W_{\Bc\cup\{d_{u_1}\},\Jc\setminus \{k_1\}}$. 
   \item If $i = d_{u_1}$, since we proved that user $k_1$ can recover all sub-blocks in $C_{\Jc,\Bc}$ except  $W_{\Bc\cup\{d_{u_1}\},\Jc\setminus \{k_1\}}$, we  prove user $k_1$ can recover $W_{\Bc\cup\{d_{u_1}\},\Jc\setminus \{k_1\}}$ by indirectly reading off.  
   \end{itemize}
In conclusion, user $k_1$ can recover all sub-blocks in  $C_{\Jc,\Bc}$.
\end{itemize}
Hence, we proved   Lemma~\ref{lem:Jc}.Item~\ref{item:lem JC1}  for Step~$1$.

\paragraph*{Lemma~\ref{lem:Jc}.Item~\ref{item:lem JC2} }
Now for each user $k\in [\Ksf]$, if $k=u_1$, it   can recover $W_{\Sc_1,\Vc_1}$
where  $d_{k}\in \Sc_1$ and $u_{1} \in \Vc_1$, from its cache. Hence, in the following, we will prove any user $k\in ([\Ksf]\setminus \{u_1\})$ can recover each $W_{\Sc_1,\Vc_1}$
where  $d_{k}\in \Sc_1$, $u_{1} \in \Vc_1$ and $k\notin \Vc_1$, from Step~$1$. We consider two cases:
\begin{itemize}
\item $d_{u_1}\notin \Sc_1$. We can see that $W_{\Sc_1,\Vc_1}$ appears in $C_{\Vc_1\cup \{k\}, \Sc_1\setminus \{d_k\}}$. By Lemma~\ref{lem:Jc}.Item~\ref{item:lem JC1}  for Step~$1$, we prove that user $k$ can recover  $W_{\Sc_1,\Vc_1}$.
\item $d_{u_1}\in \Sc_1$. We can see that $W_{\Sc_1,\Vc_1}$ appears in $C_{\Vc_1\cup \{k\}, \Sc_1\setminus \{d_{u_1}\}}$. By Lemma~\ref{lem:Jc}.Item~\ref{item:lem JC1}  for Step~$1$, we prove that user $k$ can recover  $W_{\Sc_1,\Vc_1}$.
\end{itemize}
Hence, we proved   Lemma~\ref{lem:Jc}.Item~\ref{item:lem JC2}  for Step~$ 1$.

\paragraph*{Lemma~\ref{lem:Jc}.Item~\ref{item:lem JC3} }
We then focus on one user $q$ whose demanded file is in $[\Nsf]\setminus \{d_{u_1}\}$, and one sub-block $W_{\Sc_2,\Vc_2}$ where $\{d_{q},d_{u_1}\} \subseteq \Sc_2$ and $\{u_1,q\}\cap \Vc_2=\emptyset$.
In $C_{\Vc_2\cup \{u_1\}, \Sc_2\setminus \{d_{u_1}\}}$, all sub-blocks are desired by user $q$ while only one of them is desired by user $u_1$ (which is  $W_{\Sc_2,\Vc_2}$) and the others are cached by user $u_1$. From  Lemma~\ref{lem:Jc}.Item~\ref{item:lem JC2}  for Step~$1$, user $q$ has recovered all desired sub-blocks which are cached by user $u_1$, and thus user $q$ can recover $W_{\Sc_2,\Vc_2}$ from  $C_{\Vc_2\cup \{u_1\}, \Sc_2\setminus \{d_{u_1}\}}$.
Hence, we proved   Lemma~\ref{lem:Jc}.Item~\ref{item:lem JC3}  for Step~$1$.
 
In summary, we proved Lemma~\ref{lem:Jc} for Step~$1$.

\subsection{Step~$j$}
\label{sub:step j of proof lem jc}
We focus one $j\in [\min\{N_{\textup{e}}(\dv),\Nsf-\rsf+1,\Ksf-t\}]$ and  assume that Lemma~\ref{lem:Jc} holds for the first $j-1$ steps. In the following, we prove that Lemma~\ref{lem:Jc} holds for Step~$j$.

\paragraph*{Lemma~\ref{lem:Jc}.Item~\ref{item:lem JC1} }
We focus on one set of users $\Jc\subseteq ([\Ksf]\setminus\{u_1,\ldots,u_{j-1}\})$ where $|\Jc|=t+1$ and $u_j\in \Jc$, and one set of files $\Bc\subseteq ([\Nsf]\setminus \{d_{u_1},\ldots.d_{u_{j}}\})$ where $|\Bc|=\rsf-1$.
We will prove that from the transmission until Step~$j$, each user in $k_1\in\Jc$ can recover all sub-blocks in $C_{\Jc,\Bc}$. We consider two cases:
\begin{itemize}
\item $d_{k_1} \notin \Bc$. In this case, in $C_{\Jc,\Bc}$ user $k_1$ caches all sub-blocks except $W_{\Bc\cup\{d_{k_1}\},\Jc\setminus \{k_1\}}$. Hence, user $k_1$ can recover $W_{\Bc\cup\{d_{k_1}\},\Jc\setminus \{k_1\}}$   by directly reading off.
\item $d_{k_1} \in \Bc$. In this case, $d_{k_1}\notin \{d_{u_1},\ldots,d_{u_{j}}\}$. 
In $C_{\Jc,\Bc}$ user $k_1$ caches all sub-blocks except  $W_{\Bc\cup\{i\},\Jc\setminus \{k_1\}}$, where $i \in \Nc(\Jc) \setminus \Bc$.
\begin{itemize}
\item If $i \in \{d_{u_1},\ldots,d_{u_{j-1}}\}$, by the induction assumption, user $k_1$ has already recovered the whole block $W_{\Bc\cup\{d_{k_2}\}}$.
\item If $i \notin \{d_{u_1},\ldots,d_{u_{j}}\}$,
user $k_1$ can recover  $W_{\Bc\cup\{d_{k_2}\},\Jc\setminus \{k_1\}}$ from $C_{\Jc,(\Bc\cup \{d_{k_2}\})\setminus \{d_{k_1}\}}$ transmitted in Step~$j$ by    directly reading off.
\item If $i=d_{u_j}$, in $C_{\Jc,\Bc}$ user $k_1$ has cached or recovered all sub-blocks except $W_{\Bc\cup\{d_{u_j}\},\Jc\setminus \{k_1\}}$. Hence, user $k_1$ can recover  $W_{\Bc\cup\{d_{u_j}\},\Jc\setminus \{k_1\}}$ by indirectly reading off.
\end{itemize}
\end{itemize}
 In conclusion, user $k_1$ can recover all sub-blocks in  $C_{\Jc,\Bc}$, and thus
  we proved   Lemma~\ref{lem:Jc}.Item~\ref{item:lem JC1}  for Step~$j$.

 \paragraph*{Lemma~\ref{lem:Jc}.Item~\ref{item:lem JC2} }
Now for each user $k\in [\Ksf]$ where $d_k\notin \{d_{u_1},\dots,d_{u_{j-1}}\}$, if $k=u_j$, it   can recover $W_{\Sc_1,\Vc_1}$
where  $d_{k}\in \Sc_1$ and $u_{j} \in \Vc_1$, from its cache.
Hence, in the following, we will prove any user $k\in ([\Ksf]\setminus \{u_j\})$ where $d_k\notin \{d_{u_1},\dots,d_{u_{j-1}}\}$, can recover each $W_{\Sc_1,\Vc_1}$
where  $d_{k}\in \Sc_1$, $u_{j} \in \Vc_1$ and $\{k,u_1,\ldots,u_{j-1}\} \cap \Vc_1=\emptyset$, at the end of Step~$j$.  
 We consider two cases:
\begin{itemize}
\item $d_{u_j}\notin \Sc_1$. We can see that $W_{\Sc_1,\Vc_1}$ appears in $C_{\Vc_1\cup \{k\}, \Sc_1\setminus \{d_k\}}$ transmitted in Step~$j$. By Lemma~\ref{lem:Jc}.Item~\ref{item:lem JC1}  for Step~$j$, we prove that user $k$ can recover  $W_{\Sc_1,\Vc_1}$.
\item $d_{u_j}\in \Sc_1$. We can see that $W_{\Sc_1,\Vc_1}$ appears in $C_{\Vc_1\cup \{k\}, \Sc_1\setminus \{d_{u_j}\}}$ transmitted in Step~$j$. By Lemma~\ref{lem:Jc}.Item~\ref{item:lem JC1}  for Step~$j$, we prove that user $k$ can recover  $W_{\Sc_1,\Vc_1}$.
\end{itemize}
Hence, we proved   Lemma~\ref{lem:Jc}.Item~\ref{item:lem JC2}  for Step~$j$.

\paragraph*{Lemma~\ref{lem:Jc}.Item~\ref{item:lem JC3} }
We then focus on one user $q$ whose demanded file is in $[\Nsf]\setminus \{d_{u_1},\ldots,d_{u_j}\}$, and one sub-block $W_{\Sc_2,\Vc_2}$ where $\{d_{q},d_{u_j}\} \subseteq \Sc_2$, $\{d_{u_1},\ldots,d_{u_{j-1}}\}\cap\Sc_2=\emptyset$, and $\{q,u_1,\ldots,u_j\}\cap \Vc_2=\emptyset$.
In $C_{\Vc_2\cup \{u_j\}, \Sc_2\setminus \{d_{u_j}\}}$ transmitted in Step~$j$, all sub-blocks are desired by user $q$ while only one of them is desired by user $u_j$ (which is  $W_{\Sc_2,\Vc_2}$) and the others are cached by user $u_1$. From  Lemma~\ref{lem:Jc}.Item~\ref{item:lem JC2}  for Step~$j$, user $q$ has recovered all desired sub-blocks which are cached by user $u_j$, and thus user $q$ can recover $W_{\Sc_2,\Vc_2}$ from  $C_{\Vc_2\cup \{u_1\}, \Sc_2\setminus \{d_{u_1}\}}$.
Hence, we proved   Lemma~\ref{lem:Jc}.Item~\ref{item:lem JC3}  for Step~$j$.

In conclusion, we proved Lemma~\ref{lem:Jc}.

\section{Proof of Lemma~\ref{lem:additional decoding}}
\label{sec:proof of general scheme lemma}
In Step~$j$, we focus on one $\Jc^{\prime}\subseteq ([\Ksf]\setminus\{u_1,\ldots,u_{N_{\textup{e}}(\dv)}\})$ where $|\Jc^{\prime}|=t-1$,   
 and one $\Bc^{\prime}\subseteq ([\Nsf]\setminus \{d_{u_1},\ldots,d_{u_{q}}\})$ where $|\Bc^{\prime}|=\rsf-2$ and $\Bc^{\prime}\cap \Nc([\Ksf])\neq \emptyset$, and in the following  we prove
$C_{\Jc^{\prime}\cup\{u_j,u_q\},\Bc^{\prime}}$ can be recovered by each user $k$ demanding $F_{d_{u_j}}$.

Given $\Jc^{\prime}$, we define a family of sets $\mathbb{S}(\Jc^{\prime})\ni \Jc^{\prime}$ as follows. We divide the users $\Jc^{\prime}$ into groups, where each group is corresponding to one file in $\Nc(\Jc^{\prime})$ and it contains all users in  $\Jc^{\prime}$ demanding this file.
 Each time, we choose one or zero user in each group which is not corresponding to the file in $\{d_{u_j},d_{u_q}\}$, 
and replace this user by the leader who demands the file corresponding to this group.  For example, $\Jc^{\prime}=\{5,6,7,8\}$ where $d_{u_j}=1$, $d_{u_q}=2$, $d_5=d_6=3$, $d_7=4$, and $d_8=2$. The leader user demanding $F_{3}$ is user $3$ while the leader user demanding $F_4$ is user $4$. We first choose user $5$ in the first group and replace it by user $3$, and choose user $7$ in the second group and replace it by user $4$. Hence, we have the set of users $\{3,4,6,8\}\in \mathbb{S}(\Jc^{\prime})$. Similarly, in this example we have 
$$
\mathbb{S}(\Jc^{\prime})=\big\{\{3,4,6,8\}, \{3,4,5,8\}, \{3,6,7,8\}, \{3,5,7,8\},\{4,5,6,8\},\{5,6,7,8\}  \big\}.
$$

For each $\Jc\in \mathbb{S}(\Jc^{\prime})$, with a slight abuse of notation, we let
\begin{align}
Q_{\Jc}:=\underset{k_2\in \Jc}{\oplus } \underset{\substack{\Sc\subseteq (\Nc(\Jc)\cup \Bc^{\prime}\setminus \{d_{u_j},d_{u_q}\}):\\ \Bc^{\prime}\subseteq \Sc, d_{k_2}\in \Sc }}{\oplus } W_{\Sc,\Jc\cup\{u_j,u_q\}\setminus\{k_2\}}.\label{eq:QJ}
\end{align}
In other words, $Q_{\Jc}$ is obtained by removing all sub-blocks from the blocks desired by user $u_j$ or $u_q$ in $C_{\Jc\cup\{u_j,u_q\},\Bc^{\prime}}$.

For each  $\Jc\in  \mathbb{S}(\Jc^{\prime})$, by the definitions, we have 
 \begin{align}
 C_{\Jc \cup\{u_j,u_q\},\Bc^{\prime}}\oplus C_{\Jc \cup\{u_j,u_q\},\Bc^{\prime}\cup \{u_q\}}\oplus \Qc_{\Jc } =    \underset{k_3\in \Jc \cup\{u_j\}}{\oplus } \underset{\substack{\Sc\subseteq (\Nc(\Jc \cup\{u_j\})\cup \Bc^{\prime})\setminus \{d_{u_q}\} :\\ \Bc^{\prime}\subseteq \Sc, \{d_{k_3},d_{u_j}\}\subseteq \Sc   }}{\oplus } W_{\Sc,\Jc \cup\{u_j,u_q\}\setminus\{k_3\}}.\label{eq:three sum}
 \end{align}
In~\eqref{eq:three sum}, if $k_3\neq u_j$, $W_{\Sc,\Jc\cup\{u_j,u_q\}\setminus\{k_3\}}$ is cached by $u_j$ and from Lemma~\ref{lem:Jc}.Item~\ref{item:lem JC2} , user $k$ can recover $W_{\Sc,\Jc\cup\{u_j,u_q\}\setminus\{k_3\}}$. We then focus on $k_3= u_j$. Since $u_q\notin \Sc$, by Remark~\ref{rem:no rem}, it can be seen that $W_{\Sc,\Jc \cup\{u_j,u_q\}\setminus\{k_3\}}$ can be recovered by user $k$. Hence, user $k$ can reconstruct the RHS of~\eqref{eq:three sum}.

For each $\Jc_1\in \mathbb{S}(\Jc^{\prime})$ where $\Jc_1\neq \Jc^{\prime}$,   since there exists at least one leader in $\Jc_1$, it can be seen that $C_{\Jc \cup\{u_j,u_q\},\Bc^{\prime}\cup \{u_q\}}$ and $C_{\Jc_1 \cup\{u_j,u_q\},\Bc^{\prime}}$ 
are transmitted in  Step~$j$ of the first and second sub-phases, respectively. Hence, user $k$ can reconstruct $\Qc_{\Jc_1}$ from~\eqref{eq:three sum}.

At the end of this proof, we will prove the following equation.
\begin{align}
  \underset{\Jc\in \mathbb{S}(\Jc^{\prime})}{\oplus }Q_{\Jc}=0.\label{eq:two sum}
\end{align}
 In~\eqref{eq:two sum}, all the messages except $\Qc_{\Jc^{\prime}}$  are recovered by user $k$ such that each user can reconstruct  $\Qc_{\Jc^{\prime}}$. In addition, $ C_{\Jc^{\prime} \cup\{u_j,u_q\},\Bc^{\prime}\cup \{u_q\}}$ is transmitted in Step~$j$ of the first sub-phase. Hence, from~\eqref{eq:three sum}, user $k$ can reconstruct $ C_{\Jc^{\prime} \cup\{u_j,u_q\},\Bc^{\prime} }$.
 
 Finally, we will prove~\eqref{eq:two sum}. We focus on one sub-block in~\eqref{eq:two sum} and  
assume that $W_{\Sc,\Vc}$ is in $\Qc_{\Jc}$, which is desired by user $k_5$. Hence, $\Vc=\Jc\cup\{u_j,u_q\}\setminus \{k_5\}$ and $d_{k_5}\notin \{d_{u_j},d_{u_q}\}$. By the construction of $ \mathbb{S}(\Jc^{\prime})$,
 there exists only one user in $\Jc^{\prime}\cup \{u_{g(d_{k_5})}\}$ demanding $d_{k_5}$, who is not in $\Jc$. We assume this user is $k_6$. It can be seen that  $W_{\Sc,\Vc}$ desired by $k_6$, is also in $\Qc_{\Jc_2}$ where $\Jc_2=\Jc\cup\{k_6\}\setminus \{k_5\}$. In addition, except $\Jc$ and $\Jc_2$, there does not exist other $\Jc_3\in \mathbb{S}(\Jc^{\prime})$ such that $\Qc_{\Jc_3}$ contains $W_{\Sc,\Vc}$ (because $\Vc\setminus \{d_{u_j},d_{u_q}\}=\Jc\setminus  \{k_5\}$ can not be subset of $\Jc_3$).
Hence,  $W_{\Sc,\Vc}$ appears twice in~\eqref{eq:two sum} and we prove~\eqref{eq:two sum}.

\section{Proof of Lemma~\ref{lem:lemma for interference alignment}}
\label{sec:proof of transition lemma}

\subsection{Proof of~\eqref{eq:interference transition}}
\label{sub:proof of eq:interference transition}
To prove~\eqref{eq:interference transition}, it is equivalent to prove that 
\begin{align}
 \underset{k\in \Rc}{\oplus } C_{(\Jc\setminus \{k\})\cup  \{u_{g(i)}\},(\Bc\setminus\{i\})\cup \{d_k\}}=0. \label{eq:equivalent1}
\end{align}
where we assume $\Rc=\Jc\cup \{u_{g(i)}\}$. Since $u_{g(i)}\notin \Jc$ and $|\Jc|=t+1$, we have $|\Rc|=t+2$.  Any  $C_{\Tc,\Hc}$ in~\eqref{eq:equivalent1}, should satisfy $\Tc\subseteq \Rc$ and $|\Rc\setminus \Tc|=1$. For the user in $(\Rc\setminus \Tc)$, its desired file is in $\Hc$. 
In addition, if $C_{\Tc_1,\Hc_1}$  and $C_{\Tc_2,\Hc_2}$ are in~\eqref{eq:equivalent1}, we can see that $\Tc_1\neq \Tc_2$. 

 We focus one sub-block $W_{\Sc,\Vc}$    in~\eqref{eq:equivalent1} and assume that   $C_{\Tc,\Hc}$ contains  $W_{\Sc,\Vc}$. It directly indicates that $\Sc\subseteq \Nc(\Tc)\cup \Hc$,
and that $\Tc \supseteq \Vc$, $|\Tc\setminus \Vc|=1$, the user in  $\Tc\setminus \Vc$ (assumed to be user $k^{\prime}$) desires the sub-block $W_{\Sc,\Vc}$. In addition, since $k^{\prime}\in \Tc \subseteq \Rc$ and $|\Rc\setminus \Tc|=1$, assuming $k_1\in (\Rc\setminus \Tc)$, we have $d_{k_1}\in\Hc$ and thus $W_{\Sc,\Vc}$ is also desired by user $k_1$.
Hence, 
 it can be seen that $C_{\Vc\cup\{k_1\}, \Hc\setminus \{d_{k_1}\}\cup \{d_{k^{\prime}}\} }$ is also in~\eqref{eq:equivalent1}, and   $W_{\Sc,\Vc}$ desired by user $k_1$ is in $C_{\Vc\cup\{k_1\}, \Hc\setminus \{d_{k_1}\}\cup \{d_{k^{\prime}}\} }$.
Except  $C_{\Tc,\Hc}$ and $C_{\Vc\cup\{k_1\}, \Hc\setminus \{d_{k_1}\}\cup \{d_{k^{\prime}}\} }$, there does not exist any other $C_{\Tc_1,\Hc_1}$   in~\eqref{eq:equivalent1} containing  $W_{\Sc,\Vc}$ because there is no other $\Tc_1\subseteq \Rc$   where $|\Tc_1|=|\Rc|-1$ and $\Vc\subseteq \Tc_1 $ (noticing that $\Vc\subseteq \Rc$ and $|\Vc|=|\Rc|-2$).

In conclusion, each sub-block in~\eqref{eq:equivalent1} appears twice in~\eqref{eq:equivalent1},
and thus we prove~\eqref{eq:equivalent1}.

\subsection{Proof of~\eqref{eq:interference transition2}}
\label{sub:proof of eq:interference transition2}
To prove~\eqref{eq:interference transition2}, it is equivalent to prove that 
\begin{align}
  \underset{i\in (\Nc_{\Jc_1}\setminus \Bc_1)\cup \{i_1\}}{\oplus } C_{\Jc_1,(\Bc_1\setminus\{i_1\})\cup \{i\}}=0. \label{eq:equivalent2}
\end{align}
If  $C_{\Jc_1,\Hc }$ appears in~\eqref{eq:equivalent2},  since  $i_1 \in\Nc_{\Jc_1}\cap \Bc_1 $, 
we have $(\Bc_1\setminus\{i_1\})\subseteq  \Hc$ and $|\Hc\setminus (\Bc_1\setminus\{i_1\})|=1$. For the file in $\Hc\setminus (\Bc_1\setminus\{i_1\})$, it is also in $(\Nc_{\Jc_1}\setminus \Bc_1)\cup \{i_1\}$.

 We focus one sub-block $W_{\Sc,\Vc}$    in~\eqref{eq:equivalent2} and assume that   $C_{\Jc_1,\Hc}$ contains  $W_{\Sc,\Vc}$. It directly indicates that $\Hc\subseteq \Sc$
and $|\Sc\setminus \Hc|=1$ (we assume the file in  $\Sc\setminus \Hc$ is $i^{\prime}$). 
In addition, we have $(\Bc_1\setminus\{i_1\})\subseteq \Hc$ and  $|\Hc\setminus (\Bc_1\setminus\{i_1\})|=1$ (we assume the file in $\Hc\setminus (\Bc_1\setminus\{i_1\})$ is $i_2$). As described before, $i_2$ is $(\Nc_{\Jc_1}\setminus \Bc_1)\cup \{i_1\}$ and thus file $i_2$ is demanded by some user in $\Jc_1$.
Hence, it can be seen that  $W_{\Sc,\Vc}$ is also in $C_{\Jc_1,\Hc\setminus \{i_2\}\cup \{i^{\prime}\}}$. Except  $C_{\Jc_1,\Hc}$ and $C_{\Jc_1,\Hc\setminus \{i_2\}\cup \{i^{\prime}\}}$, there does not exist any other $C_{\Jc_1,\Hc_1}$   in~\eqref{eq:equivalent2} containing  $W_{\Sc,\Vc}$ because there is no other $\Hc_1 \subseteq \Sc$ where $(\Bc_1\setminus\{i_1\})\subseteq \Hc_1$ and $|\Hc_1|=|\Sc|-1$ (noticing that $(\Bc_1\setminus\{i_1\})\subseteq \Sc$ and $|\Bc_1\setminus\{i_1\}|=|\Sc|-2$).

In conclusion, each sub-block in~\eqref{eq:equivalent2} appears twice in~\eqref{eq:equivalent2},
and thus we prove~\eqref{eq:equivalent2}.

\section{Proof of Lemma~\ref{lem:lemma for Bc}}
\label{sec:proof of lemma Bc}
We use the induction method to prove Lemma~\ref{lem:lemma for Bc}.

{\it $j=1$.}
By~\eqref{eq:interference transition2} in Lemma~\ref{lem:lemma for interference alignment} (with $i_1=d_{u_1}$), we have 
\begin{align}
C_{\Jc_2\cup\{u_1\},\Bc_2\cup\{d_{u_1}\}}=\underset{i_2\in \Nc_{\Jc_2}\setminus (\Bc_2\cup\{d_{u_1}\})}{\oplus } C_{\Jc_2\cup\{u_1\},\Bc_2\cup \{i_2\}},\label{eq:j=1 by i1=du1}
\end{align}
where each $C_{\Jc_2\cup\{u_1\},\Bc_2\cup \{i_2\}}$ is transmitted in Step~$1$ of the first sub-phase. Hence, each user can reconstruct $C_{\Jc_2\cup\{u_1\},\Bc_2\cup\{d_{u_1}\}}$.

{\it $j\in [2:\min\{N_{\textup{e}}(\dv),\Nsf-\rsf+1,\Ksf-t\}]$.}
We first focus on $C_{\Jc_2\cup\{u_j\},\Bc_2\cup\{i\}}$ where $i\in \{d_{u_1},\ldots,d_{u_{j-1}}\}$. 
By~\eqref{eq:interference transition} in Lemma~\ref{lem:lemma for interference alignment}, we have
\begin{align}
C_{\Jc_2\cup\{u_j\},\Bc_2\cup\{i\}}=\underset{k\in\Jc_2\cup\{u_j\}}{\oplus } C_{ \Jc_2\cup\{u_j,u_{g(i)}\}\setminus \{k\} ,\Bc_2\cup \{d_k\}}. \label{eq:j by i=i}
\end{align}
If $d_k\in \{d_{u_1},\ldots,d_{u_{j-1}}\}$, each user can reconstruct  $C_{ \Jc_2\cup\{u_j,u_{g(i)}\}\setminus \{k\} ,\Bc_2\cup \{d_k\}}$ by the induction assumption; else if $d_k\notin \Bc_2$, $C_{ \Jc_2\cup\{u_j,u_{g(i)}\}\setminus \{k\} ,\Bc_2\cup \{d_k\}}$ is transmitted in Step~$g(i)$ of the first sub-phase; else, we consider $d_k\in \Bc_2$. Since  $\Nc(\Jc_2\cap \Lc)\setminus \Bc_2 \neq \emptyset$ (recall that $\Lc$ is the set of leaders), we can see that there exists one leader in $\Jc_2\setminus \{k\}$ whose demanded file is not in $\Bc_2$.  Thus in this case, $C_{ \Jc_2\cup\{u_j,u_{g(i)}\}\setminus \{k\} ,\Bc_2\cup \{d_k\}}$ is transmitted in Step~$g(i)$ of the second sub-phase.

We then focus on $C_{\Jc_2\cup\{u_j\},\Bc_2\cup\{d_{u_j}\}}$. By~\eqref{eq:interference transition2} in Lemma~\ref{lem:lemma for interference alignment} (with $i_1=d_{u_j}$), we have 
\begin{align}
C_{\Jc_2\cup\{u_j\},\Bc_2\cup\{d_{u_j}\}}=\underset{i_2\in \Nc(\Jc_2)\setminus (\Bc_2\cup\{d_{u_j}\})}{\oplus } C_{\Jc_2\cup\{u_j\},\Bc_2\cup \{i_2\}}.\label{eq:j  by i1=duj}
\end{align}
In~\eqref{eq:j  by i1=duj}, if $i_2\in \{d_{u_1},\ldots,d_{u_{j-1}}\}$, we have proved $C_{\Jc_2\cup\{u_j\},\Bc_2\cup \{i_2\}}$ can be reconstructed by each user from~\eqref{eq:j by i=i}; otherwise, $C_{\Jc_2\cup\{u_j\},\Bc_2\cup \{i_2\}}$ is transmitted in Step~$j$ of the first sub-phase. 

%\begin{rem}
%\label{rem:r=2 lemma3}
%It can be seen that, to prove Lemma~\ref{lem:lemma for Bc}, the transmission in the second sub-phase is only used when $d_k\in \Bc_2$. When $\rsf=2$, we have $|\Bc_2|=0$ and $d_k$ cannot be in $\Bc_2$. Hence, when $\rsf=2$, to prove %Lemma~\ref{lem:lemma for Bc}, the transmission in the second sub-phase is not used. 
%\end{rem}

\begin{rem}
\label{rem:equivalent proof}
Notice that to prove Lemma~\ref{lem:lemma for Bc} the transmission in the second sub-phase is only used when there exists some user in $\Jc_2\cup\{u_j\}$ whose demanded file is in  $\Bc_2$ (i.e., $d_k\in \Bc_2$ in~\eqref{eq:j by i=i}).
 Hence,  if $\Nc(\Jc_2)  \cap \Bc_2=\emptyset $, to reconstruct $C_{\Jc_2\cup\{u_j\},\Bc_2\cup\{i\}}$, each user only needs the transmission in the first sub-phase. 
 
 Formally, 
for each $j \in [\min\{N_{\textup{e}}(\dv),\Nsf-\rsf+1,\Ksf-t\}]$ and each $i \in\{d_{u_1},\ldots,d_{u_{j} }\}$,
any non-leader $k\in[\Ksf]$ can reconstruct $C_{\Jc_2\cup\{u_{j }\},\Bc_2\cup\{i \}}$ where $\Jc_2\subseteq  [\Ksf]\setminus \{u_{1},\ldots,u_{j }\}$, $|\Jc_2|=t$, $\Bc_2\subseteq ([\Nsf]\setminus \{d_{u_1},\ldots,d_{u_{j }}\})$, $|\Bc_2|=\rsf-2$, and $\Nc(\Jc_2)  \cap \Bc_2=\emptyset $, 
from the transmission of the first sub-phase.

\end{rem}
\section{Proof of Decodability of the General Scheme in Section~\ref{sub:scheme 2}}
\label{sec:decodability general scheme}
Now we are ready to prove the decodability of each non-leader $k$. In other words, we want to prove that it can decode $W_{\Sc_2,\Vc_2}$ where $d_k \in \Sc_2$, $\{d_{u_1},\ldots,d_{u_{g(d_k)-1}}\} \cap \Sc_2=\emptyset$ and 
 $\{k,u_1,\ldots,u_{g(d_k)}\}\cap \Vc_2 = \emptyset$ (in Lemma~\ref{lem:Jc} we showed that the other desired sub-blocks could  be decoded by user $k$ from transmission of the first sub-phase). We consider two cases, $|\Sc_2\cap \Nc([\Ksf])|>1$ and $|\Sc_2\cap \Nc([\Ksf])|=1$.
 %where in the first case  $|\Sc_2\cap \Nc(\dv)|>1$ and $|\Vc_2\cap \{u_1,\ldots,u_{N_{\textup{e}}(\dv)}\}|>0$, in the second case $|\Sc_2\cap \Nc(\dv)|=1$ and in the third case $|\Vc_2\cap \{u_1,\ldots,u_{N_{\textup{e}}(\dv)}\}|=0$.

\subsection{ $|\Sc_2\cap \Nc([\Ksf])|>1$  } 
\label{sub:case 1 deco} 
Among all desired sub-blocks in this case,
we use the induction method to prove for each $j\in [g(d_k)+1:N_{\textup{e}}(\dv)]$, user $k$ can recover its desired sub-block $W_{\Sc_2,\Vc_2}$ where $d_{u_j}\in \Sc_2$ or $u_{j}\in \Vc_2$.
 
{\it Induction on $j=g(d_k)+1$.}
We consider three cases:
\begin{itemize}
\item  $u_{j}\in \Vc_2$ and $d_{u_j}\notin \Sc_2$.  In $C_{\Vc_2\cup\{k\},\Sc_2\setminus\{d_k\}}$ transmitted in Step~$j$ of the first sub-phase, user $k$ caches all sub-blocks except $W_{\Sc_2,\Vc_2}$ and thus it can  recover  $W_{\Sc_2,\Vc_2}$   by directly reading off.
\item $u_{j}\in \Vc_2$ and $d_{u_j}\in \Sc_2$. 
Since $u_{j}\in \Vc_2$, from Lemma~\ref{lem:lemma for Bc} it can be seen that user $k$ can reconstruct $C_{\Vc_2\cup \{u_{g(d_k)}\},\Sc_2\setminus \{d_{u_j}\}}$. 

In $C_{\Vc_2\cup \{u_{g(d_k)}\},\Sc_2\setminus \{d_{u_j}\}}$, all   sub-blocks are desired by user $k$. In addition, all     sub-blocks   desired by user $k$ which are cached by user $u_{g(d_k)}$, can be recovered by user $k$ from Lemma~\ref{lem:Jc}.\ref{item:lem JC2}.

 The sub-blocks in $C_{\Vc_2\cup \{u_{g(d_k)}\},\Sc\setminus \{d_{u_j}\}}$ which are not cached by user $u_{g(d_k)}$, are all cached by user $u_j$ (because $u_j\in \Vc_2$). 
 For any file $i\in \Nc(\Vc_2)\setminus (\Sc_2\setminus \{d_{u_j}\})$,  the sub-block $W_{\Sc_2\setminus \{d_{u_j}\}\cup \{i\}, \Vc_2 }$ is in $C_{\Vc_2\cup \{u_{g(d_k)}\},\Sc_2\setminus \{d_{u_j}\}}$ which is desired (and not cached) by user $u_{g(d_k)}$.
 % any user $k_1\in \Vc_2$ where $d_{k_1}\notin (\Sc_2\setminus \{d_{u_j}\})$, the sub-block $W_{\Sc_2\setminus \{d_{u_j}\}\cup \{d_{k_1}\}, \Vc_2 }$ is in $C_{\Vc_2\cup \{u_{g(d_k)}\},\Sc_2\setminus \{d_{u_j}\}}$ which is not cached by user $u_{g(d_k)}$. 
  If $i \neq d_{u_j}$, 
since $d_{u_j}\notin(\Sc_2\setminus \{d_{u_j}\}\cup \{i\})$ and  $u_{j}\in \Vc_2$, we   proved in the first case that $W_{\Sc_2\setminus \{d_{u_j}\}\cup \{i\}, \Vc_2 }$   can be recovered by user $k$; otherwise, the sub-block $W_{\Sc_2\setminus \{d_{u_j}\}\cup \{i\}, \Vc_2 }$ is  $W_{\Sc_2,\Vc_2}$.
 Hence, in $C_{\Vc_2\cup \{u_{g(d_k)}\},\Sc_2\setminus \{d_{u_j}\}}$, only sub-block  $W_{\Sc_2,\Vc_2}$ is not recovered by user $k$, such that user $k$ can recover  $W_{\Sc_2,\Vc_2}$.
\item   $u_{j}\notin \Vc_2$ and  $d_{u_j}\in \Sc_2$.  We first prove that user $k$ can reconstruct $C_{\Vc_2\cup \{u_{j}\},\Sc_2\setminus \{d_{u_{j}}\}}$. From~\eqref{eq:interference transition} in Lemma~\ref{lem:lemma for interference alignment}, we have 
\begin{align}
C_{\Vc_2\cup \{u_{j}\},\Sc_2\setminus \{d_{u_{j}}\}}=\underset{k_2 \in (\Vc_2\cup \{u_{j}\})}{\oplus } C_{\Vc_2\cup \{u_{j},u_{g(d_k)}\}\setminus \{k_2\},(\Sc_2\setminus \{d_{u_{j}},d_k\})\cup\{d_{k_2} \}}.\label{eq:case 1 C to be proved}
\end{align}
For each $k_2 \in (\Vc_2\cup \{u_{j}\})$ in~\eqref{eq:case 1 C to be proved}, 
\begin{itemize}
\item if $k_2=u_j$, we have  
$$ 
C_{\Vc_2\cup \{u_{j},u_{g(d_k)}\}\setminus \{k_2\},(\Sc_2\setminus \{d_{u_{j}},d_k\})\cup\{d_{k_2} \}}=C_{\Vc_2\cup\{u_{g(d_k)}\},\Sc_2\setminus \{d_k\}},
$$
 which is transmitted     in Step~$g(d_k)$ of the first sub-phase;
\item if $k_2\neq u_j$ and $d_{k_2}\notin \{d_{u_1},\ldots,d_{u_{g(d_k)}}\}$, 
it can be seen that  $ C_{\Vc_2\cup \{u_{j},u_{g(d_k)}\}\setminus \{k_2\},(\Sc_2\setminus \{d_{u_{j}},d_k\})\cup\{d_{k_2} \}}$ is transmitted either in Step~$g(d_k)$ of the first sub-phase (if $|(\Sc_2\setminus \{d_{u_{j}},d_k\})\cup\{d_{k_2}\}|=\rsf-1$) or Step~$g(d_k)$ of the second sub-phase (if $|(\Sc_2\setminus \{d_{u_{j}},d_k\})\cup\{d_{k_2}\}|=\rsf-2$ and $(\Vc_2\setminus\{k_2\})\cap \Nc([\Ksf])\neq \emptyset$) or Step~$g(d_k)$ in Lemma~\ref{lem:additional decoding} (if $|(\Sc_2\setminus \{d_{u_{j}},d_k\})\cup\{d_{k_2}\}|=\rsf-2$ and $(\Vc_2\setminus\{k_2\})\cap \Nc([\Ksf])=\emptyset$); 
\item if  $k_2\neq u_j$ and $d_{k_2}\in \{d_{u_1},\ldots,d_{u_{g(d_k)}}\}$, 
by Lemma~\ref{lem:lemma for Bc}, $ C_{\Vc_2\cup \{u_{j},u_{g(d_k)}\}\setminus \{k_2\},(\Sc_2\setminus \{d_{u_{j}},d_k\})\cup\{d_{k_2} \}}$ can be reconstructed by user $k$.
\end{itemize}
Hence, user $k$ can recover each message on the RHS of~\eqref{eq:case 1 C to be proved} and thus it can reconstruct $C_{\Vc_2\cup \{u_{j}\},\Sc_2\setminus \{d_{u_{j}}\}}$. 

In $C_{\Vc_2\cup \{u_{j}\},\Sc_2\setminus \{d_{u_{j}}\}}$, all sub-blocks are desired by user $k$.
For each $k^{\prime}\in (\Vc_2\cup \{u_{j}\})$, if $k^{\prime}\neq u_j$, the desired sub-blocks  in  $C_{\Vc_2\cup \{u_{j}\},\Sc_2\setminus \{d_{u_{j}}\}}$ by user $k^{\prime}$ are stored by user $u_j$, which can be recovered by user $k$ from the transmission of the first sub-phase (as we proved above for the case $u_{j}\in \Vc_2$ and $d_{u_j}\notin \Sc_2$). If $k^{\prime}=u_j$, the desired sub-block by user $k^{\prime}$ is $W_{\Sc_2,\Vc_2}$. Hence, user $k$ can recover  $W_{\Sc_2,\Vc_2}$. 
 
\end{itemize}
 
{\it Induction on $j\in [g(d_k)+1:N_{\textup{e}}(\dv)]$.} If there exists  $j^{\prime}\in [g(d_k)+1:j-1]$, where  $u_{j^{\prime}}\in \Vc_2$ or $d_{u_{j^{\prime}}}\in \Sc_2$, by the induction assumption,  user $k$ can recover  $W_{\Sc_2,\Vc_2}$; otherwise, we can use the similar proof  by dividing into three cases and using the induction assumption, to prove   user $k$ can recover  $W_{\Sc_2,\Vc_2}$ (for the sake of simplicity, we do not repeat).

\begin{rem}
\label{rem:no rem}
If there exists one leader in $\Vc_2$ (assumed to be $k^{\prime}$) such that $d_{k^{\prime}}\notin \Sc_2$, we can prove user $k$ can recover $W_{\Sc_2,\Vc_2}$ without using Lemma~\ref{lem:additional decoding}. More precisely, we focus on the case $u_{j}\in \Vc_2$ and $d_{u_j}\in \Sc_2$, where Lemma~\ref{lem:additional decoding} may be needed. 
 In~\eqref{eq:case 1 C to be proved}, for each $k_2\in \Vc_2$ where $d_{k_2}\notin \{d_{u_1},\ldots,d_{u_{g(d_k)}}\}$, if $k_2\neq k^{\prime}$, it can be seen that $(\Vc_2\setminus\{k_2\})\cap \Nc([\Ksf])\neq \emptyset$ and thus Lemma~\ref{lem:additional decoding} is not  needed; otherwise, we have $k_2=k^{\prime}$ and $|(\Sc_2\setminus \{d_{u_{j}},d_k\})\cup\{d_{k^{\prime}}\}|=\rsf-1$ such that Lemma~\ref{lem:additional decoding} is not  needed.
\end{rem}

\subsection{$|\Sc_2\cap \Nc([\Ksf])|=1$} 
\label{sub:case 2 deco} 
We can gather all blocks $W_{\Sc}$ where $\Sc\cap ([\Nsf]\setminus \Nc([\Ksf]))=\Sc_2\cap ([\Nsf]\setminus \Nc([\Ksf]))$. The transmission for these blocks is equivalent to the MAN caching problem in~\cite{dvbt2fundamental} and thus from the transmission of the first sub-phase on these blocks which is equivalent to the optimal caching scheme  in~\cite{exactrateuncoded},
each non leader can recover $W_{\Sc_2,\Vc_2}$.
 
\subsection{Proof of Observations}
\label{sub:proof of observation remark}

If  $|\Sc_2\cap \Nc([\Ksf])|=1$, it has been proved that only the first sub-phase is needed. Hence, in the following we consider  $|\Sc_2\cap \Nc([\Ksf])|>1$.
 We focus on each non-leader $k$ and the induction Step~$j\in [g(d_k)+1:N_{\textup{e}}(\dv)]$  in the proof of the decodability in Appendix~\ref{sub:case 1 deco}.

\paragraph*{Proof of Observation~\ref{item:obs1}}  
The proof of  Observation~\ref{item:obs1} is trivial and directly given in  Section~\ref{sub:scheme 3 r=2}.

\paragraph*{Proof of Observation~\ref{item:obs2}}  
Other steps of the second sub-phase may be needed only when we use Lemma~\ref{lem:lemma for Bc} to show that  user $k$ can reconstruct
$$ 
C_{\Vc_2\cup \{u_{j},u_{g(d_k)}\}\setminus \{k_2\},(\Sc_2\setminus \{d_{u_{j}},d_k\})\cup\{d_{k_2} \}}=C_{\Vc_2\cup \{u_{j},u_{g(d_k)}\}\setminus \{k_2\},(\Sc_2\setminus \{d_{u_{j}}  \}},
$$ 
where  $d_{k_2}=d_k$. 
However, in   induction Step~$g(d_k)$ of the proof of Lemma~\ref{lem:lemma for Bc} with    $i=d_k$,   the first sub-phase and Step~$g(d_k)$ of the second sub-phase are only needed   if there is no user in $\Vc_2$ whose demanded file is in $\{d_{u_1},\ldots,d_{u_{g(d_k)-1}}\}$.   
Hence, we prove Observation~\ref{item:obs2}.

\paragraph*{Proof of Observation~\ref{item:obs3}}
 \begin{itemize}
 \item If $u_{j}\in \Vc_2$ and $d_{u_j}\notin \Sc_2$, from the proof in Appendix~\ref{sub:case 1 deco}, the first sub-phase is only  needed;
 \item if  $u_{j}\in \Vc_2$ and $d_{u_j}\in \Sc_2$, user $k$ needs to reconstruct  $C_{\Vc_2\cup \{u_{g(d_k)}\},\Sc_2\setminus \{d_{u_j}\}}$. Since $(\cup_{k_1\in \Vc_2} \{d_{k_1}\}) \cap (\Sc_2\setminus \{d_k\})=\emptyset$, from Remark~\ref{rem:equivalent proof} %(with $j^{\prime}=g(d_k)$, $i^{\prime}=d_k$, $\Jc_3=\Vc_2$, and $\Bc_3=\Sc_2\setminus \{d_{u_j},d_{k}\}$),
  we can see that  $C_{\Vc_2\cup \{u_{g(d_k)}\},\Sc_2\setminus \{d_{u_j}\}}$ can be reconstructed by user $k$ from the transmission of the first sub-phase;
 \item Lastly we focus on $u_{j}\notin \Vc_2$ and  $d_{u_j}\in \Sc_2$. In~\eqref{eq:case 1 C to be proved}, 
\begin{itemize}
\item if $k_2 = u_j$, the first sub-phase is only needed;
\item  if $k_2\neq u_j$ and $d_{k_2}\notin \{d_{u_1},\ldots,d_{u_{g(d_k)}}\}$, since $(\cup_{k_1\in \Vc_2} \{d_{k_1}\}) \cap (\Sc_2\setminus \{d_k\})=\emptyset$, we have $|(\Sc_2\setminus \{d_{u_{j}},d_k\})\cup\{d_{k_2}\}|=\rsf-1$ and thus we only need the first sub-phase;
\item  if  $k_2\neq u_j$ and $d_{k_2}\in \{d_{u_1},\ldots,d_{u_{g(d_k)}}\}$, user $k$ should reconstruct  $ C_{\Vc_2\cup \{u_{j},u_{g(d_k)}\}\setminus \{k_2\},(\Sc_2\setminus \{d_{u_{j}},d_k\})\cup\{d_{k_2} \}}$. Again, since $(\cup_{k_1\in \Vc_2} \{d_{k_1}\}) \cap (\Sc_2\setminus \{d_k\})=\emptyset$,  from Remark~\ref{rem:equivalent proof} we can see that  user $k$ can reconstruct $ C_{\Vc_2\cup \{u_{j},u_{g(d_k)}\}\setminus \{k_2\},(\Sc_2\setminus \{d_{u_{j}},d_k\})\cup\{d_{k_2} \}}$  from the first sub-phase. 
\end{itemize} 
 \end{itemize}
Hence, we prove Observation~\ref{item:obs3}.

\section{Proof of the Decodability for $\rsf=\Nsf-1$ or $t=2$}
\label{sec:r=N-1 t=2}
We now consider  $\rsf=\Nsf-1$ or $t=2$ and prove that each non-user $k$ can recover $W_{\Sc_2,\Vc_2}$ where $d_k \in \Sc_2$, $\{d_{u_1},\ldots,d_{u_{g(d_k)-1}}\} \cap \Sc_2=\emptyset$ and 
 $\{k,u_1,\ldots,u_{g(d_k)}\}\cap \Vc_2 = \emptyset$, by the transmission of the first sub-phase. 
If  $\Nc(\Vc_2) \cap (\Sc_2\setminus \{d_k\})=\emptyset$, by Observation~\ref{item:obs3}, user $k$ can recover $W_{\Sc_2,\Vc_2}$ from the first sub-phase.  Hence, in the following, we focus on $\Nc(\Vc_2) \cap (\Sc_2\setminus \{d_k\})\neq \emptyset$.
 We consider two cases, $\Nc(\Vc_2) \cap \{d_{u_1},\ldots,d_{u_{g(d_k)-1}}\}=\emptyset$ and  $\Nc(\Vc_2) \cap \{d_{u_1},\ldots,d_{u_{g(d_k)-1}}\}\neq \emptyset$.
 
\subsection{ $\Nc(\Vc_2) \cap \{d_{u_1},\ldots,d_{u_{g(d_k)-1}}\}=\emptyset$}
\label{sub:r=N-1 t=2 case 1}
In the following, we  prove that for each integer $q\in [g(d_k)+1: \min\{\Nsf-\rsf+2,\Ksf-t+1,N_{\textup{e}}(\dv)\}]$,
 user $k$ can reconstruct $C_{\Jc^{\prime}\cup\{u_{g(d_k)},u_q\},\Bc^{\prime}}$ from the first sub-phase, where $\Jc^{\prime}\subseteq ([\Ksf]\setminus\{u_1,\ldots,u_{q}\})$, $|\Jc^{\prime}|=t-1$, $\Bc^{\prime}\subseteq ([\Nsf]\setminus \{d_{u_1},\ldots,d_{u_{q}}\})$, $|\Bc^{\prime}|=\rsf-2$, and $\Bc^{\prime}\cap \Nc([\Ksf])\neq \emptyset$.

If there is no user in $\Jc^{\prime}$ whose demand is in $[\Nsf]\setminus (\{d_{k},d_{u_q} \}\cup \Bc^{\prime})$, it can be seen that all sub-blocks in $C_{\Jc^{\prime}\cup\{u_{g(d_k)},u_q\},\Bc^{\prime}}$ are from $W_{\Bc^{\prime}\cup \{d_{k},d_{u_q} \}}$. Hence, we have $C_{\Jc^{\prime}\cup\{u_{g(d_k)},u_q\},\Bc^{\prime}}=C_{\Jc^{\prime}\cup\{u_{g(d_k)},u_q\},\Bc^{\prime}\cup \{d_{u_q}\}}$, which is transmitted in Step~$g(d_k)$ of the first sub-phase. Hence, in the following, we consider that  there exists some user in $\Jc^{\prime}$ whose demand is in $[\Nsf]\setminus (\{d_{k},d_{u_q} \}\cup \Bc^{\prime})$.

For the case $t=2$, we have $|\Jc^{\prime}|=1$ and we assume user $k^{\prime}$ is in $\Jc^{\prime}$. For the case $\rsf=\Nsf-1$, we have $|\Bc^{\prime}|=\rsf-2=\Nsf-3$.

 Hence, when  $\rsf=\Nsf-1$ or $t=2$, all  interferences  in  $C_{\Jc^{\prime}\cup\{u_{g(d_k)},u_q\},\Bc^{\prime}}$ to user $k$ whose demands $F_{d_{u_{g(d_k)}}}$,  are from one block (assuming this block is $W_{\Bc^{\prime}\cup \{i\}}$, where $i=d_{k^{\prime}}$ for $t=2$, and $i=[\Nsf]\setminus (\Bc^{\prime}\cup\{d_{k},d_{u_q}\})$, with a slight abuse of notation). The sum of the interferences in $C_{\Jc^{\prime}\cup\{u_{g(d_k)},u_q\},\Bc^{\prime}}$ is 
\begin{align}
 I=\underset{k_2\in \Jc^{\prime}\cup \{u_q\}:d_{k_2}\neq d_k}{\oplus } W_{\Bc^{\prime}\cup\{d_{u_q},i\},\Jc^{\prime}\cup\{u_{g(d_k)},u_q\} \setminus \{k_2\}}.\label{eq:interferences}
\end{align}
\begin{itemize}
\item if $i\notin \{d_{u_1},\ldots,d_{u_{q-1}}\}$, we can see that 
\begin{align}
C_{\Jc^{\prime}\cup\{u_{g(d_k)},u_q\},\Bc^{\prime}}=C_{\Jc^{\prime}\cup\{u_{g(d_k)},u_q\},\Bc^{\prime}\cup\{d_{u_q}\}} \oplus \underset{k_3\in \Jc^{\prime}\cup \{u_{g(d_k)}\}:d_{k_3}\neq d_{u_q}}{\oplus } W_{\Bc^{\prime}\cup\{d_{k},i\},\Jc^{\prime}\cup\{u_{g(d_k)},u_q\} \setminus \{k_3\}}.\label{eq:i notin u1}
\end{align}
 In~\eqref{eq:i notin u1}, $C_{\Jc^{\prime}\cup\{u_{g(d_k)},u_q\},\Bc^{\prime}\cup\{d_{u_q}\}}$ is transmitted in Step~$g(d_k)$ of the first sub-phase.
 \begin{itemize}
 \item  If $k_3\neq u_{g(d_k)}$, the sub-block  $ W_{\Bc^{\prime}\cup\{d_{k},i\},\Jc^{\prime}\cup\{u_{g(d_k)},u_q\} \setminus \{k_3\}}$ is desired by user $k$ and cached by user $u_{g(d_k)}$. Thus by Lemma~\ref{lem:Jc}.Item~\ref{item:lem JC2} , user $k$ can recover this sub-block from the transmission of the first sub-phase;
 \item  if $k_3= u_{g(d_k)}$, $W_{\Bc^{\prime}\cup\{d_{k},i\},\Jc^{\prime}\cup\{u_q\}}$ can be recovered by user $k$ from $C_{\Jc^{\prime}\cup\{u_{g(d_k)},u_q,k\} \setminus \{k_3\},\Bc^{\prime}\cup\{i\}}$ transmitted in Step~$q$ of the first sub-phase, where in $C_{\Jc^{\prime}\cup\{u_q,k\} ,\Bc^{\prime}\cup\{i\}}$ user $k$ caches all except   $W_{\Bc^{\prime}\cup\{d_{k},i\},\Jc^{\prime}\cup\{u_q\}}$ such that it can recover this sub-block.
 \end{itemize}
Hence, user $k$ can reconstruct $C_{\Jc^{\prime}\cup\{u_{g(d_k)},u_q\},\Bc^{\prime}}$ from the transmission of the first sub-phase;
 \item  if $i\in \{d_{u_1},\ldots,d_{u_{g(d_k)-1}}\}$, by Lemma~\ref{lem:Jc}.Item~\ref{item:lem JC3} , we can see that each sub-block $ W_{\Bc^{\prime}\cup\{d_{k},i\},\Jc^{\prime}\cup\{u_{g(d_k)},u_q\} \setminus \{k_3\}}$ in~\eqref{eq:i notin u1} is from  $ W_{\Bc^{\prime}\cup\{d_{k},i\}}$, which can be recovered by user $k$ from the first sub-phase. Hence, user $k$ can reconstruct $C_{\Jc^{\prime}\cup\{u_{g(d_k)},u_q\},\Bc^{\prime}}$ from the transmission of the first sub-phase;
 \item  if  $i\in \{d_{g(d_k)+1},\ldots,d_{u_{q}}\}$, for each user $k_4\in  \Jc^{\prime}\cup \{u_q\}$ where $d_{k_4}\neq d_k$, we focus on $C_{\Jc^{\prime}\cup\{u_{g(d_k)},u_q,u_{g(i)}\}\setminus \{k_4\},\Bc^{\prime}\cup \{d_{u_q}\}}$ which is transmitted in Step~$ g(d_k) $ of the first sub-phase. In $C_{\Jc^{\prime}\cup\{u_{g(d_k)},u_q,u_{g(i)}\}\setminus \{k_4\},\Bc^{\prime}\cup \{d_{u_q}\}}$, since we have $|\Jc^{\prime}|=1$ or $|\Bc^{\prime}|=\Nsf-3$, it can be seen that all sub-blocks are from either $W_{\Bc^{\prime}\cup \{d_{u_q},d_{k}\}}$ or $W_{\Bc^{\prime}\cup \{d_{u_q},i\}}$,  and cached by  either cached by user $u_{g(d_k)}$ or by user $u_{g(i)}$. 
 
 By  Lemma~\ref{lem:Jc}.Item~\ref{item:lem JC2} , user $k $ can recover the desired sub-block cached by user  $u_{g(d_k)}$ from the first sub-phase.
 % If $i\in \{d_{u_1},\ldots,d_{u_{g(d_k)-1)}}\}$,  by  Lemma~\ref{lem:Jc}.Item~\ref{item:lem JC2} , user $k $ can recover the desired sub-block cached by user  $u_{g(i)}$ from the first sub-phase; 
 Each   sub-block of  $W_{\Bc^{\prime}\cup \{d_{u_q},d_{k}\}}$  cached by user  $u_{g(i)}$ and not by $u_{g(d_k)}$ (assumed to be $W_{\Bc^{\prime}\cup \{d_{u_q},d_{k}\},\Vc^{\prime}}$ ), can be recovered by user $k$ from $C_{\Vc^{\prime}\cup\{k\}, \Bc^{\prime}\cup \{d_{u_q} \}}$ (transmitted in Step~$g(i)$ of the first sub-phase), because all sub-blocks in  $C_{\Vc^{\prime}\cup\{k\}, \Bc^{\prime}\cup \{d_{u_q} \}}$ except  $W_{\Bc^{\prime}\cup \{d_{u_q},d_{k}\},\Vc^{\prime}}$ are cached by user $k$. Hence, in  $C_{\Jc^{\prime}\cup\{u_{g(d_k)},u_q,u_{g(i)}\}\setminus \{k_4\},\Bc^{\prime}\cup \{d_{u_q}\}}$, user $k$ can recover all sub-blocks  of $W_{\Bc^{\prime}\cup \{d_{u_q},d_{k}\}}$. 
 
 So user $k$ can recover the   sum of the sub-blocks of  $W_{\Bc^{\prime}\cup \{d_{u_q},i\}}$ in $C_{\Jc^{\prime}\cup\{u_{g(d_k)},u_q,u_{g(i)}\}\setminus \{k_4\},\Bc^{\prime}\cup \{d_{u_q}\}}$ from the transmission of the first sub-phase,  
 \begin{align}
 I(k_4)=\underset{k_5\in \Jc^{\prime}\cup\{ u_q,u_{g(i)}\}\setminus \{k_4\}:d_{k_5}\neq d_k}{\oplus } W_{\Bc^{\prime}\cup\{d_{u_q},i\},\Jc^{\prime}\cup\{u_{g(d_k)},u_q,u_{g(i)}\}\setminus \{k_4,k_5\}}.
 \end{align}

By the similar proof as~\eqref{eq:equivalent1} and~\eqref{eq:equivalent2}, we can prove that  
\begin{align}
I\oplus \underset{k_4\in  \Jc^{\prime}\cup \{u_q\}:d_{k_4}\neq d_k}{\oplus } I(k_4)=0,\label{eq:I sum}
\end{align}
from the fact that each sub-block in~\eqref{eq:I sum} appears twice in~\eqref{eq:I sum}. Hence, user $k$ can recover $I$ from the transmission of the first sub-phase. In addition, by the definition, we have 
\begin{align}
C_{\Jc^{\prime}\cup\{u_{g(d_k)},u_q\},\Bc^{\prime}}=C_{\Jc^{\prime}\cup\{u_{g(d_k)},u_q\},\Bc^{\prime}\cup\{i\}} \oplus C_{\Jc^{\prime}\cup\{u_{g(d_k)},u_q\},\Bc^{\prime}\cup \{d_{u_q}\}} \oplus I,\label{eq:last case i}
\end{align}
 where $C_{\Jc^{\prime}\cup\{u_{g(d_k)},u_q\},\Bc^{\prime}\cup\{i\}}$ and $C_{\Jc^{\prime}\cup\{u_{g(d_k)},u_q\},\Bc^{\prime}\cup \{d_{u_q}\}}$ are transmitted in Step~$g(d_k)$ of the first sub-phase. Hence, user $k$ can reconstruct $C_{\Jc^{\prime}\cup\{u_{g(d_k)},u_q\},\Bc^{\prime}}$ from the transmission of the first sub-phase.
\end{itemize}
 In conclusion,  we prove that from the transmission of the first sub-phase, user $k$ can reconstruct $C_{\Jc^{\prime}\cup\{u_{g(d_k)},u_q\},\Bc^{\prime}}$, for each integer $q\in [g(d_k)+1: \min\{\Nsf-\rsf+2,\Ksf-t+1,N_{\textup{e}}(\dv)\}]$,
 each  $\Jc^{\prime}\subseteq ([\Ksf]\setminus\{u_1,\ldots,u_{q}\})$ where $|\Jc^{\prime}|=t-1$, and each $\Bc^{\prime}\subseteq ([\Nsf]\setminus \{d_{u_1},\ldots,d_{u_{q}}\})$, where $|\Bc^{\prime}|=\rsf-2$ and $\Bc^{\prime}\cap \Nc([\Ksf])\neq \emptyset$. 
 
 Hence, from Observation~\ref{item:obs2}, user $k$ can recover $W_{\Sc_2,\Vc_2}$ where $\Nc(\Vc_2)  \cap \{d_{u_1},\ldots,d_{u_{g(d_k)-1}}\}=\emptyset$, from the transmission of the first sub-phase.

\subsection{ $\Nc(\Vc_2) \cap \{d_{u_1},\ldots,d_{u_{g(d_k)-1}}\}\neq \emptyset$}
\label{sub:r=N-1 t=2 case 2}
For the case $t=2$, since $\Nc(\Vc_2) \cap (\Sc_2\setminus \{d_k\})\neq \emptyset$ and $\Nc(\Vc_2) \cap \{d_{u_1},\ldots,d_{u_{g(d_k)-1}}\}\neq \emptyset$  where $\{d_{u_1},\ldots,d_{u_{g(d_k)-1}}\} \cap \Sc_2=\emptyset$ and $|\Vc_2|=2$, it can be seen that  $|\Nc(\Vc_2) \setminus \Sc_2|=1$ and $d_{k} \notin (\Nc(\Vc_2) \setminus \Sc_2)$. For the case $\rsf=\Nsf-1$, since  $\Nc(\Vc_2)\cap \{d_{u_1},\ldots,d_{u_{g(d_k)-1}}\}\neq \emptyset$ where $\{d_{u_1},\ldots,d_{u_{g(d_k)-1}}\} \cap \Sc_2=\emptyset$ and $|\Sc_2|=\rsf=\Nsf-1$, it can also be seen that  $|\Nc(\Vc_2) \setminus \Sc_2|=1$ and $d_{k} \notin (\Nc(\Vc_2) \setminus \Sc_2)$. 

Hence,  when $t=2$ or $\rsf=\Nsf-1$, 
  the interferences in $C_{\Vc_2\cup \{u_{g(d_k)}\}, \Sc_2\setminus \{d_k\}}$  (transmitted in Step~$g(d_k)$ of the first sub-phase) to user $k$ are all from the block $W_{\Sc_2\cup \{i\}\setminus \{d_k\}}$, where $i$ is the element in  $\Nc(\Vc_2) \setminus \Sc_2$. The sum of the interferences in $C_{\Vc_2\cup \{u_{g(d_k)}\}, \Sc_2\setminus \{d_k\}}$ is 
\begin{align}
 I^{\prime}=\underset{k_2\in \Vc_2:d_{k_2}\neq d_k}{\oplus } W_{\Sc_2\cup \{i\}\setminus \{d_k\},\Vc_2\cup \{u_{g(d_k)}\}\setminus\{k_2\}  }.\label{eq:interferences case 2}
\end{align}

For each user $k_4\in \Vc_2 $  where $d_{k_4}\neq d_k$, we focus on $C_{\Vc_2\setminus \{k_4\}\cup \{u_{g(d_k)},u_{g(i)}\}, \Sc_2\setminus \{d_k\}}$ which is transmitted in Step~$i$ of the first sub-phase. In  $C_{\Vc_2\setminus \{k_4\}\cup \{u_{g(d_k)},u_{g(i)}\}, \Sc_2\setminus \{d_k\}}$,  since $|\Vc_2|=2$ or $| \Sc_2|=\Nsf-1$, it can be seen that
all sub-blocks are from either $W_{ \Sc_2}$ or $W_{ \Sc_2\setminus \{d_k\} \cup\{i\}}$. Each sub-block from $W_{ \Sc_2}$ is either cached by user $u_{g(d_k)}$ or by user $u_{g(i)}$, which can be recovered by user $k $ from the first sub-phase, by Lemma~\ref{lem:Jc}.Item~\ref{item:lem JC2} . Hence, user $k$ can recover the sum of the sub-blocks of $W_{ \Sc_2\setminus \{d_k\} \cup\{i\}}$ in $C_{\Vc_2\setminus \{k_4\}\cup \{u_{g(d_k)},u_{g(i)}\}, \Sc_2\setminus \{d_k\}}$   as follows,
\begin{align}
I^{\prime}(k_4)=\underset{k_2\in\Vc_2\setminus \{k_4\}\cup \{u_{g(i)}\}:d_{k_2}\neq d_k}{\oplus } W_{ \Sc_2\setminus \{d_k\} \cup\{i\},\Vc_2\setminus \{k_4,k_2\}\cup \{u_{g(d_k)},u_{g(i)}\} }.\label{eq:interferences Iprime k4 case 2}
\end{align}

By the similar proof as~\eqref{eq:equivalent1} and~\eqref{eq:equivalent2}, we can prove that  
\begin{align}
I^{\prime}\oplus \underset{k_4\in \Vc_2 :d_{k_4}\neq d_k}{\oplus } I^{\prime}(k_4)=0,\label{eq:I prime sum}
\end{align}
from the fact that each sub-block in~\eqref{eq:I prime sum} appears twice in~\eqref{eq:I prime sum}. Hence, user $k$ can reconstruct the sum of all interferences  $I^{\prime}$  in $C_{\Vc_2\cup \{u_{g(d_k)}\}, \Sc_2\setminus \{d_k\}}$. Other  sub-blocks in  $C_{\Vc_2\cup \{u_{g(d_k)}\}, \Sc_2\setminus \{d_k\}}$ are from the block $W_{\Sc_2}$  which is  desired by user $k$. In addition, all   these sub-blocks are cached by user $u_{g(d_k)}$ except $W_{\Sc_2,\Vc_2}$. By Lemma~\ref{lem:Jc}.Item~\ref{item:lem JC2} , from the transmission of first sub-phase user $k$ 
  can recover  the sub-blocks of $W_{\Sc_2 }$ which are cached by user   $u_{g(d_k)}$. Hence, user $k$ can also recover  $W_{\Sc_2,\Vc_2}$ in the first sub-phase.

\section{Codes for Extension to Caching with Multiple Requests}
\label{sec:codes for extension}
For the caching problem with multiple requests considered in~\cite{multireqWei2017} where each user demands $\Lsf$ uncorrelated and equal-length files,  the proposed delivery scheme  in~\cite{multireqWei2017} was proved to be optimal under the constraint of the MAN placement for most demands with $\Ksf \leq 4$, $\Msf=\Nsf/\Ksf$, and $\Lsf=2$, except one demand for  $\Ksf=3$ and three demands for $\Ksf=4$. 
Different from the considered problem in this paper, the demands are not generally symmetric for the caching problem with multiple requests. Hence, for the caching problem with multiple requests, we pick a set of leaders such that each leader has at least one specific demanded file which is not demanded by other leaders, and the union set of demanded files by the leaders should be equal to   the union set of demanded files by all users. In addition, the number of leaders should be as small as possible. We can then extend the proposed scheme for $t=1$ in order to  achieve the optimality for those four exceptional demands, by satisfying the demands of leaders subsequently and aligning the interferences to non-leaders simultaneously.

\begin{enumerate}
\item $d_1=\{F_1,F_2\}$, $d_2=\{F_1,F_3\}$, and $d_3=\{F_2,F_3\}$ (case $D_7$ in~\cite{multireqWei2017}). We use the MAN placement and divide each file $F_i$ where $i\in[\Nsf]$ into $\binom{\Ksf}{t}$ non-overlapping and equal-length subfiles, $F_{i}=\{ F_{i,\Wc}:\Wc\subseteq [\Ksf],|\Wc|=t\}$, where $t=\Ksf\Msf/\Nsf=1$. It can be seen this case is equivalent to our considered $(\Nsf,\Ksf,\Msf,\rsf)=(3,3,1,2)$ shared-link caching problem with correlated files. Hence, we can directly use  the proposed delivery phase in this paper to transmit the linear combinations (with leader permutation $(1,2)$)
 \begin{align*}
  \text{Step~$1$: } & F_{1,\{2\}}\oplus F_{1,\{1\}}, \ F_{1,\{3\}}\oplus F_{3,\{1\}}, \ F_{2,\{2\}}\oplus F_{3,\{1\}}, \ F_{2,\{3\}}\oplus F_{2,\{1\}};\\
   \text{Step~$2$: } & F_{3,\{3\}}\oplus F_{3,\{2\}}.
 \end{align*}
Hence, the load is $5/3$ which coincides with the converse bound under the constraint of MAN placement in~\cite{multireqWei2017}, while the proposed caching scheme in~\cite{multireqWei2017} achieves $2$.
\item $d_1=\{F_1,F_2\}$, $d_2=\{F_1,F_3\}$, $d_3=\{F_2,F_3\}$, and $d_4=\{F_4,F_5\}$ (case $D^{\prime}_{15}$ in~\cite{multireqWei2017}). It can be seen that 
if we only focus on the demands of users $1,2,3$, it is equivalent to our considered $(\Nsf,\Ksf,\Msf,\rsf)=(3,3,1,2)$ shared-link caching problem with correlated files. In addition, the demanded file by user $4$ are independent to any demanded file by   users $1,2,3$.
Hence, we first satisfy the demands of user $4$ and then use the codes for  our considered $(\Nsf,\Ksf,\Msf,\rsf)=(3,3,1,2)$ shared-link caching problem with correlated files. Thus we transmit (with leader permutation $(4,1,2)$)
 \begin{align*}
  \text{Step~$1$: } &F_{4,\{1\}}\oplus F_{1,\{4\}}, \ F_{4,\{2\}}\oplus F_{1,\{4\}}, \ F_{4,\{3\}}\oplus F_{3,\{4\}},\\
     &F_{5,\{1\}}\oplus F_{2,\{4\}}, \ F_{5,\{2\}}\oplus F_{3,\{4\}}, \ F_{5,\{3\}}\oplus F_{2,\{4\}};\\
     \text{Step~$2$: } &F_{1,\{2\}}\oplus F_{1,\{1\}}, \ F_{1,\{3\}}\oplus F_{3,\{1\}}, \ F_{2,\{2\}}\oplus F_{3,\{1\}}, \ F_{2,\{3\}}\oplus F_{2,\{1\}};\\
     \text{Step~$3$: } & F_{3,\{3\}}\oplus F_{3,\{2\}}.
 \end{align*}
Hence, the load is $11/4$ which coincides with the converse bound under the constraint of MAN placement in~\cite{multireqWei2017}, while the proposed caching scheme in~\cite{multireqWei2017} achieves $3$.
\item $d_1=\{F_1,F_2\}$, $d_2=\{F_1,F_3\}$, $d_3=\{F_1,F_4\}$, and $d_4=\{F_2,F_3\}$ (case $D^{\prime}_{17}$ in~\cite{multireqWei2017}). 
We choose the leader set as $\{3,4\}$ and the chose permutation is $(3,4)$. Inspired from proposed scheme for $t=1$, the delivery contains two steps where in the first and second steps, we satisfy the demands of users $3$ and $4$, respectively. 

In Step~$1$, we first let user $3$ recover $F_1$. For each user $k \in \{1,2,4\}$, if $F_1$ is demanded by user $k$, we transmit $F_{1,\{k\}} \oplus F_{1,\{3\}}$; otherwise, we pick one demanded file by user $k$ which is not $F_4$ (assumed to be $F_i$), and transmit $F_{1,\{k\}} \oplus F_{i,\{3\}}$. 

We then let user $3$ recover $F_4$. For each user $k \in \{1,2,4\}$, if $F_4$ is demanded by user $k$, we transmit $F_{4,\{k\}} \oplus F_{4,\{3\}}$; otherwise, we pick one demanded file by user $k$ which is not $F_i$ nor $F_1$ (assumed to be $F_{i^{\prime}}$), and transmit $F_{4,\{k\}} \oplus F_{i^{\prime},\{3\}}$.

By this way,  we transmit in the the steps  (with leader permutation $(3,4)$)
 \begin{align*}
  \text{Step~$1$: } &F_{1,\{1\}}\oplus F_{1,\{3\}}, \ F_{1,\{2\}}\oplus F_{1,\{3\}}, \ F_{1,\{4\}}\oplus F_{3,\{3\}},\\
&F_{4,\{1\}}\oplus F_{2,\{3\}}, \ F_{4,\{2\}}\oplus F_{3,\{3\}}, \ F_{4,\{4\}}\oplus F_{2,\{3\}};\\
     \text{Step~$2$: } &F_{3,\{1\}}\oplus F_{1,\{4\}}, \ F_{3,\{2\}}\oplus F_{3,\{4\}}, \ F_{2,\{1\}}\oplus F_{2,\{4\}},\ F_{2,\{2\}}\oplus F_{1,\{4\}}.
 \end{align*}
Hence, the load is $10/4$ which coincides with the converse bound under the constraint of MAN placement in~\cite{multireqWei2017}, while the proposed caching scheme in~\cite{multireqWei2017} achieves $11/4$.
\item $d_1=\{F_1,F_2\}$, $d_2=\{F_1,F_2\}$, $d_3=\{F_1,F_3\}$, and $d_4=\{F_2,F_3\}$ (case $D^{\prime}_{20}$ in~\cite{multireqWei2017}). 
It can be seen this case is equivalent to our considered $(\Nsf,\Ksf,\Msf,\rsf)=(3,4,1,2)$ shared-link caching problem with correlated files. Hence, we can directly use  the proposed delivery phase in this paper to transmit the linear combinations (with leader permutation $(1,3)$)
 \begin{align*}
  \text{Step~$1$: } &F_{1,\{2\}}\oplus F_{1,\{1\}}, \ F_{1,\{3\}}\oplus F_{1,\{1\}}, \ F_{1,\{4\}}\oplus F_{3,\{1\}},\\
&F_{2,\{2\}}\oplus F_{2,\{1\}}, \ F_{2,\{3\}}\oplus F_{3,\{1\}}, \ F_{2,\{4\}}\oplus F_{2,\{1\}};\\
     \text{Step~$2$: } &F_{3,\{2\}}\oplus F_{2,\{3\}}, \ F_{3,\{4\}}\oplus F_{3,\{3\}}.
 \end{align*}
Hence, the load is $2$, which coincides with the converse bound under the constraint of MAN placement in~\cite{multireqWei2017}, while the proposed caching scheme in~\cite{multireqWei2017} achieves $9/4$.

\end{enumerate}

\bibliographystyle{IEEEtran}
\bibliography{IEEEabrv,IEEEexample}

%\begin{IEEEbiographynophoto}{Coauthor}
%Same again for the co-author, but without photo\end{IEEEbiographynophoto}

\end{document}

%% file: macros.tex
\long\def\comment#1{}

\newcommand{\dv}{{\mathbf d}}

\newcommand{\tv}{{\mathbf t}}
\newcommand{\uv}{{\mathbf u}}

\newcommand{\Zm}{{\mathbf Z}}

\newcommand{\Bc}{{\mathcal B}}

\newcommand{\Dc}{{\mathcal D}}

\newcommand{\Hc}{{\mathcal H}}

\newcommand{\Jc}{{\mathcal J}}

\newcommand{\Lc}{{\mathcal L}}

\newcommand{\Nc}{{\mathcal N}}

\newcommand{\Qc}{{\mathcal Q}}
\newcommand{\Rc}{{\mathcal R}}
\newcommand{\Sc}{{\mathcal S}}
\newcommand{\Tc}{{\mathcal T}}

\newcommand{\Wc}{{\mathcal W}}
\newcommand{\Vc}{{\mathcal V}}

\newcommand{\psf}{{\mathsf p}}

\newcommand{\rsf}{{\mathsf r}}

\newcommand{\usf}{{\mathsf u}}

\newcommand{\Bsf}{{\mathsf B}}

\newcommand{\Ksf}{{\mathsf K}}
\newcommand{\Lsf}{{\mathsf L}}
\newcommand{\Msf}{{\mathsf M}}
\newcommand{\Nsf}{{\mathsf N}}

\newcommand{\Rsf}{{\mathsf R}}

\newtheorem{thm}{Theorem}%[section]
\newtheorem{cor}{Corollary}
\newtheorem{lem}{Lemma}

\newtheorem{rem}{Remark}

\providecommand{\definitionname}{Definition}